\documentclass[fleqn,usenatbib]{mnras}

\usepackage{newtxtext,newtxmath}

\usepackage[T1]{fontenc}
\usepackage{ae,aecompl}

\usepackage{graphicx}	
\usepackage{amsmath}	
\usepackage{amssymb}	
\title[]{Spatial Distribution of Metals in the ICM: Evolution of the Iron Excess in Relaxed Galaxy Clusters}
\author[A. Liu et al.]{
Ang Liu,$^{1,2,3}$\thanks{E-mail: liuang@arcetri.astro.it (A. Liu)}
Paolo Tozzi,$^{1}$\thanks{E-mail: ptozzi@arcetri.astro.it (P. Tozzi)}
Heng Yu$^4$, Sabrina De Grandi$^5$, Stefano Ettori$^{6,7}$
\\
$^{1}$INAF - Osservatorio Astrofisico di Arcetri, Largo E. Fermi, I-50122 Firenze, Italy\\
$^{2}$Department of Physics, Sapienza University of Rome, I-00185 Rome, Italy\\
$^{3}$Department of Physics, University of Rome Tor Vergata, I-00133, Rome, Italy\\
$^{4}$Department of Astronomy, Beijing Normal University, Beijing, China 100875\\
$^{5}$INAF - Osservatorio Astronomico di Brera, Via E. Bianchi, 46, I-23807 Merate (LC), Italy\\
$^{6}$INAF - Osservatorio di Astrofisica e Scienza dello Spazio, via Pietro Gobetti 93/3, 40129 Bologna, Italy\\
$^{7}$INFN, Sezione di Bologna, viale Berti Pichat 6/2, I-40127 Bologna, Italy
}

\date{Accepted XXX. Received YYY; in original form ZZZ}

\pubyear{2018}

\begin{document}


\maketitle

\begin{abstract}
We investigate the spatial distribution of iron in the intra-cluster medium in a selected sample of 41 relaxed clusters in
the redshift range $0.05 < z < 1.03$ using {\sl Chandra} archival data. We compute the azimuthally-averaged, deprojected $Z_{\rm Fe}$ profile of each cluster out to $\sim 0.4r_{500}$, and identify a peak in the distribution of iron followed by a flatter distribution at larger radii.  Due to the steep gradient both in gas density and abundance, we find that the emission-weighted iron abundance within $0.2r_{500}$, which entirely includes the iron peak in most of the cases, is on average $\sim$25\% higher than the mass-weighted value, showing that spatially resolved analysis and accurate deprojection are key to study the evolution of iron enrichment in the very central regions of cool core clusters. We quantify the extent of the iron distribution in each cluster with a normalized scale parameter $r_{\rm Fe}$, defined as the radius where the iron abundance excess is half of its peak value. We find that $r_{\rm Fe}$ increases by a factor of $\sim 3$ from $z\sim 1$ to $z\sim 0.1$, suggesting that the spatial distribution of iron in the ICM extends with time, possibly due to the mixing with the mechanical-mode feedback from the central galaxy. We also find that the iron mass excess within $0.3r_{500}$, when normalized to the total baryonic mass within the same region, does not evolve significantly, showing that this iron mass component is already established at $z\sim 1$.

\end{abstract}



\section{Introduction}

Clusters of galaxies are the largest gravitationally-bound systems in the universe.  Growing through gravitational instability
from the fluctuations in the primordial density field \citep[see][and references therein]{kravtsov2012}, not only can their abundance
be used to trace the cosmic evolution thanks to a strong dependence on the cosmological parameters
\citep{allen2011,Mantz_2016b}, but they can also be considered approximately as closed boxes which retain the imprints of the
evolution of the member galaxies and the surrounding intracluster medium \citep[ICM, see, e.g.,][]{2010_Bohringer}.
The X-ray emitting ICM is by far the dominant baryonic component across the full range of virial masses
of galaxy groups and clusters \citep{2012Lin}, and is constituted of a hot, diffuse, optically-thin plasma in collisional equilibrium,
enriched by heavy elements produced mainly by supernovae (SNe) explosions.
The abundance of heavy elements can be directly measured through the equivalent width of the corresponding emission lines
in the X-ray energy range, thanks to the assumption of collisional equilibrium and the low optical depth of the ICM. Self-absorption of the most prominent resonant lines has been shown to be an important diagnostic in high S/N X-ray spectra, but it has noticeable effect on the line emission only in the high-density central regions, as recently measured in the Hitomi observation of the innermost
$\sim 30$ kpc of the Perseus cluster  for the $K_\alpha$ line of He-like iron \citep{Hitomi_RS_2017}.  Despite the
simple framework outlined here, the relative abundance of the different heavy elements, the cosmic evolution of their abundance
and of their spatial distribution are shaped by complex physics which are not fully understood yet.

In the most massive, hence most luminous clusters, given the high temperatures reached in the ICM
($kT > 3$ keV), most of the heavy elements are highly or fully ionized.  The most prominent feature
from heavy elements is the $K_\alpha$ emission line complex from He-like and H-like iron at 6.7--6.9
keV rest-frame.  The detection of other elements is much easier at lower
temperatures ($kT < 3 $ keV) and low redshifts, and typically requires high S/N spectra
\citep[see][]{degrandi2009,tamura2009,sanders2011,mernier2017}. For these reasons, iron is the only heavy element that can be
detected in galaxy clusters up to $z \sim 1.6$ \citep{rosati2009,tozzi2013,2015Tozzi}, with some tentative detection at $z \sim 2$
\citep[see][where iron emission is detected at $2.6 \sigma$ confidence level]{2017Mantz}. Therefore, several studies in the last
ten years focused on the cosmic evolution of the global iron abundance. Previous studies of evolution in the metallicity of the ICM were consistent with a significant evolution of a factor of $\sim 2$
in the range $0<z<1.3$, at least in {\sl Chandra} data
\citep[e.g.][]{2007Balestra,maughan2008,anderson2009}, although more recent results  are consistent with little or no evolution \citep{2015Ettori,mcdonald2016}.

The spatial distribution of heavy elements, typically azimuthally averaged at each radius, became an important aspect to be taken into account in evolutionary studies,
given that the relative abundance can strongly vary with radius.  The strongest gradients are observed in the cool cores, defined as the central cluster regions where
the gas cooling time is shorter than a given reference value \citep[for an extensive discussion of the definition of cool core see][]{hudson2010}.  Ideally, one would aim at
resolving the chemical evolution as a function of the radius.
Recently, \citet{2017Mantz} found negative evolution with redshift in the intermediate regions
($0.1r_{500} < r < 0.5r_{500}$), while at smaller ($r < 0.1r_{500}$) and larger radii
 ($r > 0.5r_{500}$), the ICM is consistent with a constant iron abundance at least up to $z\sim 1.2$. On the other hand, systematic investigations of the
spatially resolved evolution of iron abundance in the ICM with XMM-{\sl Newton} data support the general trend of a negative evolution
with redshift in the central regions of cool-core clusters, and an almost constant behavior at radii larger than
$0.4r_{500}$ \citep{2015Ettori}. Therefore, we can conclude that studies dominated by X-ray selected clusters, as those
previously mentioned, agree on a general early ($z>1.5$) enrichment, and on the presence of some moderate evolution in the iron
abundance, but do not agree on the radial range where this evolution is taking place.  In addition, the same presence of evolution
of ICM enrichment in the cluster population has been recently challenged by studies based on an SPT-SZ selected sample at
$z < 1.5$, where no significant evolution has been found \citep{mcdonald2016}.  This result may be at least partially reconciled
with the previous results on the basis of a much lower fraction of cool cores, which are more enriched in metals, among SZ-
selected clusters \citep{rossetti2017}. However, this hypothesis still needs to be tested across a combined cluster sample free of
any selection bias.

To summarize, there are significant differences in several relevant aspects among recent studies, and we are far
from having a coherent description of the cosmic evolution of iron in the ICM.  Not only we do not know which is the radial range where most of the iron evolution takes place,
but we do not even know whether some amount of evolution does actually take place, nor whether it depends on the cluster selection itself.  A general lesson from the literature, is that it is
mandatory to follow the spatial distribution of the iron abundance as a function of redshift, halo mass, and thermodynamical
properties of the ICM to successfully constrain a physical model of its chemical enrichment.

This goal can be barely achieved on the basis of present-day data and X-ray facilities.
Currently, due to the limited statistics of high-$z$ cluster samples, the photon-starved X-ray
follow-up observation of
high-$z$ clusters, and the frustrating perspective of X-ray astronomy (with {\sl Chandra}, the only
high-resolution instrument,
rapidly losing efficiency in the soft band), it is a hard task to improve the measurement of the
evolution of the iron
abundance in the ICM.  Despite this, the cosmic evolution of heavy element enrichment of
the ICM across the cosmic epochs is gaining increasing interest. In particular, the relative ratio of
the abundance of various ions
provides important clues on the ratio of type Ia supernovae (SNIa) and core-collapse supernovae
(SNcc), which eject different amount of heavy elements \citep{werner2008,madau2014,maoz2017}. Thus
the evolution of ICM enrichment can be used, in principle, to
constrain the SNe rates in cluster galaxies, once absolute yields are robustly constrained.
In turn, the prediction of absolute yields constitutes a key aspect which is still highly debated
\citep{finoguenov2000,bohringer2004,degrandi2009,matsushita2013}, but is far beyond the goal of this paper.

For these reasons we adopt an approach that begins by exploiting nearby and bright clusters, where we can successfully constrain the distribution of the heavy elements
as a function of the cluster radius, and eventually extend our investigation to higher-z targets.
The general properties of the iron distribution at low redshifts are well known.
\citet{degrandi2001} and \citet{degrandi2004} investigated the projected iron profiles for a
sample of 17 low-$z$ clusters observed by {\sl BeppoSAX}, and clearly showed that non-cool-core
clusters had flat iron profiles while cool-core clusters show a strong iron enhancement towards the
center.  This property is now commonly observed in all the regular/relaxed clusters, suggesting a
physical link between the processes that shape the thermodynamics of the ICM and its chemical
enrichment. In particular, in high S/N data, it is possible to identify a well defined peak in the
iron distribution above the average abundance level, which allows one to measure a relative {\sl
excess} of iron with respect to the global iron distribution.  This excess may be associated with
relatively recent star formation events in the brightest cluster galaxy \citep[BCG, see][]
{degrandi2004}, but its origin and evolution are still unclear. The
shape of the iron excess is clearly very sensitive to the many complex physical processes occurring
in the center of galaxy clusters, such as gas motions driven by outflows of the central AGN
\citep{sijacki2007,roediger2007,fabian2012}, the sloshing of cool cores
\citep{markevitch2007,ghizzardi2014}, stochastic gas motions \citep{rebusco2005}, sinking of highly
enriched low-entropy gas \citep{cora2006,cora2008}, and galactic winds
\citep{tornatore2004,romeo2006}. In \citet{degrandi2014}, our group has already shown that, at least
in the case of WARPJ1415.1+3612 (the brightest cool-core cluster at $z\sim 1$), the peak in the iron
distribution is significantly narrower than in local clusters, when compared to the stellar light
distribution of the underlying BCG.

We argue that in this respect, the exquisite angular resolution of {\sl Chandra} may play a key role
in providing well defined abundance profiles not only at low redshift. Therefore, in this work, we
start a systematic investigation of the spatial distribution of iron abundance in the ICM at
different epochs (up to $z \sim 1$) with a limited but ideal sample of relaxed clusters observed
with {\sl Chandra}. Massive, relaxed clusters constitute the best targets where we can attempt to
disentangle different components in the iron distribution.  Our goal here is to extend the few works
currently available in the literature
\citep{2012Baldi,2015Ettori,2017Mantz}, putting most of the emphasis on the spatial distribution and its physical implications.
Our final aim is to build a universal physical model for the iron distribution and use it to extend our study to all the clusters with X-ray detections, irrespective of their S/N.

The paper is organized as follows. In Section 2, we describe the sample selection, the dataset extracted from the {\sl Chandra}
archive, and the data reduction. In Section 3, we briefly discuss the global properties of the sample.  In Section 4,
we present the deprojected spatially-resolved analysis and derive the deprojected $Z_{\rm Fe}$  profiles. In Section 5, we discuss the correlation between pseudo entropy, gas cooling time and iron abundance. In Section 6, we discuss
our results on the central iron excess and the width of the iron peak. Our conclusions
are presented in Section 7. Throughout this paper, we adopt the seven-year WMAP cosmology
($\Omega_{\Lambda} =0.73 $, $\Omega_m =0.27$, and $H_0 = 70.4 $ km s$^{-1}$ Mpc$^{-1}$ \citep{2011Komatsu}.
Quoted error bars correspond to a 1 $\sigma$ confidence level unless noted otherwise.

\section{Sample selection and data reduction}

To achieve our science goal, the selection criteria include both the physical properties of the targets and the data quality.
In this work we do not aim at measuring the global evolution of iron as a function of cosmic epoch, and we use an optimally selected cluster sample to fully exploit the power of spatially resolved spectroscopy.

First, we set our requirements on the physics of the targets.
To constrain the shape of the iron distribution in the ICM as a function of the radial distance from the center, we are required to
select clusters for which an azimuthally averaged $Z_{\rm Fe}$ value as a function of the cluster radius is well defined,
which implies an approximate spherical symmetry and a relaxed dynamical state.  Obviously, any
major merger event makes the temperature and abundance distribution highly asymmetric and patchy, undermining any attempt to
define a meaningful radial distribution of $Z_{\rm Fe}$. Therefore, this requirement forces us to select
relaxed, round-shaped clusters classified as such on the basis of morphological information.

\begin{table*}
 \caption{The sample and data we used in this work. The coordinate is the position of the peak of X-ray emission. }
 \begin{tabular}{lccclc}
\hline
Cluster   & $z_{\rm opt}$ & RA(J2000) & Dec(J2000) & ObsID/Detector(ACIS-I/-S) & Exptime (ks)   \\
\hline
Hydra-A           & 0.055 & 09:18:05.7  & -12:05:43.9  & 575I                         & 20.9    \\
Abell2029         & 0.077 & 15:10:56.1  & +05:44:41.3  & 891S,4977S,6101I             & 106.7   \\
Abell2597         & 0.083 & 23:25:19.8  & -12:07:26.7  & 6934S,7329S                  & 110.2   \\
Abell478          & 0.088 & 04:13:25.1  & +10:27:57.2  & 1669S,6102I                  & 51.5    \\
PKS0745-191       & 0.103 & 07:47:31.3  & -19:17:39.2  & 12881S                       & 117.8   \\
RXJ1524.2-3154    & 0.103 & 15:24:12.9  & -31:54:22.5  & 9401S                        & 40.8    \\
Abell1068         & 0.138 & 10:40:44.5  & +39:57:11.5  & 1652S                        & 26.8    \\
Abell2204         & 0.152 & 16:32:46.9  & +05:34:32.0  & 499S,6104I,7940I             & 96.6    \\
Abell1204         & 0.171 & 11:13:20.5  & +17:35:40.9  & 2205I                        & 23.6    \\
Abell383          & 0.188 & 02:48:03.4  & -03:31:46.1  & 524I,2320I,2321S             & 48.7    \\
RXJ0439.0+0520    & 0.208 & 04:39:02.2  & +05:20:43.6  & 527I,9369I,9761I             & 37.9    \\
ZwCL2701          & 0.214 & 09:52:49.1  & +51:53:06.0  & 12903S                       & 95.6    \\
RXJ1504.1-0248    & 0.215 & 15:04:07.5  & -02:48:16.8  & 4935I, 5793I                 & 52.3    \\
ZwCL2089          & 0.235 & 09:00:36.8  & +20:53:39.9  & 10463S                       & 40.5    \\
RXJ2129.6+0005    & 0.235 & 21:29:39.9  & +00:05:21.0  & 552I,9370I                   & 39.4    \\
RXJ1459.4-1811    & 0.236 & 14:59:28.8  & -18:10:45.2  & 9428S                        & 39.3    \\
Abell1835         & 0.253 & 14:01:01.9  & +02:52:44.2  & 6880I,6881I,7370I,495S,496S  & 221.5   \\
Abell3444         & 0.253 & 10:23:50.2  & -27:15:23.1  & 9400S                        & 36.3    \\
MS1455.0+2232     & 0.258 & 14:57:15.1  & +22:20:34.3  & 7709I,543I,4192I             & 107.8   \\
MS2137.3-2353     & 0.313 & 21:40:15.2  & -23:39:40.2  & 928S,4974S,5250S             & 119.5   \\
MACSJ2229.7-2755  & 0.324 & 22:29:45.2  & -27:55:36.7  & 3286S,9374S                  & 30.3    \\
MACSJ0947.2+7623  & 0.345 & 09:47:12.7  & +76:23:13.9  & 7902S                        & 38.3    \\
MACSJ1931.8-2634  & 0.352 & 19:31:49.6  & -26:34:33.8  & 9382I                        & 97.6    \\
MACSJ1115.8+0129  & 0.355 & 11:15:51.9  & +01:29:55.9  & 3275I,9375I                  & 53.1    \\
RXJ1532.9+3021    & 0.362 & 15:32:53.8  & +30:20:59.3  & 14009S                       & 88.2    \\
MACSJ0011.7-1523  & 0.378 & 00:11:42.9  & -15:23:21.2  & 3261I,6105I                  & 58.8    \\
MACSJ1720.2+3536  & 0.391 & 17:20:16.8  & +35:36:25.5  & 3280I,6107I,7718I            & 60.8    \\
MACSJ0429.6-0253  & 0.399 & 04:29:36.1  & -02:53:08.4  & 3271I                        & 22.4    \\
MACSJ0159.8-0849  & 0.404 & 01:59:49.3  & -08:49:58.2  & 3265I,6106I,9376I            & 72.5    \\
MACSJ2046.0-3430  & 0.423 & 20:46:00.5  & -34:30:18.2  & 5816I,9377I                  & 49.2    \\
IRAS09104+4109    & 0.442 & 09:13:45.5  & +40:56:28.6  & 10445I                       & 75.9    \\
MACSJ0329.6-0211  & 0.450 & 03:29:41.6  & -02:11:46.8  & 3257I,3582I,6108I,7719I      & 76.1    \\
MACSJ1621.3+3810  & 0.463 & 16:21:24.8  & +38:10:08.8  & [3254I,3594I,6109I,6172I,    & 161.1   \\
                  &       &             &              & 7720I,9379I,10785I]          &         \\
3C295             & 0.464 & 14:11:20.5  & +52:12:09.9  & 2254I                        & 87.2    \\
MACSJ1423.8+2404  & 0.543 & 14:23:47.9  & +24:04:42.6  & 4195S                        & 38.9    \\
SPT-CLJ2331-5051  & 0.576 & 23:31:51.1  & -50:51:54.0  & 9333I,11738I                 & 31.8    \\
SPT-CLJ2344-4242  & 0.596 & 23:44:44.0  & -42:43:12.4  & 13401I,16135I,16545I         & 118.0   \\
SPT-CLJ0000-5748  & 0.702 & 00:01:00.0  & -57:48:33.1  & 9335I                        & 28.4    \\
SPT-CLJ2043-5035  & 0.723 & 20:43:17.6  & -50:35:32.2  & 13478I                       & 73.3    \\
                  &       &             &              & [14017I,14018I,14349I,14350I &         \\
PLCKG 266.6       & 0.940 & 06:15:51.8  & -57:46:47.3  & 14351I,14437I,15572I,15574I  & 240.6   \\
                  &       &             &              & 15579I,15582I,15588I,15589I] &         \\
CLJ1415+3612      & 1.030 & 14:15:11.1  & +36:12:02.7  & 12255S,12256S,13118S,13119S  & 276.5   \\

  \hline
 \end{tabular}
\label{sample}
\end{table*}

We start from the sample presented in \citet{mantz2015}, where the symmetry($s$)-peakiness($p$)-alignment($a$) (SPA) criterion
is used to select relaxed clusters. By applying the criterion $s > 0.87, \,\, p > -0.82$, and $a > 1.00$ to a sample of 361
clusters, they identify 57 clusters in the redshift range $0.01<z<1.03$ as relaxed \citep[see][for details]{mantz2015}.

We check the 0.5--7 keV {\sl Chandra} images of these clusters, obtained by merging all the useful observations. We exclude from
the sample the clusters which show clear signatures of non-equilibrium previously missed by the SPA test, such as obvious
substructures in X-ray surface brightness distribution (e.g., A133 and RXJ1347.5-1145). There are 52 clusters surviving these
criteria. We also include a relaxed cluster which passes the SPA criterion but is not included in \citet{mantz2015}:
PLCKG266.6-27.3 at z=0.940, a remarkable cluster with a high S/N and high redshift \citep[see][]{2017Bartalucci}.

We require a number of net counts $\geq 6000$ in the 0.5--7~keV energy band and within the extraction radius of $\sim0.4r_{500}$ to have at least 6 independent annuli with more than
$\sim$1000 net counts each.  This threshold is required to achieve typical errors on the iron abundance of the order of 30\% or less
in each ring \citep[see][]{2011Yu}.
We also require that $\sim0.4r_{500}$ (our maximum extraction radius) be included entirely within the
ACIS field of view. Depending on the position of the aimpoint, the maximum radius from the cluster center covered by ACIS is about 8 arcmin, which corresponds to $\sim$400 kpc at $z\sim 0.04$ in our seven-year WMAP cosmology. If we assume a typical $r_{500}$ of 1 Mpc, this threshold thus excludes some nearby clusters (e.g. Perseus).
There are 41 clusters left after applying the data quality threshold. Our final sample
of the 41 relaxed clusters in the redshift range $0.05<z<1.03$ is listed in Table \ref{sample}, where we show the redshift and position of X-ray emission peak of each cluster, and the details of the
{\sl Chandra} data we used in this work. The X-ray emission peak is determined as the position of the brightest pixel of the point source extracted image in 0.5--7~keV band, smoothed with a Gaussian function with FWHM = 1.5$\arcsec$.

We performed a standard data
reduction starting from the level=1 event files, using the {\tt CIAO 4.9} software package, with the most recent version (at the time) of the
{\sl Chandra} Calibration Database ({\tt CALDB 4.7.8}). When observations are taken in the VFAINT mode, we run the task {\tt
acis$\_$process$\_$events} to flag background events that are most likely associated with cosmic rays and remove them.  With
this procedure, the ACIS particle background can be significantly reduced compared to the standard grade selection.  The
data are filtered to include only the standard event grades 0, 2, 3, 4 and 6.  We checked visually for hot columns left after the
standard reduction.  For exposures taken in VFAINT mode (the large majority of our dataset), there are practically
no hot columns or flickering pixels left after filtering out bad events.  We also apply CTI correction to ACIS-I data.  We finally
filter time intervals with high background by performing a $3\sigma$ clipping of the background level using the script {\tt
analyze\_ltcrv}.  The final effective exposure times are generally very close to the original observing time.

When the concentric annuli for spectral analysis are selected, we extract the full spectrum after
masking unresolved sources, which are previously identified with {\tt wavdetect} and eventually
checked manually to identify faint sources missed by the detection algorithm due to the dominating ICM
emission. For clusters with multiple observations, we extract the spectrum and compute the response
matrix file and ancillary response file for each Obsid separately, and fit the spectra with
linked parameters.  Background spectra are extracted from a selection of regions far from the ICM
emission in each Obsid. When the ICM emission fills the entire field of view, we use the background
generated from the ``blank sky" files with the {\tt blanksky} script in {\tt CIAO}, which finds the correct blank sky files, reprojects them to match the data, and properly determines the scaling. The ``blank sky" background is used only in one case: ObsID 575 for Hydra-A.  The spectra fitting is done with {\tt Xspec 12.9.0}. The {\tt apec} thermal plasma emission model \citep{smith2001} is used to fit the ICM spectrum, with abundance
relative to the solar values of \citet{asplund2009}. C-statistics \citep{cash1979} are used in the spectra fitting. Galactic absorption is described by the model
{\tt tbabs} \citep{2000Wilms}, where the Galactic column density $NH_{Gal}$ is frozen to the value
corresponding to the cluster position in the HI survey of \citet{2005Kalberla}

\section{Global properties}

\begin{figure}
\includegraphics[width=0.49\textwidth, trim=100 240 100 240, clip]{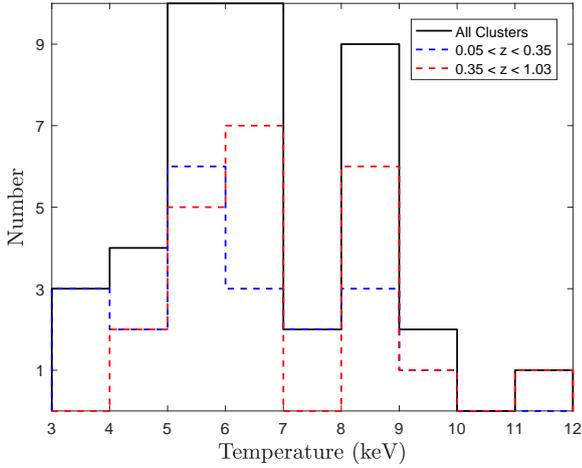}
\caption{Histogram of the values of the global temperature $\langle kT \rangle$ measured in the radial
range $0.1r_{500} < r < 0.4r_{500}$ with a single-temperature {\tt apec} model.}
\label{thisto}
\end{figure}

\begin{figure}
 \includegraphics[width=0.49\textwidth, trim=100 240 100 240, clip]{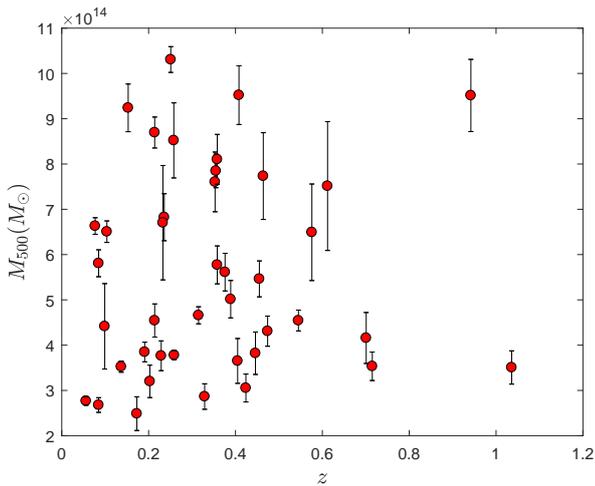}
\caption{The $M_{500}$ obtained from the scale relation $M_{500} \propto \langle kT \rangle^{3/2}/E(z)$ plotted against redshift for
our sample. }
\label{massproxy}
\end{figure}

\begin{figure}
\includegraphics[width=0.49\textwidth, trim=100 240 100 240, clip]{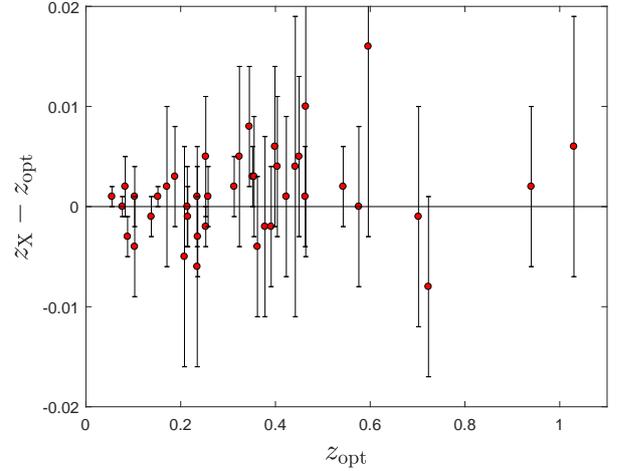}
\caption{The difference between X-ray redshift and optical redshift of all clusters. }
\label{redshift}
\end{figure}

In this section we derive the global properties of each cluster, namely the X-ray redshift, the
emission weighted global temperature, and the radius $r_{500}$.  The global properties will be used
uniquely to characterize the sample, and will not be
used in our analysis except for the normalization of the radii to the value of $r_{500}$.

We measure the global temperature from the cumulative spectrum extracted in the region
$0.1r_{500} < r <\sim 0.4r_{500}$. This choice is often adopted in the literature to
obtain temperature values that more closely trace the virial value, avoiding the effect of the cool
core when present.  We use a single-temperature {\tt apec} model, therefore $\langle kT\rangle$ is an
emission weighted value resulting from the range of temperatures present in the explored radial range.
To estimate $r_{500}$, we use the average relation described in
\citet{vikhlinin2006b}:

\begin{equation}
 r_{500} = \frac{0.792}{hE(z)} \left( \frac{\langle kT \rangle}{5~{\rm KeV}} \right)^{0.53} {\rm Mpc},
\label{r500}
\end{equation}

where $E(z) = (\Omega_{\rm m}(1+z)^{3}+\Omega_{\rm \Lambda})^{0.5} $. The global temperatures
$\langle kT \rangle$ and $r_{500}$ are evaluated iteratively until we obtain a stable temperature. As
shown in column 3 of Table \ref{table2}, our sample spans a large range of temperatures of
$ 3 <\langle kT \rangle < 12$ keV with a peak at 6 keV.   As we show in Figure \ref{thisto},
we find higher temperatures at higher redshift, however this does not necessarily imply
significantly larger masses. We check the mass $M_{500}$ with the self-similar model $M_{500} \propto \langle kT \rangle^{3/2}/E(z)$, with the normalization measured in \citet{vikhlinin2006a}, and plot $M_{500}$ as a function of redshift in figure \ref{massproxy}. We notice that the mass range spanned by our sample is not significantly changing with redshift. In particular, the mass range (spanning a factor of $\sim 5$) is more or less equally populated up to redshift $z\sim 0.6$, with only four clusters
at $z>0.6$.

We note that, in principle, we can measure mass and $r_{500}$ directly from our spectrally resolved
analysis, by measuring the total mass from the hydrostatic equation, and computing the radius
corresponding to the average overdensity $\Delta =500$ with respect to the critical density at the
cluster redshift.  However, in order to do this we should measure robust density and temperature profiles
and therefore sample the ICM emission carefully at radii larger than $r_{500}$, which is beyond the goal of
this paper. We perform a check on the four clusters in our sample with the highest $S/N$ (Abell2597,
PKS0745-191, MACSJ2229.7-2755 and MACSJ1423.8+2404) and confirm that the extrapolated value of $r_{500}$
obtained from our hydrostatic mass profile is consistent with that obtained from Equation \ref{r500} within
5\%, and that the mass proxy we used in Figure \ref{massproxy} is accurate within 8\%.

Since the measurement of iron abundance is sensitive to the X-ray redshift, we also investigate
possible discrepancies between the optical redshift and the X-ray redshift, as obtained from the fit
to the global emission with an {\tt apec} model leaving the redshift parameter free.  For simplicity, we
measure the X-ray redshift of the clusters by fitting the 2.0--7.0 keV spectra so that the best-fit
redshift is determined uniquely by the H-like and He-like iron line complex at 6.7--6.9 keV rest
frame. As shown in Figure \ref{redshift}, we find that the X-ray and spectroscopic redshifts are
consistent within $\sim 1\sigma$. We fix the redshift at the best-fit global X-ray value in the following X-ray analysis.

\section{Measurement of the iron density profile}

To measure the iron density profiles (together with the temperature and density profiles), the first step is
to derive the the azimuthally-averaged, projected $Z_{\rm Fe}$ profile, as measured in a series of
concentric annuli centered in the peak of X-ray emission out to the maximum radius $\sim 0.4r_{500}$. The annuli are chosen with an adaptive criterion based on a smoothly varying S/N threshold on the
0.5--7 keV image, to ensure a roughly equal number of net counts $\geq 1000$ (0.5--7 keV band) in each bin
for the inner annuli, and a doubled number of net counts in the outermost ring.  This choice is key to keep
a comparable quality of the spectral fit in the outer regions, which are mostly affected by the
background due to the rapidly declining ICM density profile and the correspondingly larger extraction regions.
The number of independent bins per cluster ranges from 6 to 13.

The next step is the measurement of the actual 3D profile from the projected one, under the
assumption of spherical symmetry. Both the {\it projct} model in Xspec \citep{arnaud1996} and the tool {\it dsdeproj} presented in \citet{sanders2007} can perform a direct and non parametric deprojection of ICM spectra. However, {\it projct} may produce large unphysical oscillations in the 3D profiles in some situations (see \citet{fabian2006}, and examples in
\citet{2008Russell}). This instability has been claimed to
be due to departure from spherical or ellipsoidal symmetry or the presence of multiphase gas
\citep[see, e.g.,][]{fabian2006}. In this work, we use the tool {\it dsdeproj} (version 1.2) to produce 3D profiles: it deprojects the spectrum
of a shell by subtracting the rescaled count rate of the foreground and background emission.

\begin{figure}
\centering
\includegraphics[width=0.49\textwidth, trim=100 240 100 240, clip]{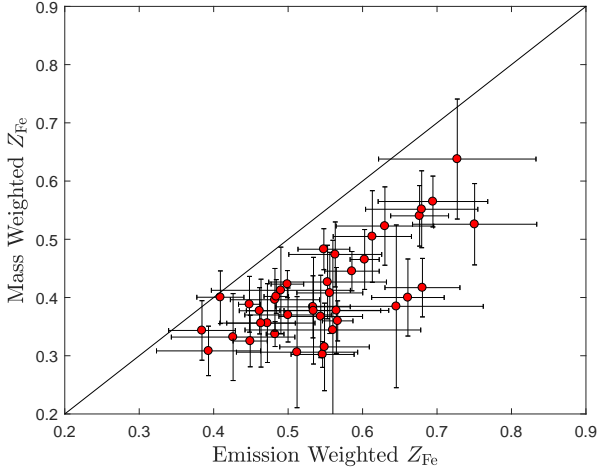}
\caption{The global iron abundance $\langle Z_{\rm Fe}\rangle\times M_{\rm gas}$ versus emission weighted iron abundance
obtained with our spatially resolved analysis in the radial range $r<0.2r_{500}$.}
\label{ew_mw}
\end{figure}

Our spectral analysis provides profiles for the temperature, iron abundance and the electron density,
which can be used to compute the gas density.  These three quantities are essentially independent: the
electron
density mainly depends on the normalization of the emission, with little effects from temperature and
abundance.  Temperature measurements are determined by spectral shape and by line ratio, and therefore
relatively independent from the line intensities.  Abundances are measured directly from the equivalent
width of the emission lines. Temperature, metallicity and electron density can become strongly coupled
in presence of strong gradients across the annulus or along the line of sight, a case in which the
spectra from single annuli can no longer be approximated with a single temperature {\tt apec} model.
Clearly, within the limitation of our data, we assume that the angular resolution (number of annuli)
in our analysis is sufficient to provide quasi-isothermal spectra.

The deprojection procedure provides us with the iron abundance, $Z_{\rm Fe}$, temperature and gas
density, hence total ICM mass,  within each spherical shell.  Therefore, we can finally calculate the
mass-weighted iron abundance $Z_{\rm mw}$ defined as $Z_{\rm mw}\equiv \sum (Z_{Fe}^{i}\cdot M^{i}_{\rm gas}$)/$\sum M_{\rm gas}^{i}$, where $i$ is the
index of the shell. We compare $Z_{\rm mw}$ with the average spectral abundance
$\langle Z_{\rm Fe}\rangle$ measured from a single-temperature fit of the global emission within
the same radius.  This is the quantity that is often reported in the literature, mainly because of the
difficulty in resolving the ICM profiles for low S/N data (typically medium and high-$z$ clusters).
The average abundance can be well approximated with the emission-weighted value $Z_{\rm ew}$,
defined as $\int^R_0 \Lambda(T,Z) n_e^2 Z_{\rm Fe}/\int^R_0 \Lambda(T,Z) n_e^2$, where $R$ is the
extraction radius.  In Figure \ref{ew_mw}, we show $Z_{\rm mw}$ versus $\langle Z_{\rm Fe}\rangle $
for all the clusters, computed within $0.2r_{500}$, which entirely includes the iron peak in most of the cases, and
therefore the difference between $Z_{\rm mw}$ and $\langle Z_{\rm Fe}\rangle $ is maximized.
This radial range usually is described
with at least 4-6 shells, since it is the region with the brightest emission. We find that the average
abundance is about 25\% higher than the mass weighted value. Qualitatively, the result is expected,
since $\langle Z_{\rm Fe}\rangle \sim Z_{\rm ew}$ is weighted by $\Lambda(kT,Z) n_{e}^2$,
and higher $Z_{\rm Fe}$ is ubiquitously associated with much higher density and slightly lower
temperature, therefore corresponds to higher emission weights.

We note that the quantity $Z_{\rm mw}\times M_{\rm gas}$ is by construction different
from $\langle Z_{\rm Fe}\rangle\times M_{\rm gas}$. The use of $\langle Z_{\rm Fe}\rangle\times M_{\rm gas}$ as a proxy for the
mass of iron in the ICM may provide a significant artificial
increase of the iron mass. In addition, any cosmic evolution
of temperature and density gradients in the core of clusters would impact also on the observed
evolution of the average iron abundance.  Although this effect is expected to be smaller
when including regions beyond $0.2r_{500}$ (therefore for bright clusters at low and medium redshift),
for high-$z$ clusters with strong cool cores we do expect to obtain abundance values $\sim$25\% higher
from single measurements based on global emission.  The overall impact on iron evolution clearly depends on the
quality of the sample and the evolution of the cool-core clusters, and it is beyond the goal of this paper.
We claim that the study of the evolution of chemical
properties of ICM should be based on mass-weighted quantities in all cases, obtained
directly from spatially resolved spectral analysis, when possible, or by physically motivated models
of the ICM profiles that allow one to associate mass-weighted quantities to global quantities.
In this way, the uncertainties will be directly associated with the adopted models, allowing a better
control on the predictions of iron abundance distributions for the low S/N groups and cluster
population.

\section{Metallicity-pseudo entropy and Metallicity-cooling time relations}

\begin{figure}
\centering
\includegraphics[width=0.50\textwidth, trim=100 210 100 210, clip]{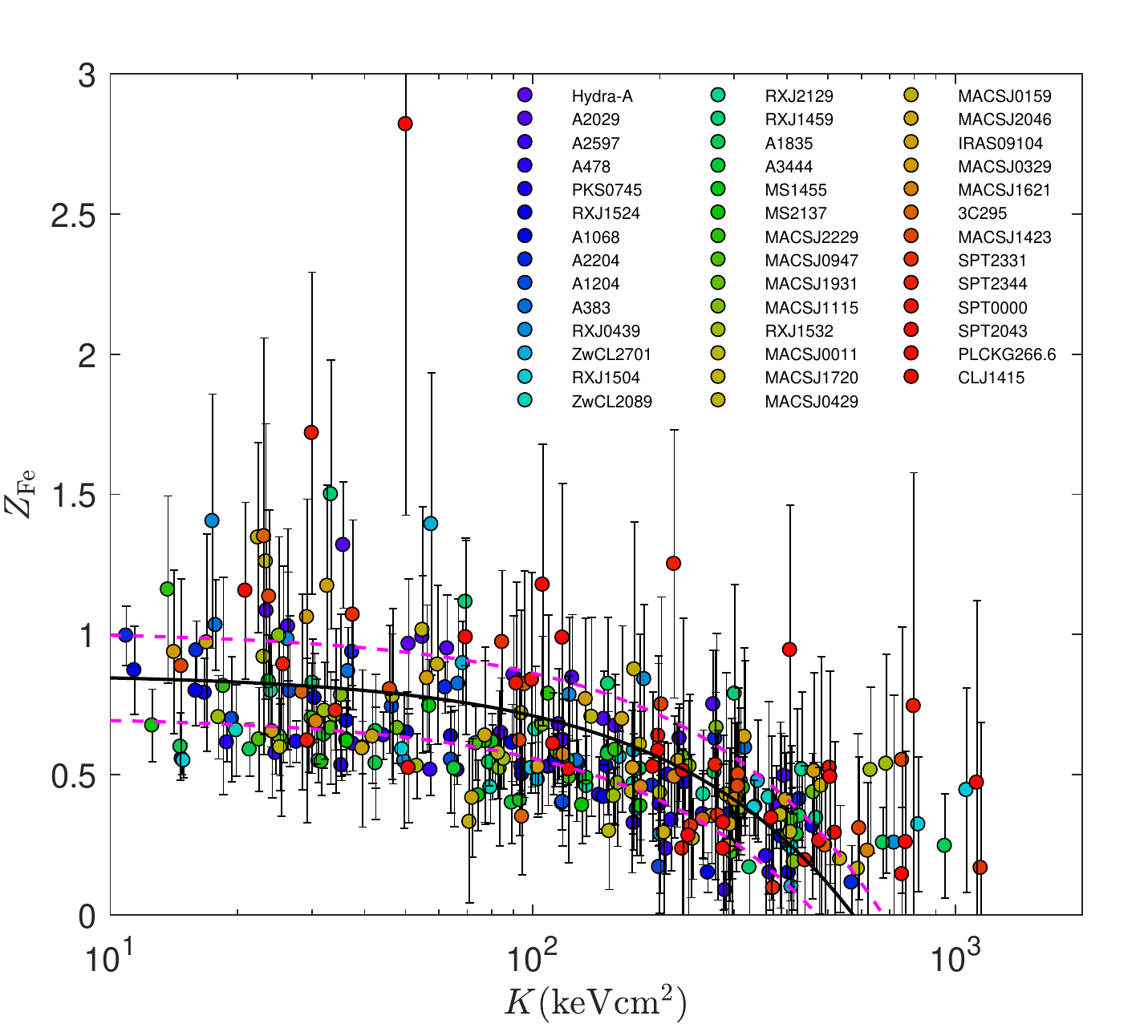}
\includegraphics[width=0.50\textwidth, trim=100 210 100 210, clip]{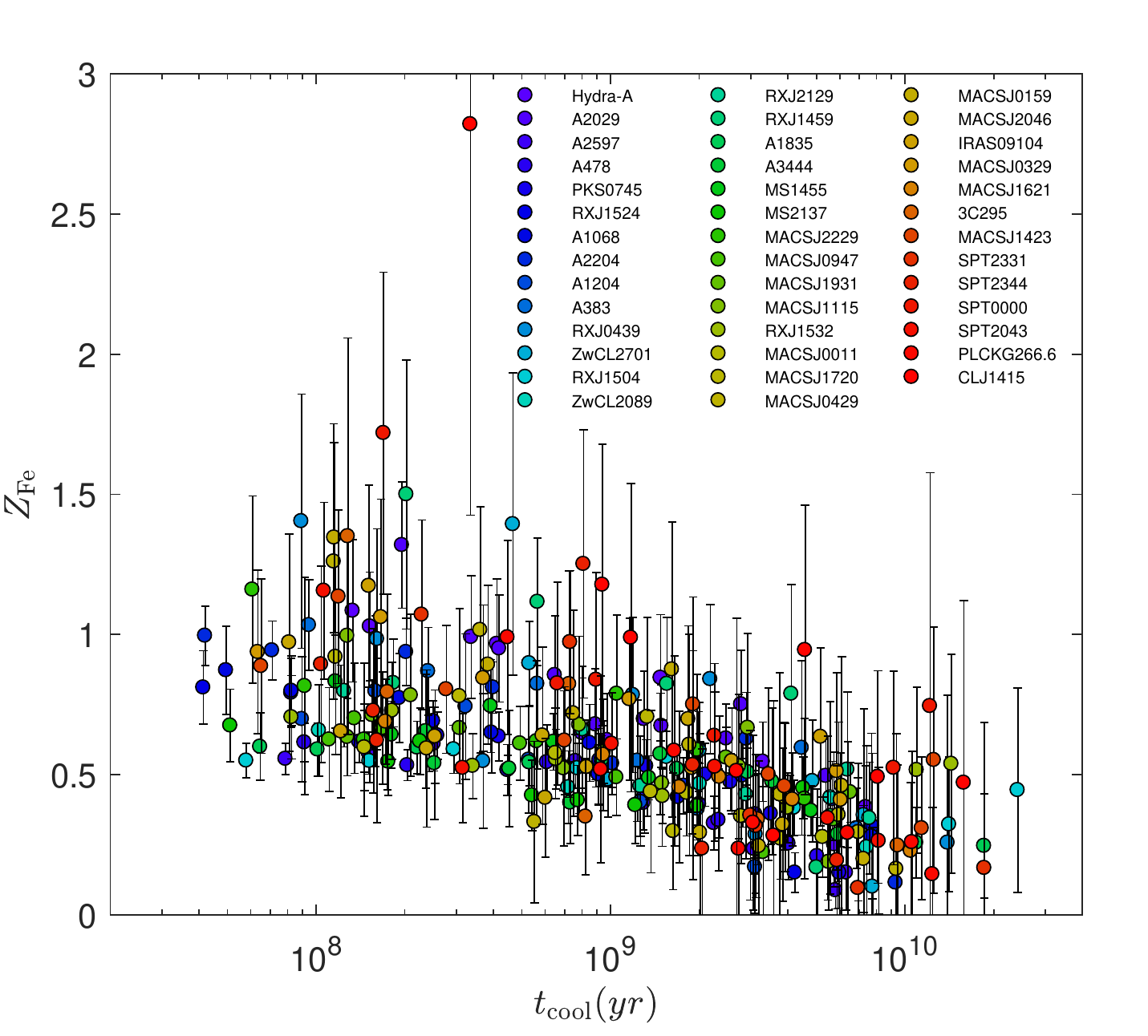}
\caption{{\it Upper panel}: The distribution of $Z_{\rm Fe}$ versus $K(kT,n_e)$ of all the measured
bins in the clusters. The color code denotes the redshift of the clusters, from the lowest (blue) to
the highest (red). The solid line is the best fit linear model. The dashed lines show the $rms$
dispersion of the distribution.
{\it Lower panel}: The distribution of $Z_{\rm Fe}$ versus the cooling time $t_{\rm cool}$.}
\label{zs}
\end{figure}

Given the lack of correlation between density and temperature, the quantity $K(kT,n_e)\equiv
kT\cdot n_{e}^{-2/3}$ as a function of the radius, often called pseudo-entropy, has been
historically used to characterize the thermal history of the ICM \citep{ponman1999}. Pseudo-entropy
stays constant during any adiabatic process, like adiabatic compression, while it may change
due to gain or loss of internal energy, mostly due to shocks, turbulence dissipation, or cooling.
Therefore, its behavior as a function of radius can be used to
identify regions dominated by shock heating \citep[with a slope $\propto r^{1.1}$, see][]{2001Tozzi},
adiabatic compression (flat profiles) and cooling.  Since the metallicity
also has a strong dependence with radius in some clusters, it has been previously claimed that metallicity and
pseudo entropy may be associated.  In particular, significant increases in metallicity are expected,
and often observed, in regions of low entropy where the cooling of the ICM may be associated with
events of star formation triggered by the gas dropping out of the cold phase.  For
example, in local clusters, $Z_{\rm Fe}$ and $K(kT,n_e)$ are found to have a negative correlation
\citep{degrandi2004,leccardi2010,ghizzardi2014}.

On these premises, we first investigate the correlation between $Z_{\rm Fe}$ and $K(kT,n_e)$ for all the
independent bins in the clusters of our sample. We fit the $Z_{\rm Fe}-K(kT,n_e)$ distribution with a linear
function $Z = Z_{0} - \alpha\cdot K/1000$, and obtain the best fit values $Z_{0}=0.86\pm0.18$ and
$\alpha=1.49\pm0.51$. The $Z_{\rm Fe}$-$K(kT,n_e)$ distribution and the best fit function are shown in
the upper panel of Figure \ref{zs}.

Similarly, we also show the distribution of $Z_{\rm Fe}$-$t_{\rm cool}$ in the lower panel of Figure \ref{zs}.
The cooling time of the gas $t_{\rm cool}$ is defined as the gas enthalpy divided by the energy loss per unit volume
\citep{peterson2006}:
\begin{equation}
t_{\rm cool} = \frac{\frac{5}{2}n_{e}kT}{n_{e}^2\Lambda(T,Z)} ,
\end{equation}
where $\Lambda(T,Z)$ is the cooling function, which is associated with the energy density emitted by a radiative cooling ICM with a given temperature and metallicity \citep{boehringer1989,sutherland1993}. We find that the $Z_{\rm Fe}$ shows a similar correlation with $t_{\rm cool}$ as with $K(kT,n_e)$, given the strong similarity of the two quantities.

The interpretation of this average, highly scattered relation, may be understood only on
the basis of a comprehensive model for chemical enrichment of the ICM through the lifetime of groups and
clusters.  The association of higher abundance values with low entropy and shorter cooling time gas may be, in fact, not simply
associated with gas cooling and star formation with consequent local chemical enrichment, but the result of
two independent process, like radiative cooling and iron production and diffusion.  These processes are both
more efficient at the cluster center, but proceed independently and with different time scales.  Namely,
the spatially resolved analysis we present in this paper is a first step towards a comprehensive model.

\section{Measuring the width of the iron excess profile}

In this section we present the characterization of the distribution of iron abundance throughout the ICM, focusing on the
innermost regions.  To do that, we fit the deprojected profile of the iron abundance with a phenomenological model with no
direct physical meaning.  There are two models which have been used in the literature to fit the profiles of ICM iron abundance:
a double-$\beta$ model \citep{santos2012}, and the empirical function provided by \citet{mernier2017}, which is a simple
power-law for $r > \sim 0.02r_{500}$ (in the innermost regions they model the possible decrease of the metallicity
by subtracting a Gaussian).  However, we often see a strong abundance gradient beyond $\sim 0.02r_{500}$, and
therefore adopt a simpler model which is preferable with respect to a composite model with many free parameters (6 in the case of a double-$\beta$ model).
Given the low number of bins we have (particularly for the high-$z$ clusters) we want to have no more than 3 free parameters.
We adopt a single $\beta$ model in the form
\begin{equation}
Z_{\rm Fe}(r) = Z_{0}\cdot (1 + (r/r_{0})^{2})^{-\beta} .
\end{equation}
The best fits are obtained minimizing the $\chi^2$ over the 3 free parameters.
The profiles and the best fit models of all the clusters are presented in the appendix.

\begin{figure}
\centering
\includegraphics[width=0.49\textwidth, trim=100 240 100 240, clip]{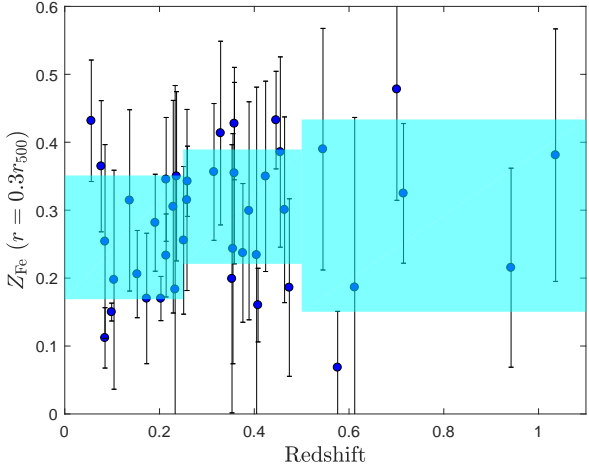} 
\caption{The iron abundance at $r=0.3r_{500}$ for all the clusters. Shaded area shows the $rms$ dispersion across three redshift bins: [0.05,0.25], [0.25,0.5], [0.5,1.03].  }
\label{abun}
\end{figure}

After fitting the profiles, we first obtain the iron abundance at $r=0.3r_{500}$ for all the clusters,
and plot $Z(0.3r_{500})$ versus redshift in figure \ref{abun}. We group the clusters into three redshift
bins: [0.05,0.25], [0.25,0.5], [0.5,1.03], and calculate the average value and $rms$ dispersion respectively.
As a result, we find that $Z(0.3r_{500})$ of the three redshift bins are 0.26, 0.30, and 0.29, with $rms$ dispersion
of 0.10, 0.08, and 0.13. This result suggests that the average iron abundance
at $0.3r_{500}$ is consistent with a constant value at $\sim 0.3Z_{\odot}$ at least up to redshift 1.

\begin{figure}
\centering
\includegraphics[width=0.49\textwidth, trim=100 240 100 240, clip]{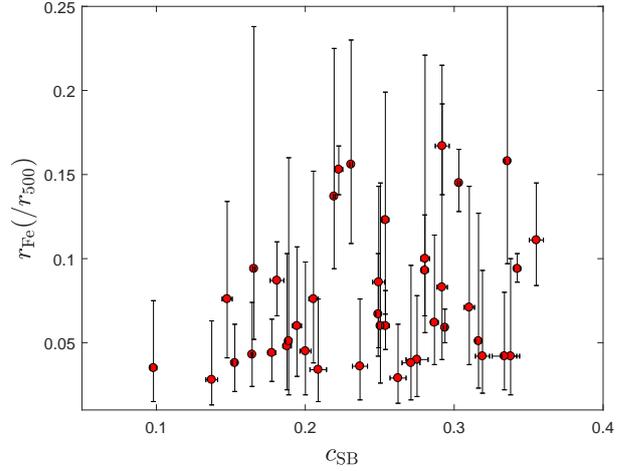} 
\caption{The scale of the iron excess peak, $r_{\rm Fe}$, plotted versus the cool core strength $c_{\rm SB}$ (as defined in
\citealt{santos2008}). }
\label{rc}
\end{figure}

The width of the iron profiles is quantified with a newly defined scale parameter $r_{\rm Fe}$, expressed in units of $r_{500}$.
The definition of $r_{\rm Fe}$ is as follows.
We calculate the excess abundance profile $Z_{\rm exc}(r) = Z(r) - Z(0.3r_{500})$, and the peak excess abundance
$Z_{\rm exc,0} = Z_{\rm peak} - Z(0.3r_{500})$, where $Z_{\rm peak}$ is the peak abundance. Rather than directly adopting
$Z_{0}$ of the best-fit beta-model as $Z_{\rm peak}$, we instead compute the average value within $0.02r_{500}$
to have a more robust estimate. Finally, we define $r_{\rm Fe}$ as the radius where $Z_{\rm exc}(r_{\rm Fe})$ is half
of $Z_{\rm exc,0}$. The distribution of $r_{\rm Fe}$ values is listed in Table \ref{table2}.

Since the distribution of iron differs
significantly in cool-core and non-cool-core clusters, the width of iron profiles may also be correlated with the strength of
cool-core. To decouple the evolution of the iron distribution from that of the cool core in the clusters, we check whether
$r_{\rm Fe}$ can quantify the broadening of the iron peak, or just shows the strength of the cool core.
As a representative parameter, we use the surface brightness concentration $c_{\rm SB}$, defined as the ratio of the flux within
40 kpc and 400 kpc \citep{santos2008}.  We plot $r_{\rm Fe}$ versus $c_{\rm SB}$ in figure \ref{rc}. Spearman's test is
performed on the distribution taking into account the uncertainties in $c_{\rm SB}$ and $r_{\rm Fe}$, with a resulting
$\rho= 0.18\pm0.11$, corresponding to a very weak correlation with the probability of null hypothesis $p=0.31\pm0.26$. This
result suggests no significant correlation between $r_{\rm Fe}$ and $c_{\rm SB}$, and hence $r_{\rm Fe}$ can quantify the width
of the iron distribution independently from the strength of the cool core.

\begin{table*}
\centering
 \caption{Main results on the clusters. Column 1: cluster name; Column 2: global redshift of the ICM; Column 3: global temperature of the ICM within $0.1r_{500}$--$0.4r_{500}$; Column 4: $r_{500}$ in Mpc measured using the global temperature in column 3; Column 5: the scale of the iron excess peak: $r_{\rm Fe}$ in $r_{500}$; Column 6: the cool core strength $c_{\rm SB}$ (as defined in \citet{santos2008}); Column 7: iron mass excess within 0.3$r_{500}$; Column 8: total gas mass within 0.3$r_{500}$.}
 \begin{tabular}{lccccccc}
\hline
Cluster   & X-ray redshift & $kT$ & $r_{500}$/Mpc & $r_{\rm Fe}/r_{500}$ &  $c_{\rm SB}$ & $M_{\rm Fe}^{\rm exc}/(10^9M_{\odot})$ & $M_{\rm gas}/(10^{13}M_{\odot})$ \\
\hline
Hydra-A             &  $ 0.056^{+0.001}_{-0.001} $ & $ 3.94^{+0.14}_{-0.13} $ & 1.00 &    $ 0.067	_{-0.025}^{+0.036}  $ & $    0.249\pm     0.002 $  & $ 0.972\pm 0.165  $  & $  0.666\pm	0.001 $ \\
Abell2029           &  $ 0.077^{+0.001}_{-0.001} $ & $ 6.89^{+0.19}_{-0.19} $ & 1.30 &    $ 0.043	_{-0.019}^{+0.031}  $ & $    0.164\pm     0.001 $  & $ 2.873\pm 0.634  $  & $  2.250\pm	0.002 $ \\
Abell2597           &  $ 0.085^{+0.001}_{-0.003} $ & $ 3.89^{+0.22}_{-0.25} $ & 0.96 &    $ 0.145	_{-0.017}^{+0.020}  $ & $    0.303\pm     0.001 $  & $ 1.612\pm 0.592  $  & $  0.690\pm	0.001 $ \\
Abell478            &  $ 0.085^{+0.002}_{-0.002} $ & $ 6.35^{+0.34}_{-0.31} $ & 1.24 &    $ 0.094	_{-0.042}^{+0.144}  $ & $    0.165\pm     0.001 $  & $ 2.175\pm 0.923  $  & $  2.389\pm	0.003 $ \\
PKS0745-191         &  $ 0.104^{+0.003}_{-0.002} $ & $ 6.86^{+0.25}_{-0.25} $ & 1.28 &    $ 0.137	_{-0.043}^{+0.088}  $ & $    0.219\pm     0.001 $  & $ 4.772\pm 3.222  $  & $  2.338\pm	0.002 $ \\
RXJ1524.2-3154      &  $ 0.099^{+0.005}_{-0.003} $ & $ 5.36^{+1.14}_{-1.15} $ & 1.12 &    $ 0.094	_{-0.008}^{+0.009}  $ & $    0.342\pm     0.002 $  & $ 1.825\pm 0.293  $  & $  0.825\pm	0.001 $ \\
Abell1068           &  $ 0.137^{+0.002}_{-0.001} $ & $ 4.70^{+0.16}_{-0.17} $ & 1.03 &    $ 0.062	_{-0.025}^{+0.052}  $ & $    0.287\pm     0.003 $  & $ 1.030\pm 0.319  $  & $  0.839\pm	0.001 $ \\
Abell2204           &  $ 0.153^{+0.001}_{-0.001} $ & $ 8.69^{+0.49}_{-0.50} $ & 1.41 &    $ 0.059	_{-0.009}^{+0.011}  $ & $    0.294\pm     0.001 $  & $ 4.186\pm 0.973  $  & $  2.596\pm	0.002 $ \\
Abell1204           &  $ 0.173^{+0.008}_{-0.008} $ & $ 3.81^{+0.57}_{-0.57} $ & 0.92 &    $ 0.111	_{-0.027}^{+0.034}  $ & $    0.355\pm     0.005 $  & $ 1.841\pm 1.344  $  & $  0.744\pm	0.003 $ \\
Abell383            &  $ 0.191^{+0.005}_{-0.005} $ & $ 5.05^{+0.26}_{-0.31} $ & 1.03 &    $ 0.060	_{-0.014}^{+0.021}  $ & $    0.254\pm     0.003 $  & $ 1.387\pm 0.297  $  & $  0.913\pm	0.002 $ \\
RXJ0439.0+0520      &  $ 0.203^{+0.011}_{-0.011} $ & $ 4.51^{+0.51}_{-0.50} $ & 0.98 &    $ 0.167	_{-0.029}^{+0.025}  $ & $    0.292\pm     0.005 $  & $ 2.967\pm 1.519  $  & $  0.759\pm	0.002 $ \\
ZwCL2701            &  $ 0.214^{+0.003}_{-0.004} $ & $ 5.65^{+0.47}_{-0.44} $ & 1.11 &    $ 0.038	_{-0.017}^{+0.023}  $ & $    0.153\pm     0.002 $  & $ 1.753\pm 0.433  $  & $  0.989\pm	0.001 $ \\
RXJ1504.1-0248      &  $ 0.214^{+0.003}_{-0.002} $ & $ 8.52^{+0.33}_{-0.34} $ & 1.35 &    $ 0.158	_{-0.061}^{+0.105}  $ & $    0.336\pm     0.002 $  & $ 3.488\pm 0.769  $  & $  2.714\pm	0.002 $ \\
ZwCL2089            &  $ 0.229^{+0.010}_{-0.010} $ & $ 5.04^{+0.44}_{-0.44} $ & 1.02 &    $ 0.083	_{-0.043}^{+0.132}  $ & $    0.292\pm     0.004 $  & $ 0.996\pm 0.390  $  & $  0.868\pm	0.002 $ \\
RXJ2129.6+0005      &  $ 0.236^{+0.005}_{-0.005} $ & $ 7.36^{+0.58}_{-0.54} $ & 1.24 &    $ 0.048	_{-0.026}^{+0.055}  $ & $    0.188\pm     0.003 $  & $ 1.336\pm 0.390  $  & $  1.662\pm	0.002 $ \\
RXJ1459.4-1811      &  $ 0.233^{+0.004}_{-0.004} $ & $ 7.27^{+1.32}_{-1.42} $ & 1.24 &    $ 0.153 _{-0.015}^{+0.014}  $ & $    0.222\pm     0.003 $  & $ 6.534\pm 3.490  $  & $  1.479\pm	0.004 $ \\
Abell1835           &  $ 0.251^{+0.002}_{-0.002} $ & $ 9.60^{+0.27}_{-0.26} $ & 1.42 &    $ 0.156	_{-0.047}^{+0.074}  $ & $    0.231\pm     0.001 $  & $ 5.157\pm 1.887  $  & $  2.955\pm 0.001 $ \\
Abell3444           &  $ 0.258^{+0.005}_{-0.006} $ & $ 8.53^{+0.81}_{-0.85} $ & 1.33 &    $ 0.051	_{-0.032}^{+0.109}  $ & $    0.189\pm     0.002 $  & $ 1.474\pm 0.492  $  & $  2.290\pm	0.004 $ \\
MS1455.0+2232       &  $ 0.259^{+0.002}_{-0.003} $ & $ 5.10^{+0.13}_{-0.14} $ & 1.02 &    $ 0.093	_{-0.027}^{+0.033}  $ & $    0.280\pm     0.002 $  & $ 1.722\pm 0.233  $  & $  1.205\pm	0.001 $ \\
MS2137.3-2353       &  $ 0.315^{+0.002}_{-0.003} $ & $ 5.93^{+0.23}_{-0.25} $ & 1.06 &    $ 0.051	_{-0.028}^{+0.076}  $ & $    0.316\pm     0.002 $  & $ 0.853\pm 0.197  $  & $  1.307\pm	0.002 $ \\
MACSJ2229.7-2755    &  $ 0.329^{+0.006}_{-0.009} $ & $ 4.38^{+0.44}_{-0.42} $ & 0.90 &    $ 0.042 _{-0.023}^{+0.058}  $ & $    0.338\pm     0.006 $  & $ 0.976\pm 0.243  $  & $  0.924\pm	0.003 $ \\
MACSJ0947.2+7623    &  $ 0.353^{+0.005}_{-0.006} $ & $ 8.19^{+0.71}_{-0.71} $ & 1.24 &    $ 0.100	_{-0.044}^{+0.121}  $ & $    0.280\pm     0.003 $  & $ 3.740\pm 2.864  $  & $  2.138\pm	0.004 $ \\
MACSJ1931.8-2634    &  $ 0.355^{+0.003}_{-0.003} $ & $ 8.36^{+0.39}_{-0.40} $ & 1.24 &    $ 0.123	_{-0.056}^{+0.076}  $ & $    0.254\pm     0.002 $  & $ 2.111\pm 0.364  $  & $  1.962\pm	0.002 $ \\
MACSJ1115.8+0129    &  $ 0.358^{+0.005}_{-0.006} $ & $ 8.54^{+0.59}_{-0.57} $ & 1.26 &    $ 0.060	_{-0.030}^{+0.047}  $ & $    0.194\pm     0.003 $  & $ 3.035\pm 1.638  $  & $  2.301\pm	0.005 $ \\
RXJ1532.9+3021      &  $ 0.358^{+0.004}_{-0.007} $ & $ 6.89^{+0.50}_{-0.50} $ & 1.13 &    $ 0.060 _{-0.034}^{+0.085}  $ & $    0.251\pm     0.002 $  & $ 1.913\pm 0.560  $  & $  2.079\pm	0.003 $ \\
MACSJ0011.7-1523    &  $ 0.376^{+0.009}_{-0.009} $ & $ 6.81^{+0.59}_{-0.42} $ & 1.11 &    $ 0.076	_{-0.035}^{+0.058}  $ & $    0.148\pm     0.003 $  & $ 1.208\pm 0.416  $  & $  1.379\pm	0.004 $ \\
MACSJ1720.2+3536    &  $ 0.389^{+0.006}_{-0.005} $ & $ 6.37^{+0.55}_{-0.50} $ & 1.05 &    $ 0.045	_{-0.026}^{+0.053}  $ & $    0.200\pm     0.004 $  & $ 1.327\pm 0.571  $  & $  1.294\pm	0.003 $ \\
MACSJ0429.6-0253    &  $ 0.405^{+0.007}_{-0.008} $ & $ 5.24^{+0.71}_{-0.71} $ & 0.95 &    $ 0.040	_{-0.022}^{+0.038}  $ & $    0.275\pm     0.007 $  & $ 1.709\pm 1.399  $  & $  1.066\pm	0.003 $ \\
MACSJ0159.8-0849    &  $ 0.408^{+0.006}_{-0.007} $ & $ 9.62^{+0.79}_{-0.52} $ & 1.30 &    $ 0.044	_{-0.017}^{+0.020}  $ & $    0.177\pm     0.003 $  & $ 3.589\pm 1.129  $  & $  2.521\pm	0.004 $ \\
MACSJ2046.0-3430    &  $ 0.424^{+0.006}_{-0.008} $ & $ 4.71^{+0.47}_{-0.48} $ & 0.88 &    $ 0.038	_{-0.022}^{+0.058}  $ & $    0.271\pm     0.006 $  & $ 0.947\pm 0.320  $  & $  0.963\pm 0.003 $ \\
IRAS09104+4109      &  $ 0.446^{+0.013}_{-0.015} $ & $ 5.47^{+0.67}_{-0.67} $ & 0.95 &    $ 0.042 _{-0.022}^{+0.051}  $ & $    0.319\pm     0.005 $  & $ 0.983\pm 0.141  $  & $  1.003\pm	0.003 $ \\
MACSJ0329.6-0211    &  $ 0.455^{+0.002}_{-0.008} $ & $ 6.88^{+0.66}_{-0.34} $ & 1.06 &    $ 0.086	_{-0.039}^{+0.057}  $ & $    0.249\pm     0.004 $  & $ 2.292\pm 0.713  $  & $  1.364\pm	0.002 $ \\
MACSJ1621.3+3810    &  $ 0.464^{+0.005}_{-0.005} $ & $ 8.60^{+1.05}_{-1.08} $ & 1.19 &    $ 0.076	_{-0.038}^{+0.076}  $ & $    0.205\pm     0.003 $  & $ 1.326\pm 0.467  $  & $  1.387\pm	0.002 $ \\
3C295               &  $ 0.474^{+0.005}_{-0.015} $ & $ 5.96^{+0.46}_{-0.46} $ & 1.22 &    $ 0.029	_{-0.015}^{+0.032}  $ & $    0.262\pm     0.005 $  & $ 1.532\pm 0.813  $  & $  1.123\pm	0.003 $ \\
MACSJ1423.8+2404    &  $ 0.545^{+0.004}_{-0.004} $ & $ 6.32^{+0.31}_{-0.33} $ & 0.97 &    $ 0.071	_{-0.034}^{+0.072}  $ & $    0.310\pm     0.004 $  & $ 1.714\pm 0.638  $  & $  1.099\pm	0.002 $ \\
SPT-CLJ2331-5051    &  $ 0.576^{+0.008}_{-0.008} $ & $ 8.01^{+1.31}_{-1.32} $ & 1.07 &    $ 0.087	_{-0.021}^{+0.023}  $ & $    0.181\pm     0.005 $  & $ 2.591\pm 2.428  $  & $  0.961\pm	0.003 $ \\
SPT-CLJ2344-4242    &  $ 0.612^{+0.009}_{-0.019} $ & $ 8.90^{+1.69}_{-1.68} $ & 1.11 &    $ 0.042	_{-0.020}^{+0.038}  $ & $    0.334\pm     0.008 $  & $ 4.691\pm 4.818  $  & $  2.693\pm	0.007 $ \\
SPT-CLJ0000-5748    &  $ 0.701^{+0.011}_{-0.009} $ & $ 6.32^{+0.86}_{-0.85} $ & 0.88 &    $ 0.034	_{-0.019}^{+0.042}  $ & $    0.209\pm     0.006 $  & $ 1.488\pm 0.434  $  & $  0.713\pm	0.002 $ \\
SPT-CLJ2043-5035    &  $ 0.715^{+0.009}_{-0.008} $ & $ 5.73^{+0.52}_{-0.50} $ & 0.83 &    $ 0.036	_{-0.020}^{+0.040}  $ & $    0.237\pm     0.005 $  & $ 0.898\pm 0.242  $  & $  0.766\pm	0.001 $ \\
PLCKG266.6          &  $ 0.942^{+0.008}_{-0.008} $ & $11.64^{+1.05}_{-0.90} $ & 1.06 &    $ 0.035	_{-0.020}^{+0.040}  $ & $    0.098\pm     0.002 $  & $ 2.415\pm 1.295  $  & $  1.922\pm	0.003 $ \\
CLJ1415+3612        &  $ 1.036^{+0.013}_{-0.013} $ & $ 6.40^{+0.69}_{-0.65} $ & 0.73 &    $ 0.028	_{-0.015}^{+0.035}  $ & $    0.137\pm     0.004 $  & $ 1.021\pm 0.387  $  & $  0.433\pm	0.001 $ \\
\hline

 \end{tabular}
 \label{table2}
\end{table*}

\begin{figure}
\centering
\includegraphics[width=0.49\textwidth, trim=100 240 100 240, clip]{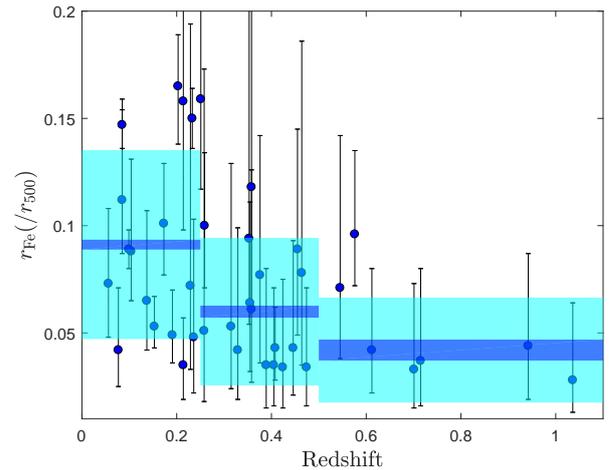}
\caption{The scale of the iron excess peak, $r_{\rm Fe}$, plotted versus redshift. Blue and cyan shaded areas show the weighted average and $rms$ in three redshift bins, respectively.}
\label{rz}
\end{figure}

In figure \ref{rz}, we show the scale of the iron excess peak $r_{\rm Fe}$  as a function of redshift. Since the scatter in
$r_{\rm Fe}$ is very large, we again group the data into three redshift bins [0.05,0.25], [0.25,0.5], and [0.5,1.03]. The
results show that $r_{\rm Fe}$ decreases significantly with redshift, despite the large scatter in the low redshift bin.
We interpret the significant increase of $r_{\rm Fe}$ from high-$z$ to low-$z$ as a clear evolution in the spatial distribution
of iron in the ICM.  Since we only investigate the most relaxed clusters in our sample, the broadening of the iron peak should not
be associated with large scale motions of the ICM, such as those due to major mergers, but it may be associated with the
turbulent mixing and uplifting due to the feedback activities of the central galaxy. Another mechanism that could contribute
to the widening of the peak is the core sloshing due to minor mergers
\citep[see the discussion of A496 in][]{ghizzardi2014}.

\begin{figure}
\centering
\includegraphics[width=0.49\textwidth, trim=100 240 100 240, clip]{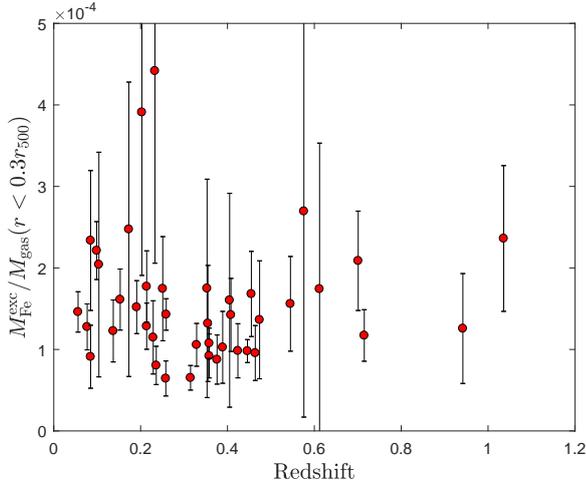}
\caption{The mass of the Fe excess, normalized to the total gas mass, within $0.3r_{500}$ as a function of redshift. }
\label{mfe_exc_z}
\end{figure}

Another important clue comes from the mass of the iron excess, computed within $0.3r_{500}$.
In Figure \ref{mfe_exc_z}, we plot
the iron mass excess normalized by the gas mass within 0.3$r_{500}$ versus redshift. Spearman's test on the distribution
gives $\rho=-0.09\pm0.12$, and the probability of no evolution $p=0.50\pm0.29$, suggesting no significant evolution of iron mass
excess with redshift.  Therefore, we conclude that the bulk of the mass in iron is already present at $z\sim 1$, and that
the increase in the quantity $r_{\rm Fe}$ should not be ascribed to an increase of the iron excess in the ICM with cosmic epoch,
but mostly to the evolution of the iron distribution.

\section{Conclusions}
We perform a systematic study on the evolution of iron distribution in the ICM with deprojected $Z_{\rm Fe}$ profiles in
a sample of 41 relaxed galaxy clusters in the redshift range $0.05 < z < 1.03$. Our conclusions are summarized as follows:

\begin{itemize}

 \item We confirm that for all of our clusters from $z\sim 0.05$ to $z\sim 1.03$, the shape of the deprojected $Z_{\rm Fe}$
 radial profiles shows a steep negative gradient followed by a roughly constant value out to $\sim 0.4r_{500}$, as commonly observed in relaxed/cool-core clusters. The average $Z_{\rm Fe}$ at $\sim 0.3r_{500}$
 is approximately $0.3 Z_{\odot}$, and shows no significant evolution with redshift.

 \item With the deprojected iron profile, we calculate the mass weighted iron abundance $Z_{\rm mw}$ within
 $0.2r_{500}$ for the clusters, and make a comparison with the average iron abundance $\langle Z_{\rm Fe}\rangle$ which
 is obtained by simply fitting the overall emission within the same radius. As a result, we find that the average value is
 always larger than the mass weighted value by $\sim$25\%, showing a potential issue when computing the cosmic evolution of the
 global enrichment of the ICM without having its spatial distribution under control.

\item We investigate the correlation between $Z_{\rm Fe}$ and the pseudo entropy $K_{T,n_e}$, and the correlation between
$Z_{\rm Fe}$ and the cooling time $t_{\rm cool}$ in all the measured bins of all clusters. We confirm that higher $Z_{\rm Fe}$
corresponds to lower $K(kT,n_e)$ and shorter $t_{\rm cool}$, with large scatters. We suggest that this association is relevant to
the interplay of the radiative cooling of the gas and the production and diffusion of iron.

 \item We quantify the width of the iron profiles with the parameter $r_{\rm Fe}$, defined as the radius where the iron
 abundance excess is half of its maximum value. We find that $r_{\rm Fe}$ decreases significantly with redshift, but not with
 the core strength, and that the total mass excess in iron is not evolving with redshift.  This shows that we are witnessing the
 evolution in the distribution of the iron mass excess in the inner region of cool cores, possibly associated with the turbulent
 mixing and uplifting of highly enriched material due to the mechanical-mode feedback from the central galaxy.

 \end{itemize}

This work is the first on a series of papers aiming at establishing a robust modeling of the iron distribution in the ICM based on a central peak and a large scale flat component. This can be obtained thanks to
a detailed analysis of bright, low and moderate $z$ clusters, as in this paper.  Eventually, the evolution of the iron abundance
across the largest cluster population observable to date in terms of mass and redshift range, can be investigated
by using these spatial distribution models.

\section*{acknowledgements}

We thank the anonymous referee for his/her helpful comments.
We acknowledge financial contribution from the agreement ASI-INAF n.2017-14-H.O.
S.E. acknowledges also financial  contribution from the contracts NARO15 ASI-INAF I/037/12/0, and ASI 2015-046-R.0.

\bibliography{iron}

\appendix
\section{The deprojected $Z_{\rm Fe}$ profiles of the clusters in our sample.}
\label{app}
\begin{figure*}
\label{profiles}
\centering
\includegraphics[width=0.245\textwidth, trim=105 240 105 240, clip]{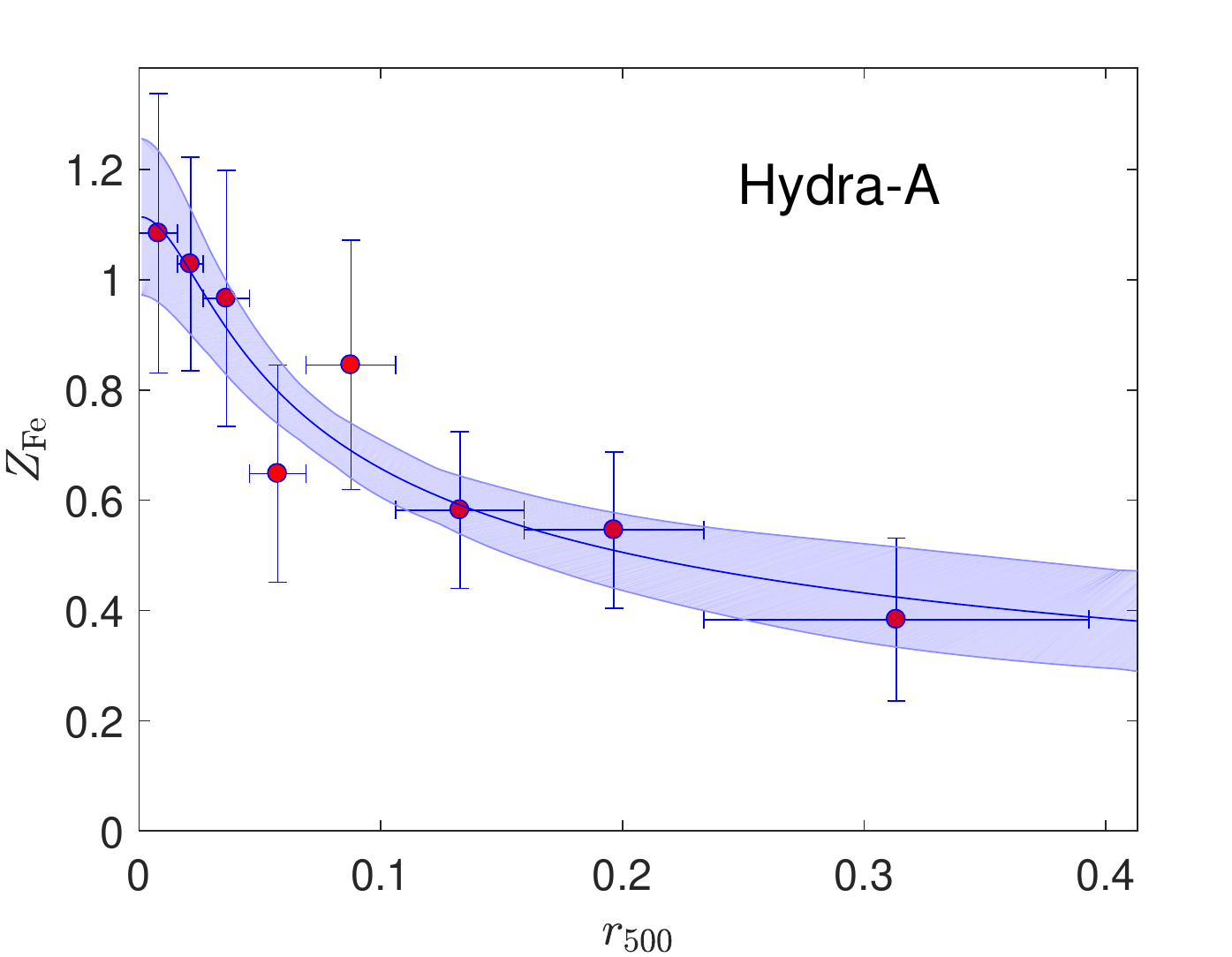}
\includegraphics[width=0.245\textwidth, trim=105 240 105 240, clip]{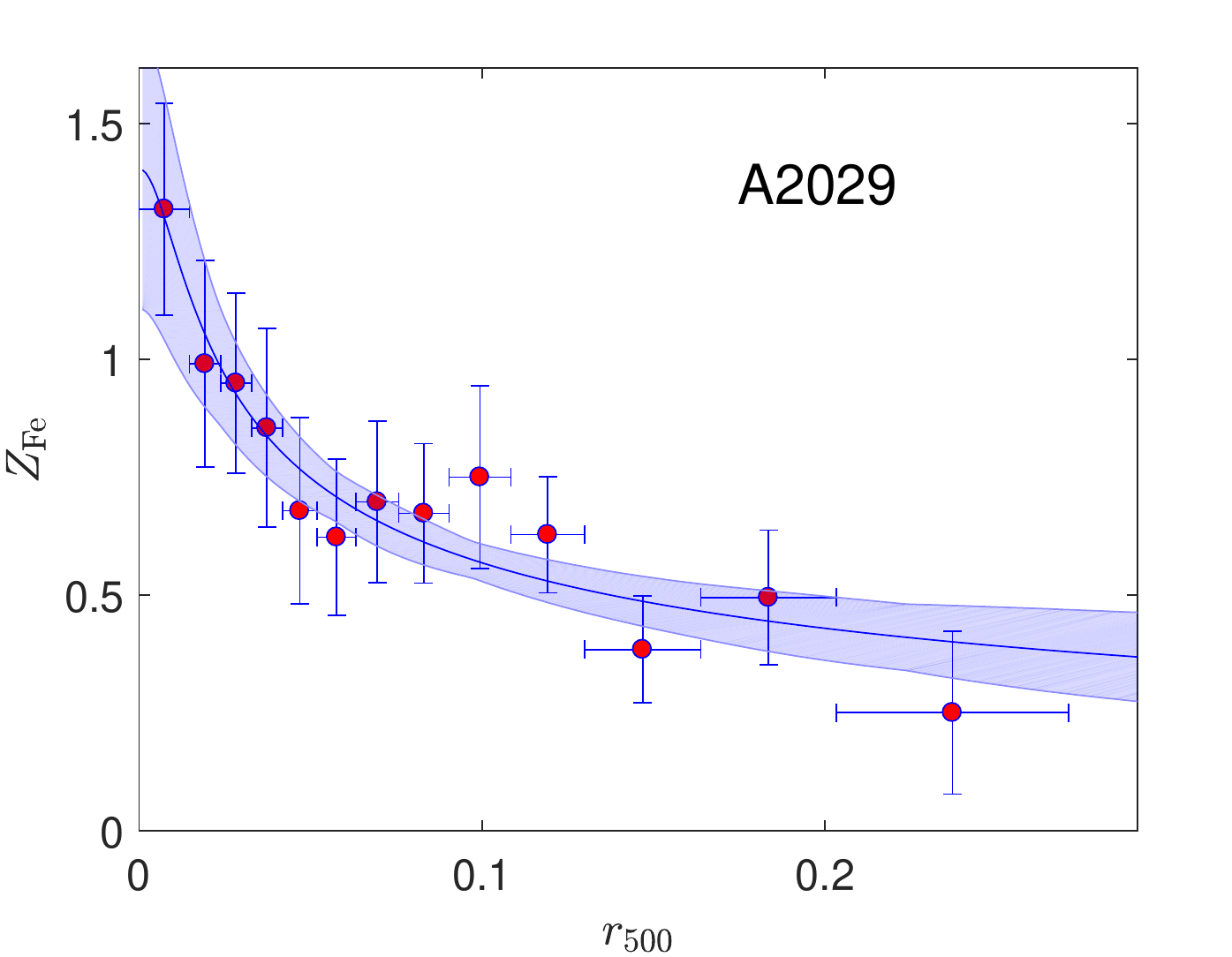}
\includegraphics[width=0.245\textwidth, trim=105 240 105 240, clip]{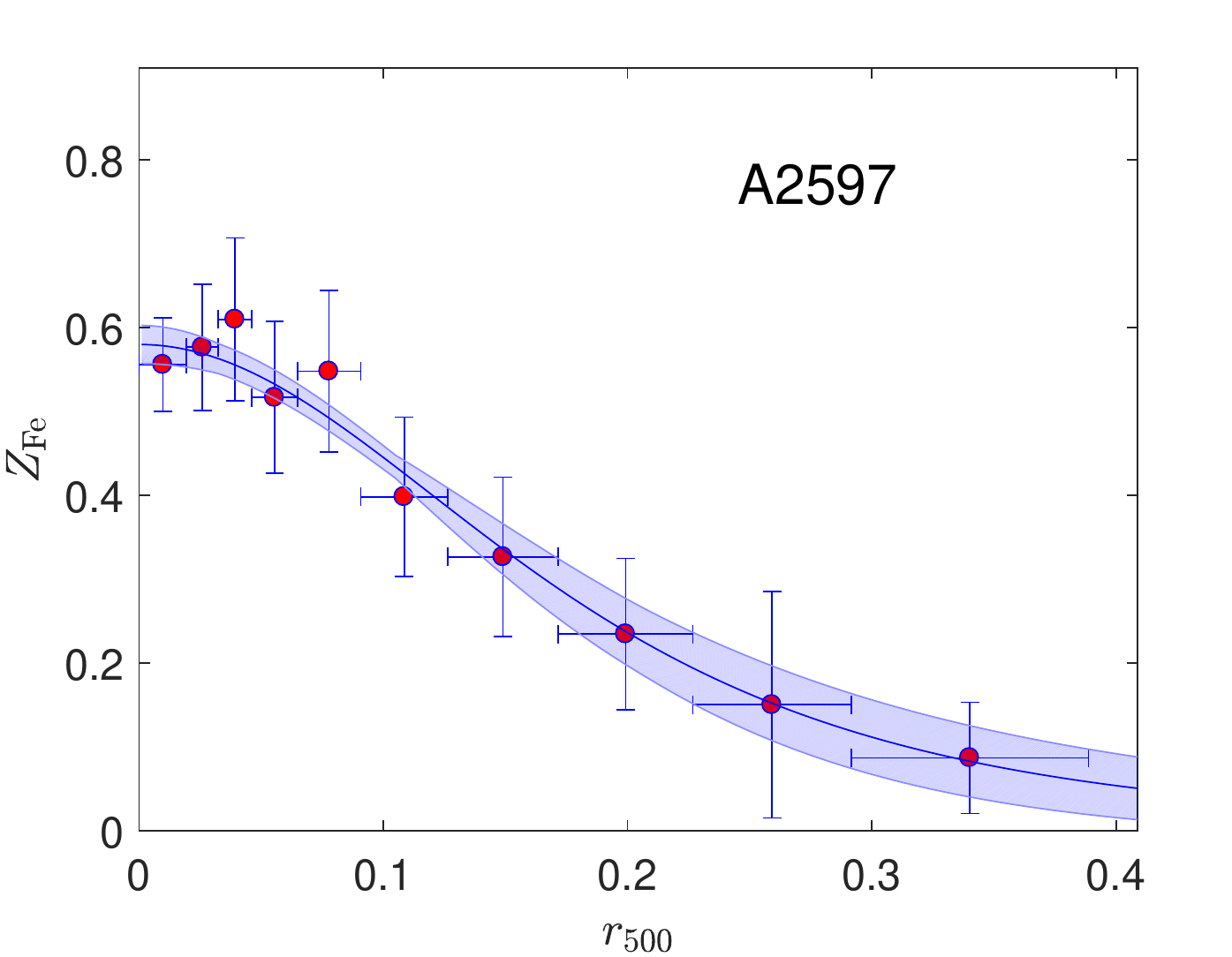}
\includegraphics[width=0.245\textwidth, trim=105 240 105 240, clip]{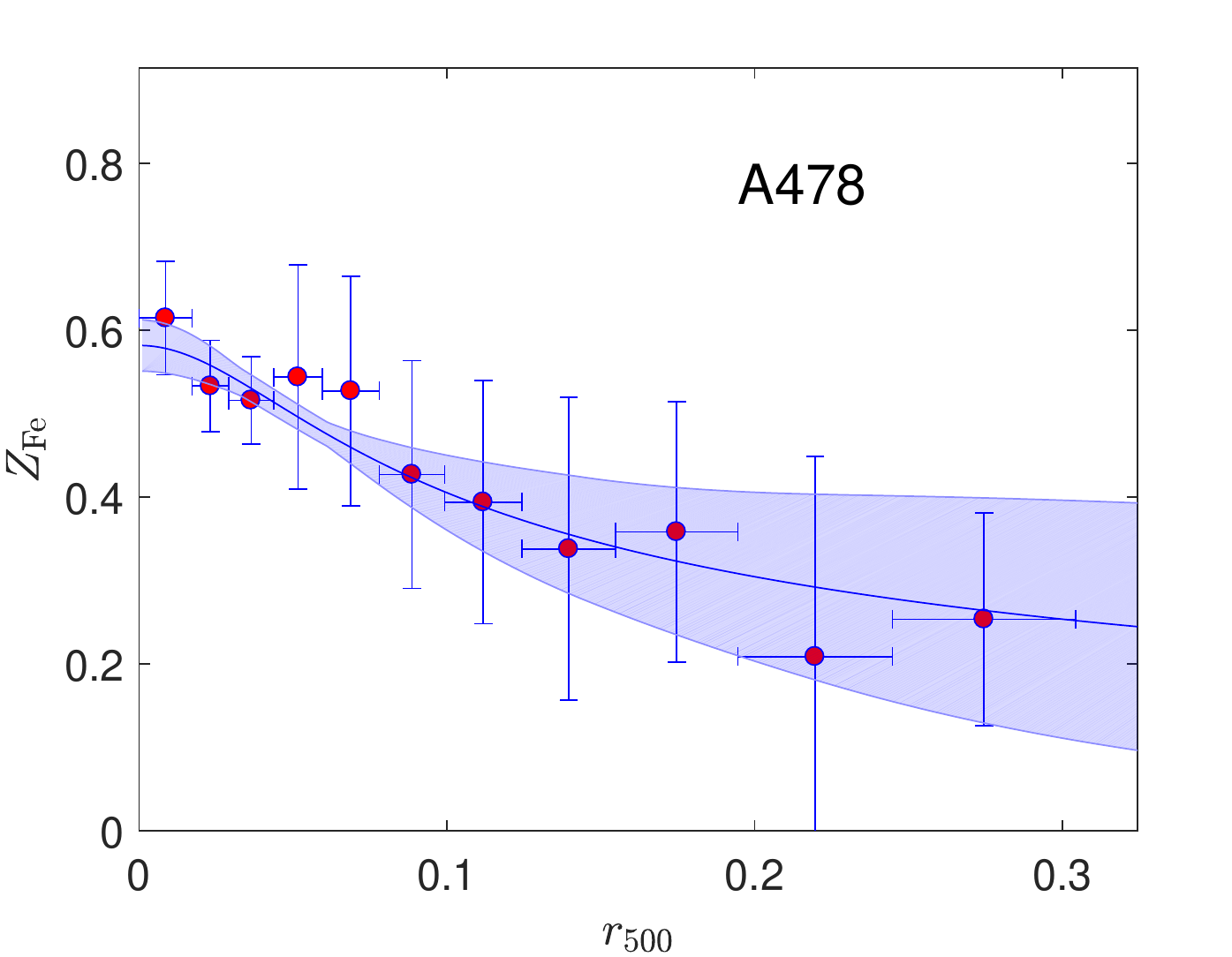}
\includegraphics[width=0.245\textwidth, trim=105 240 105 240, clip]{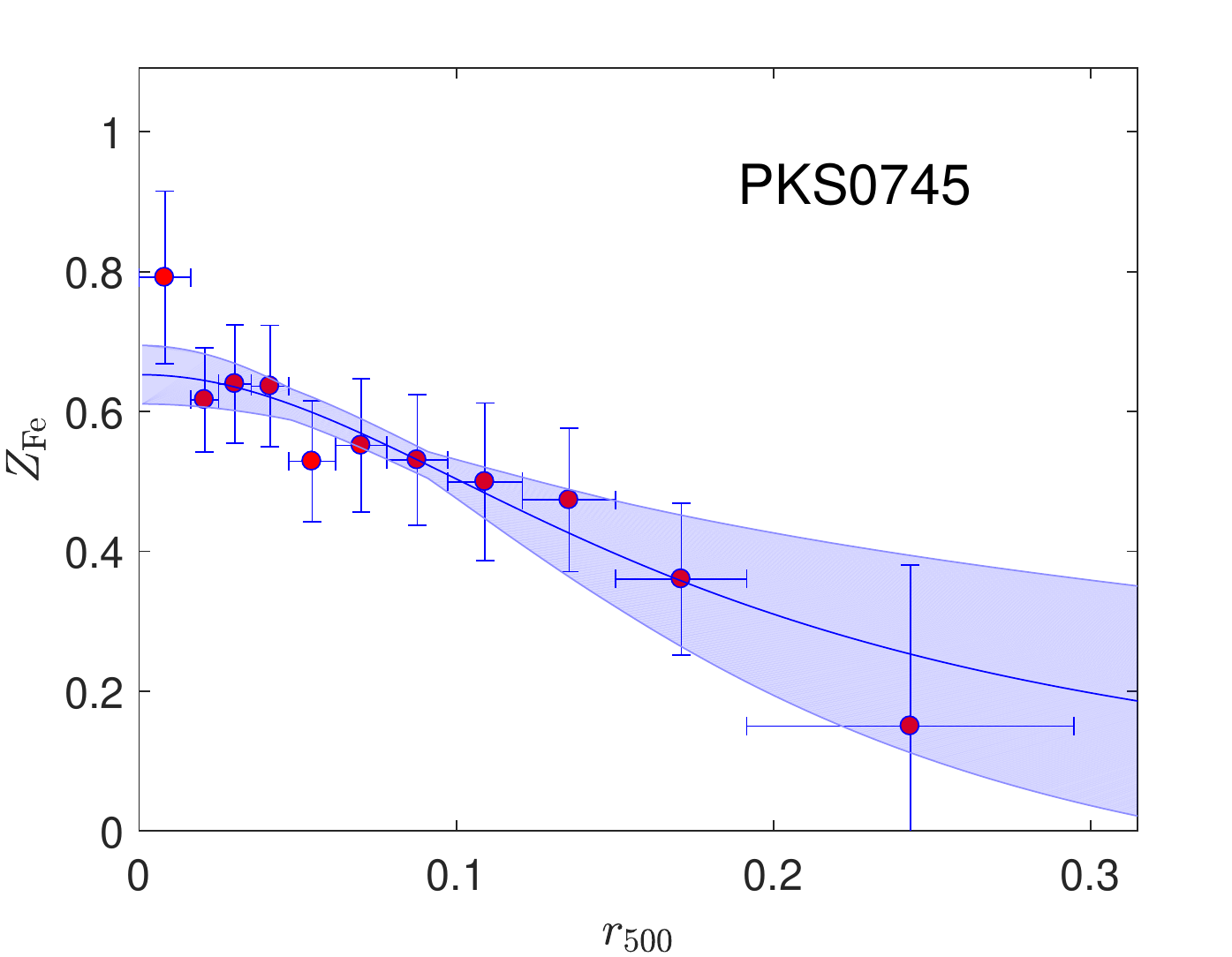}
\includegraphics[width=0.245\textwidth, trim=105 240 105 240, clip]{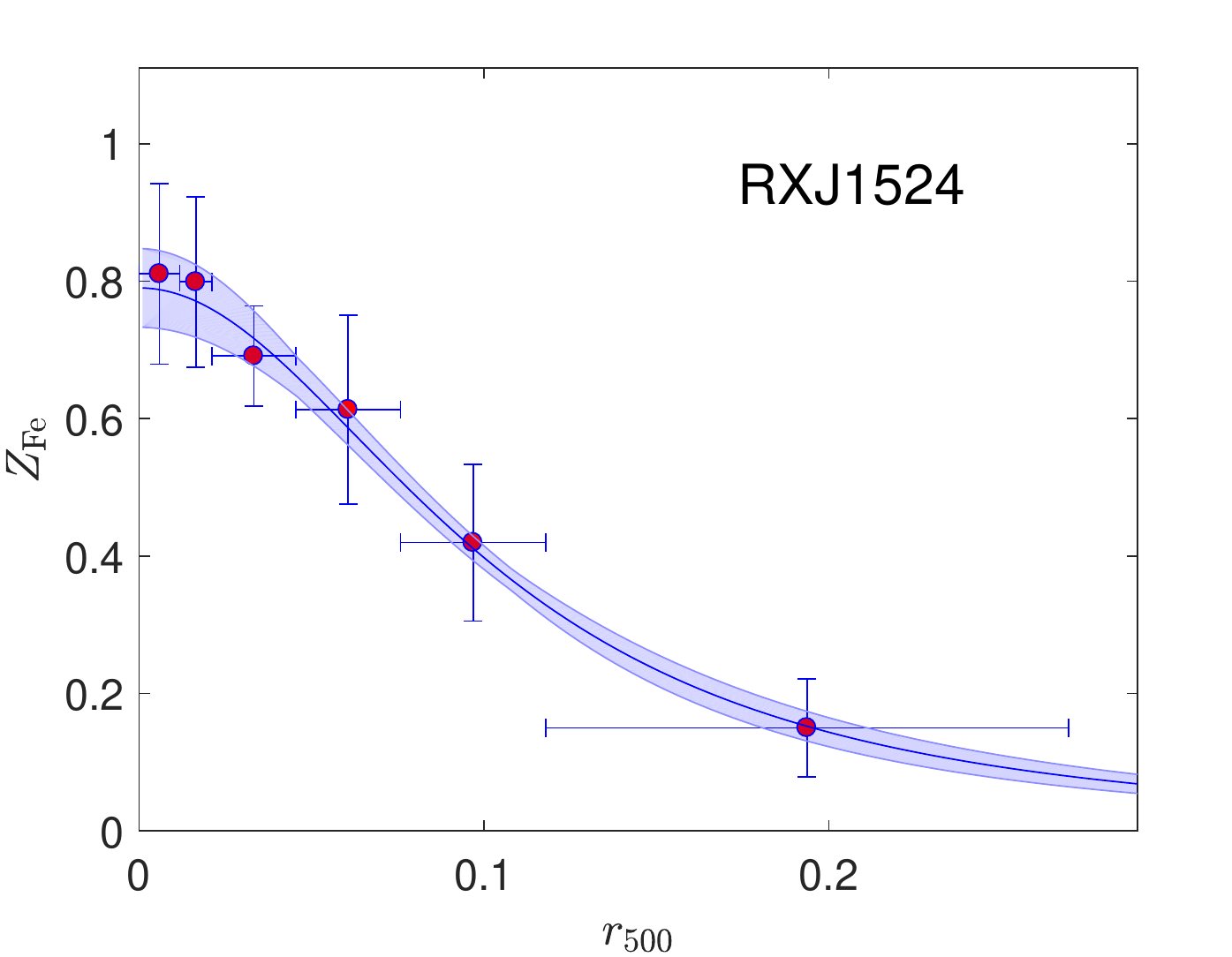}
\includegraphics[width=0.245\textwidth, trim=105 240 105 240, clip]{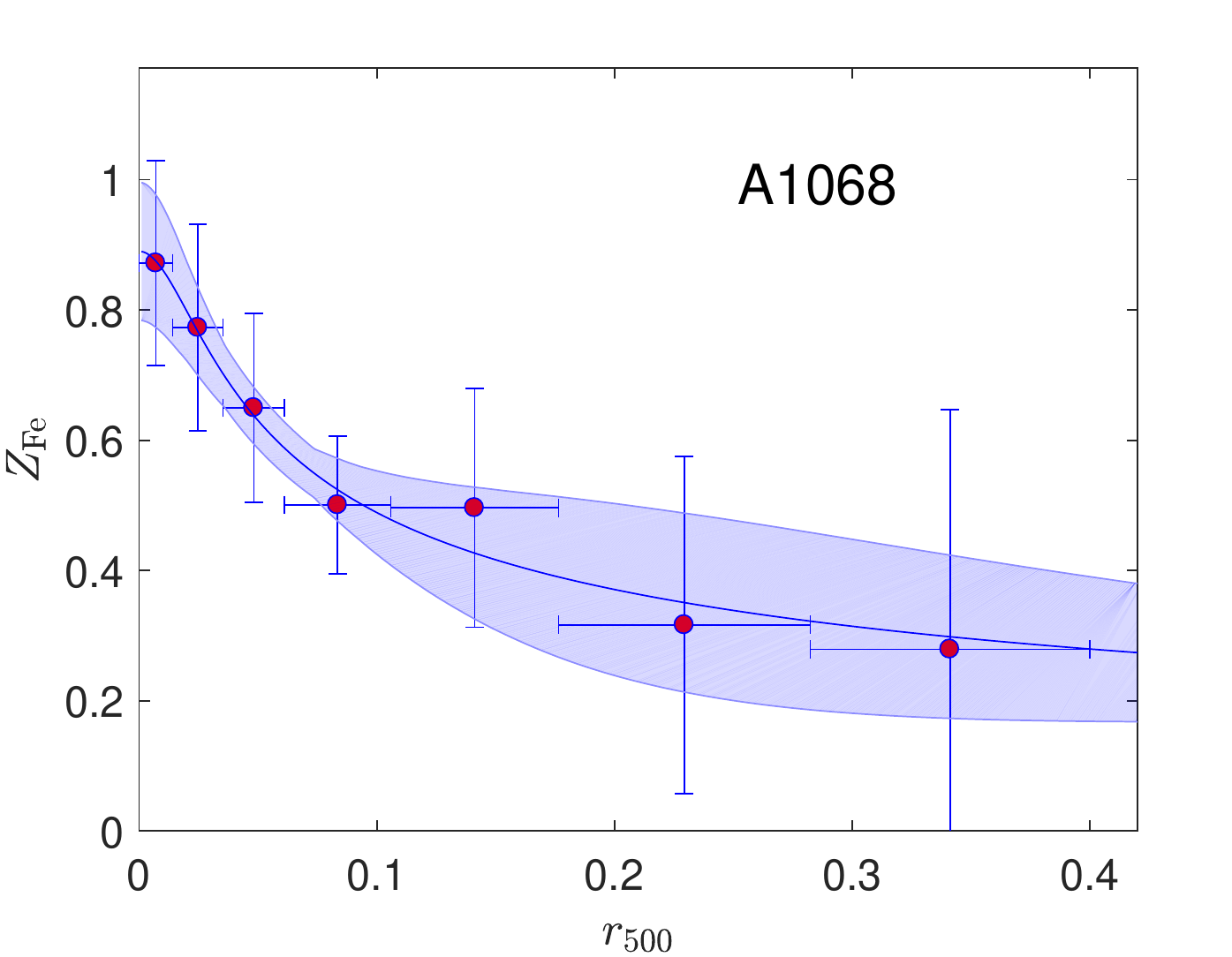}
\includegraphics[width=0.245\textwidth, trim=105 240 105 240, clip]{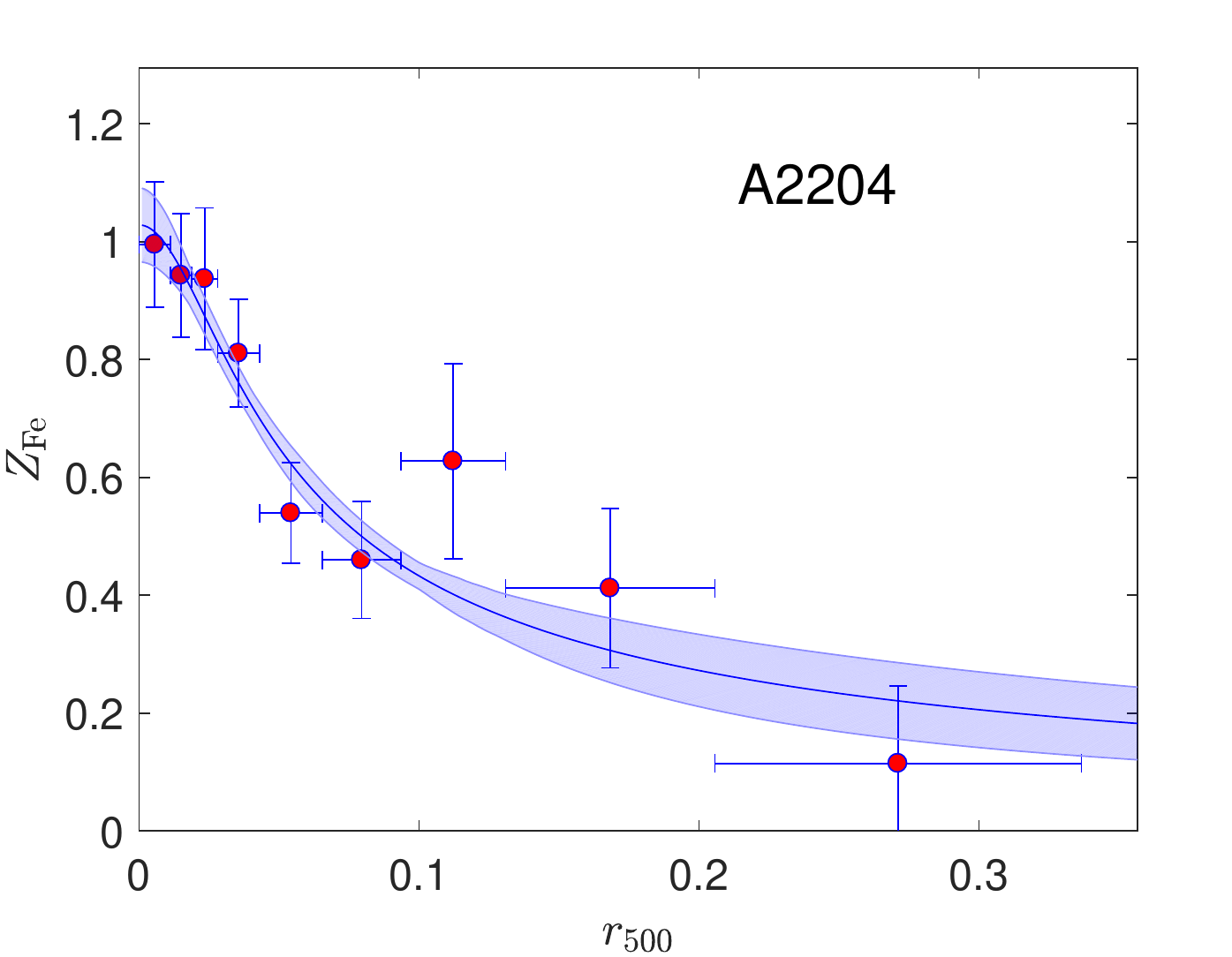}
\includegraphics[width=0.245\textwidth, trim=105 240 105 240, clip]{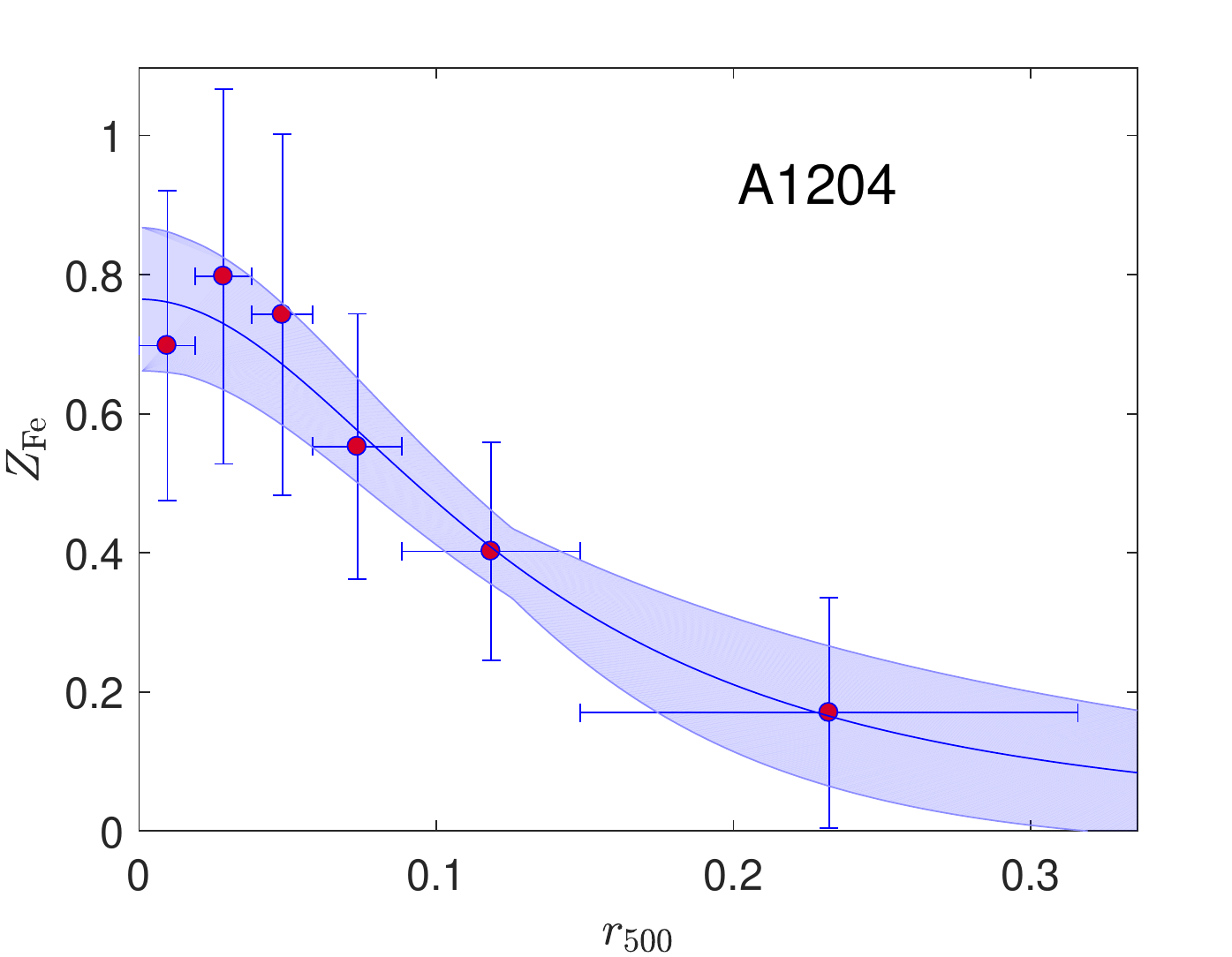}
\includegraphics[width=0.245\textwidth, trim=105 240 105 240, clip]{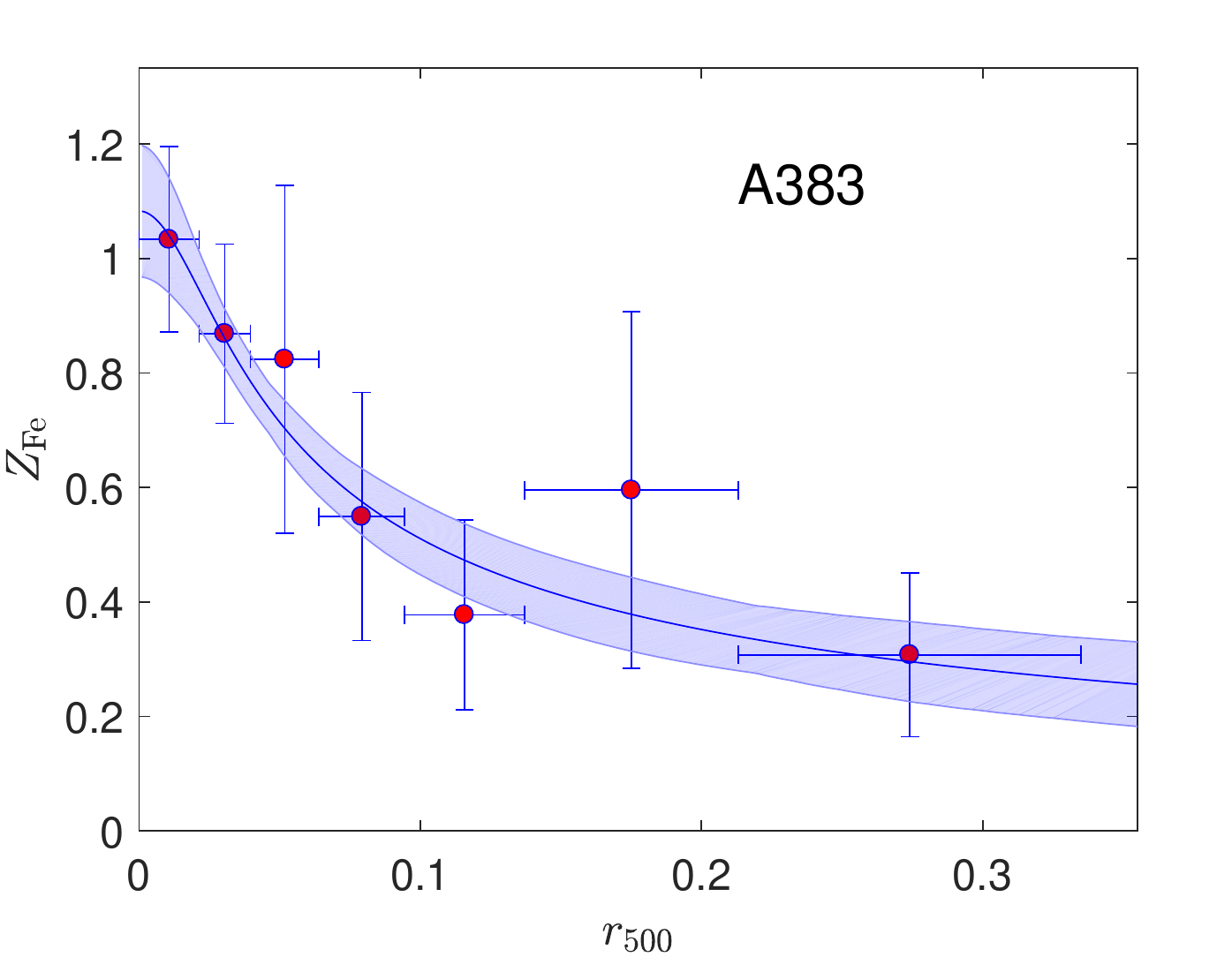}
\includegraphics[width=0.245\textwidth, trim=105 240 105 240, clip]{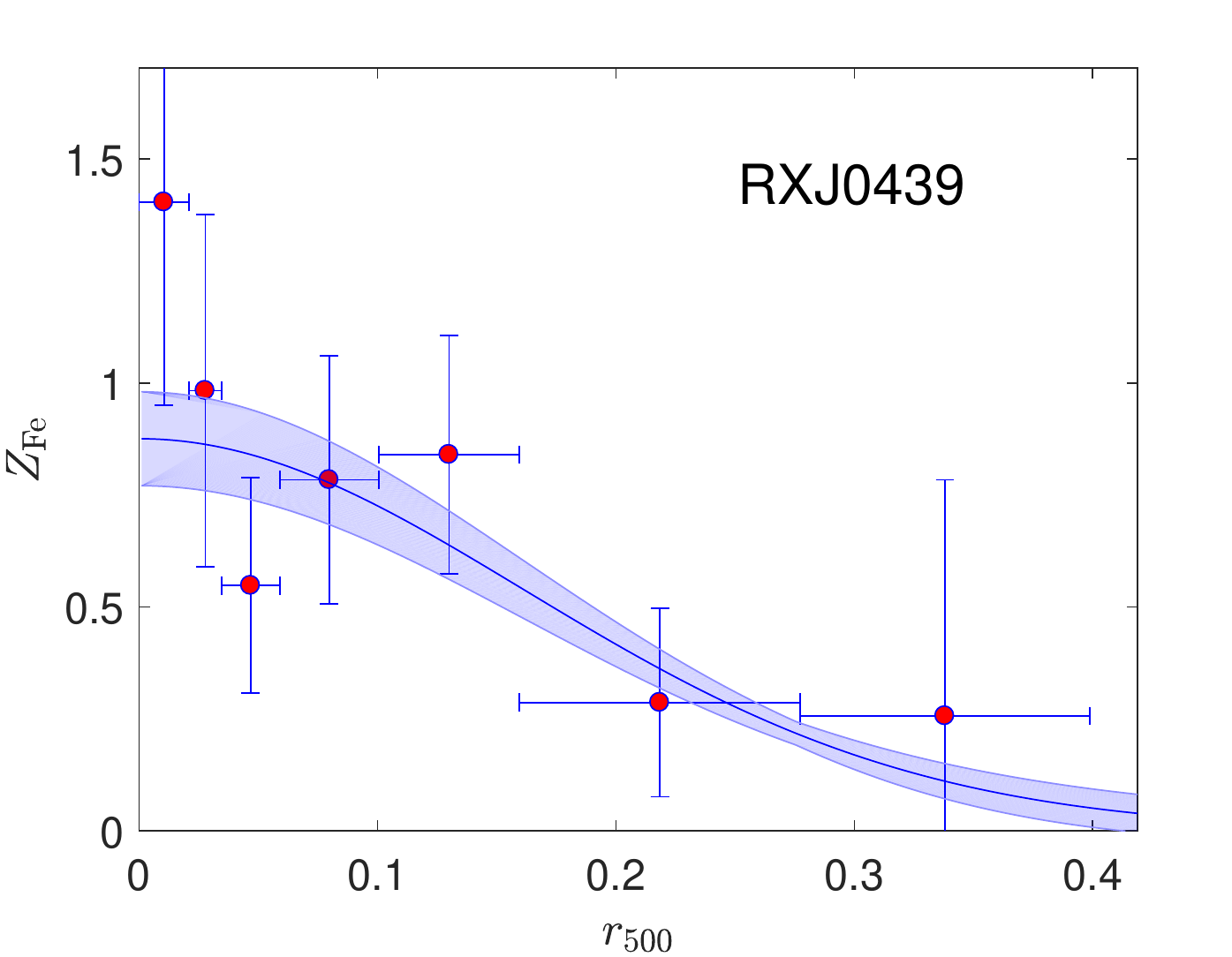}
\includegraphics[width=0.245\textwidth, trim=105 240 105 240, clip]{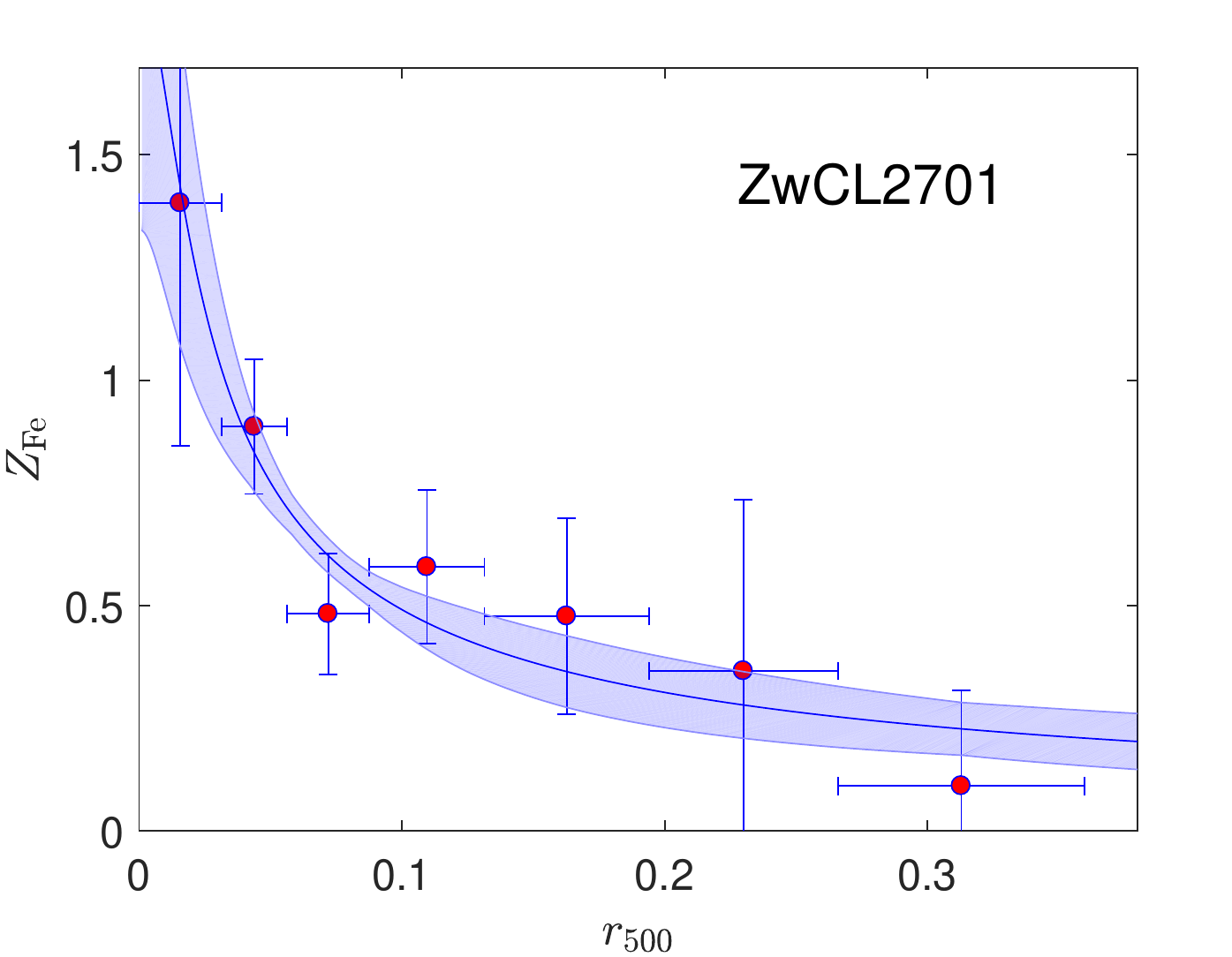}
\includegraphics[width=0.245\textwidth, trim=105 240 105 240, clip]{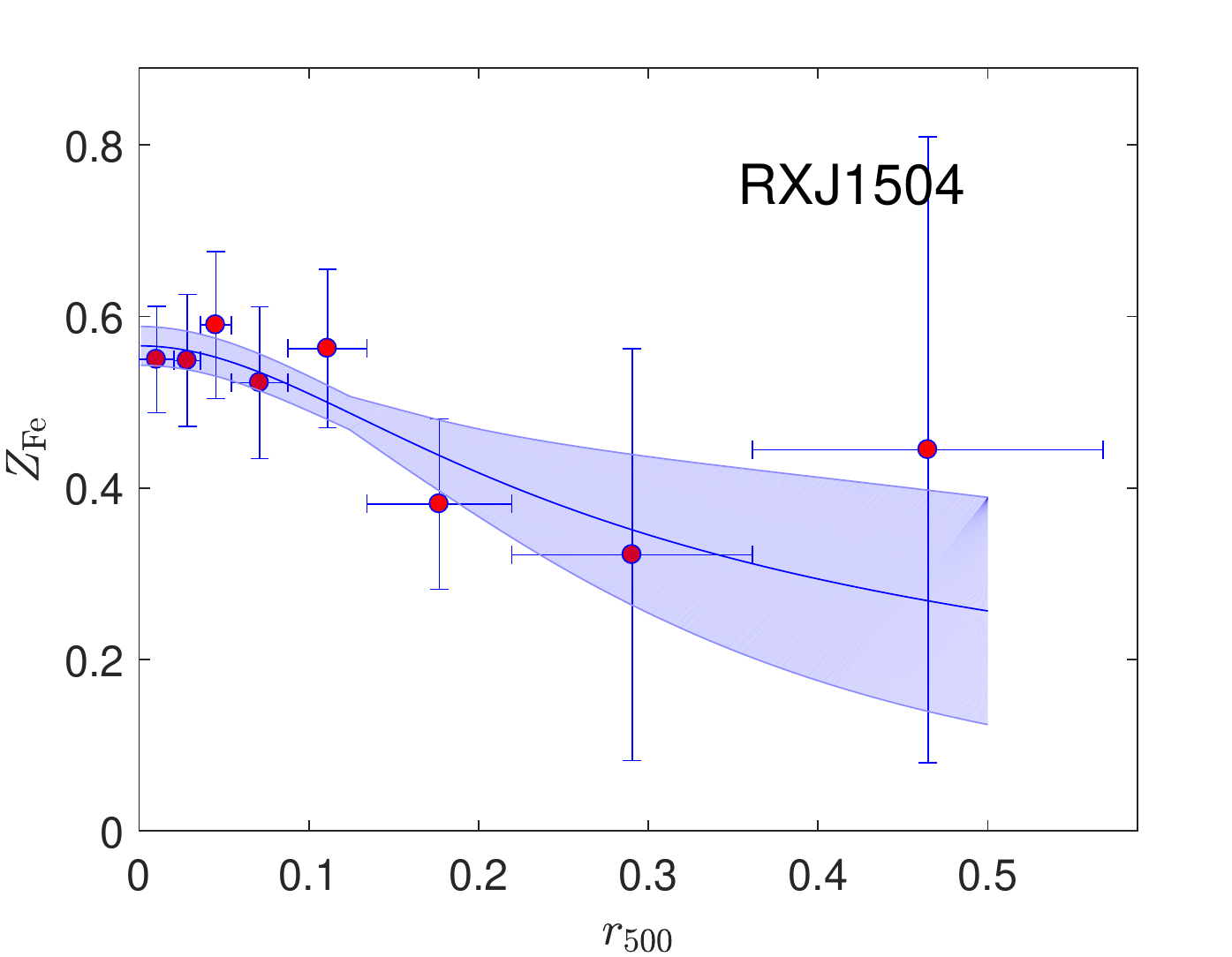}
\includegraphics[width=0.245\textwidth, trim=105 240 105 240, clip]{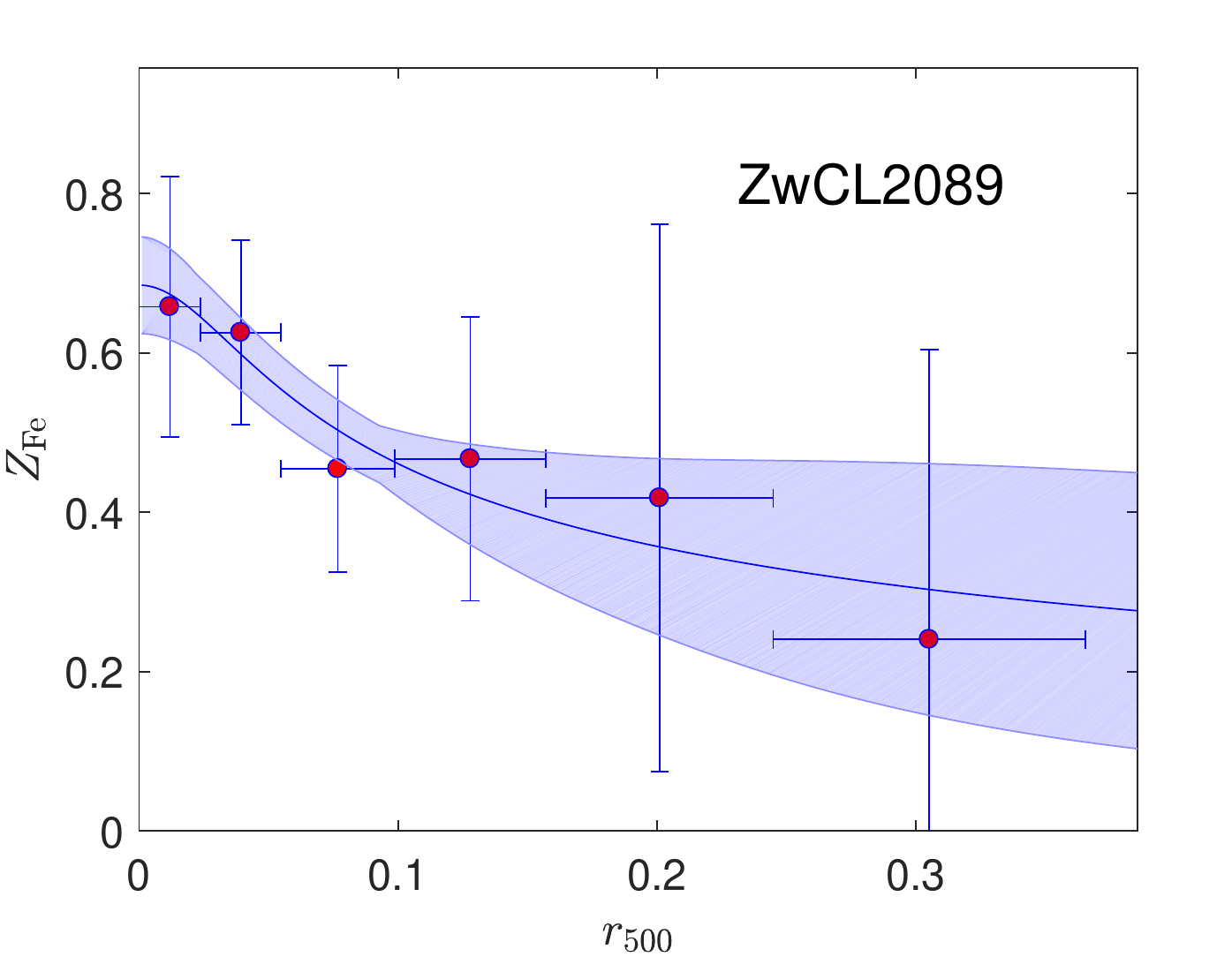}
\includegraphics[width=0.245\textwidth, trim=105 240 105 240, clip]{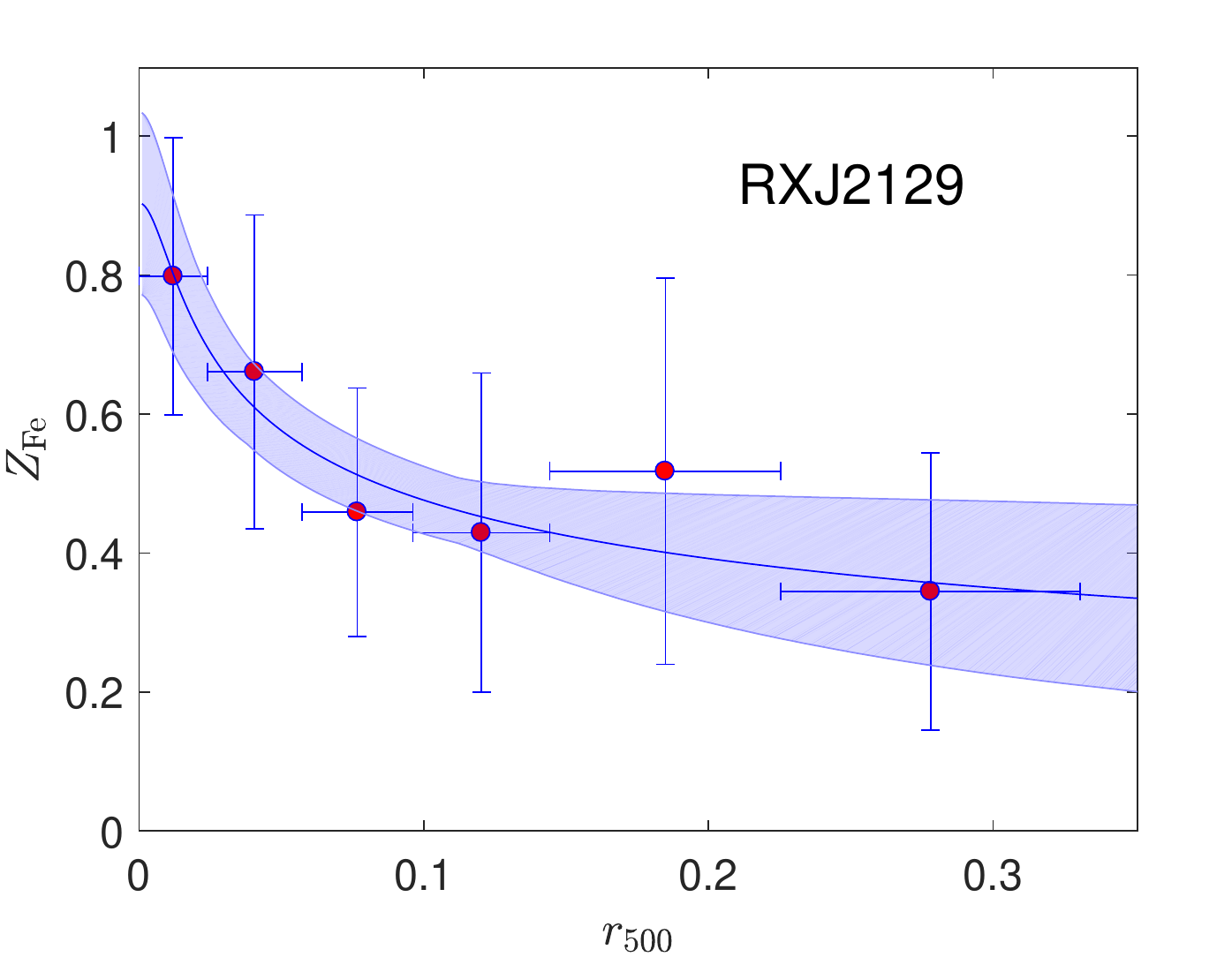}
\includegraphics[width=0.245\textwidth, trim=105 240 105 240, clip]{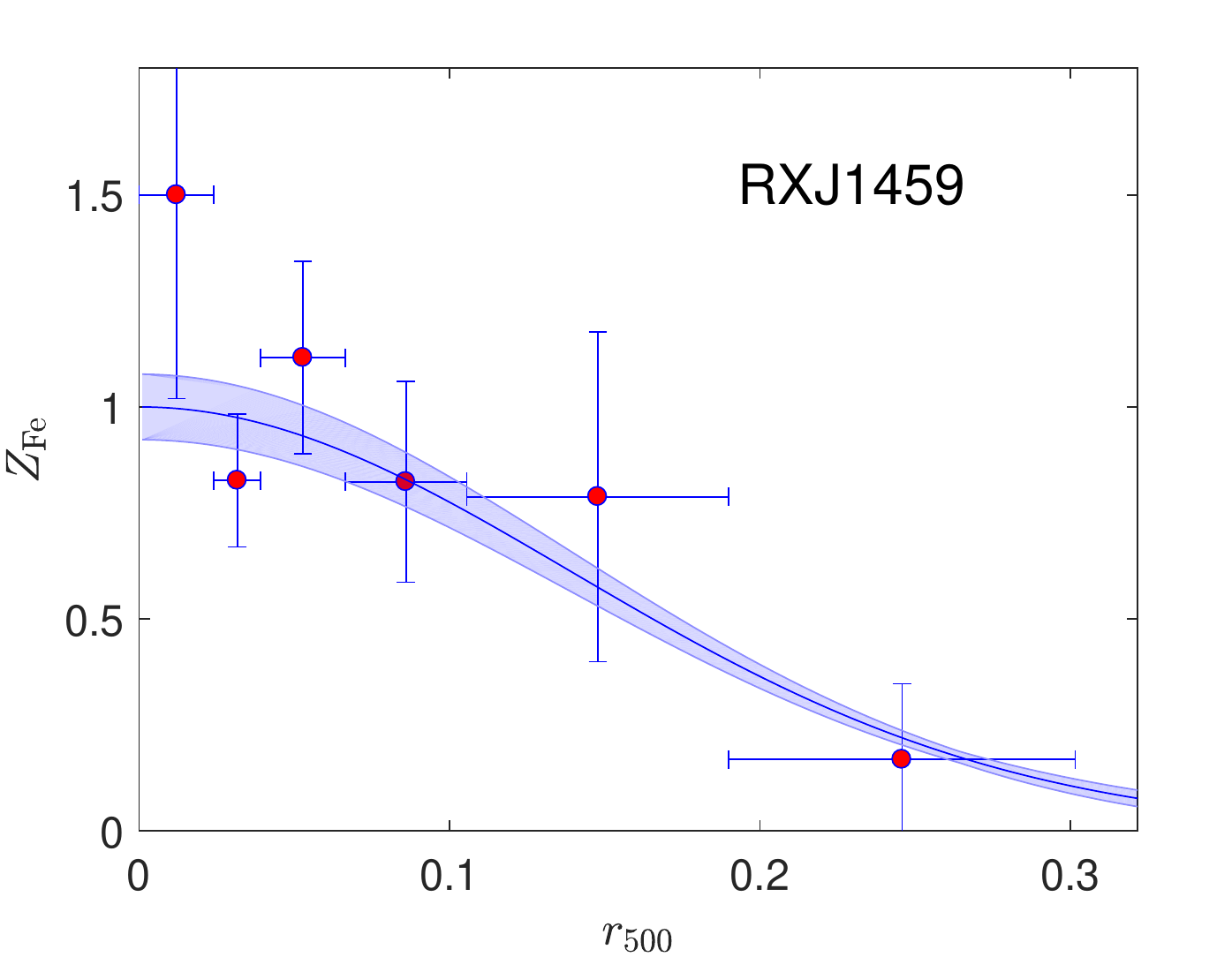}
\includegraphics[width=0.245\textwidth, trim=105 240 105 240, clip]{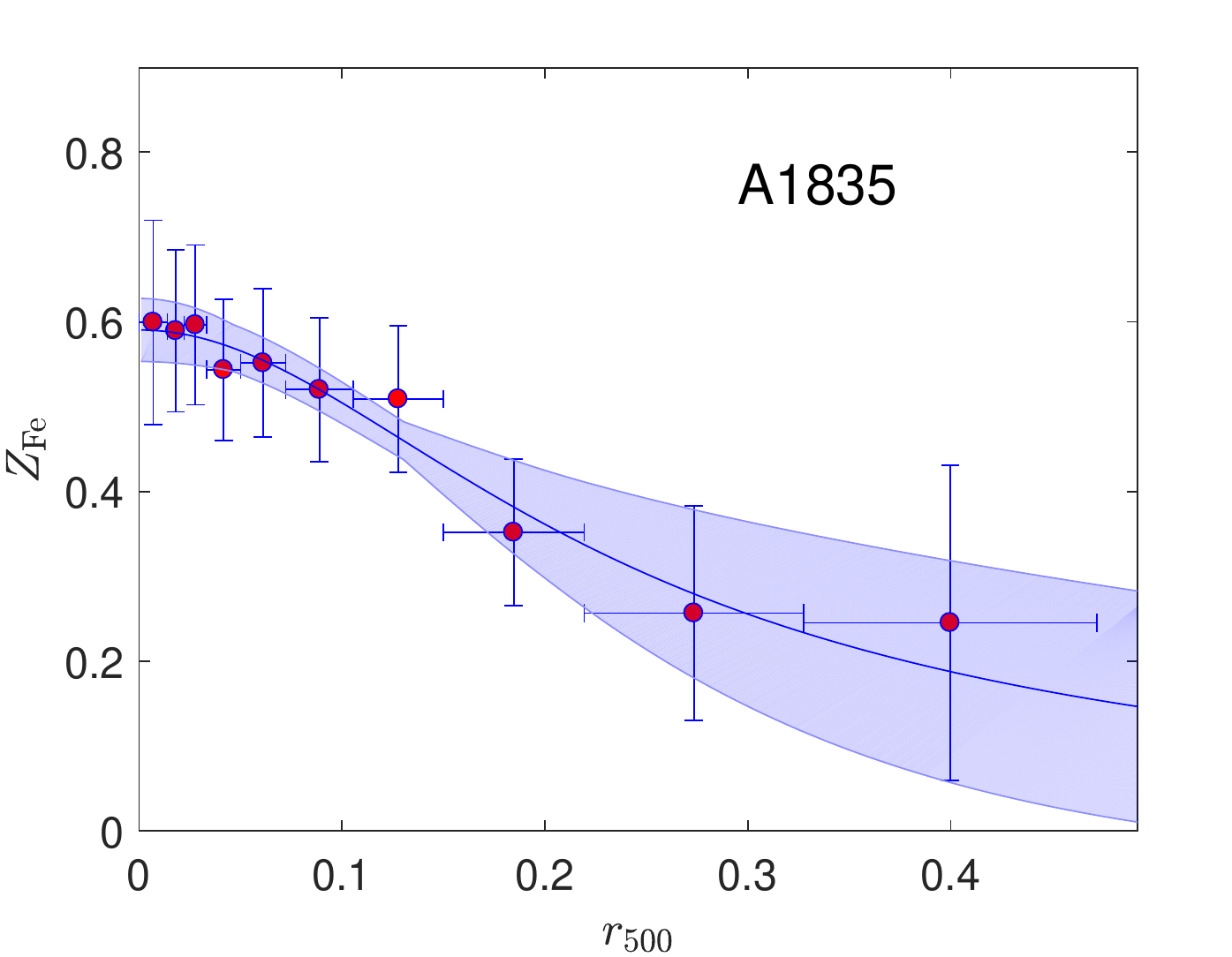}
\includegraphics[width=0.245\textwidth, trim=105 240 105 240, clip]{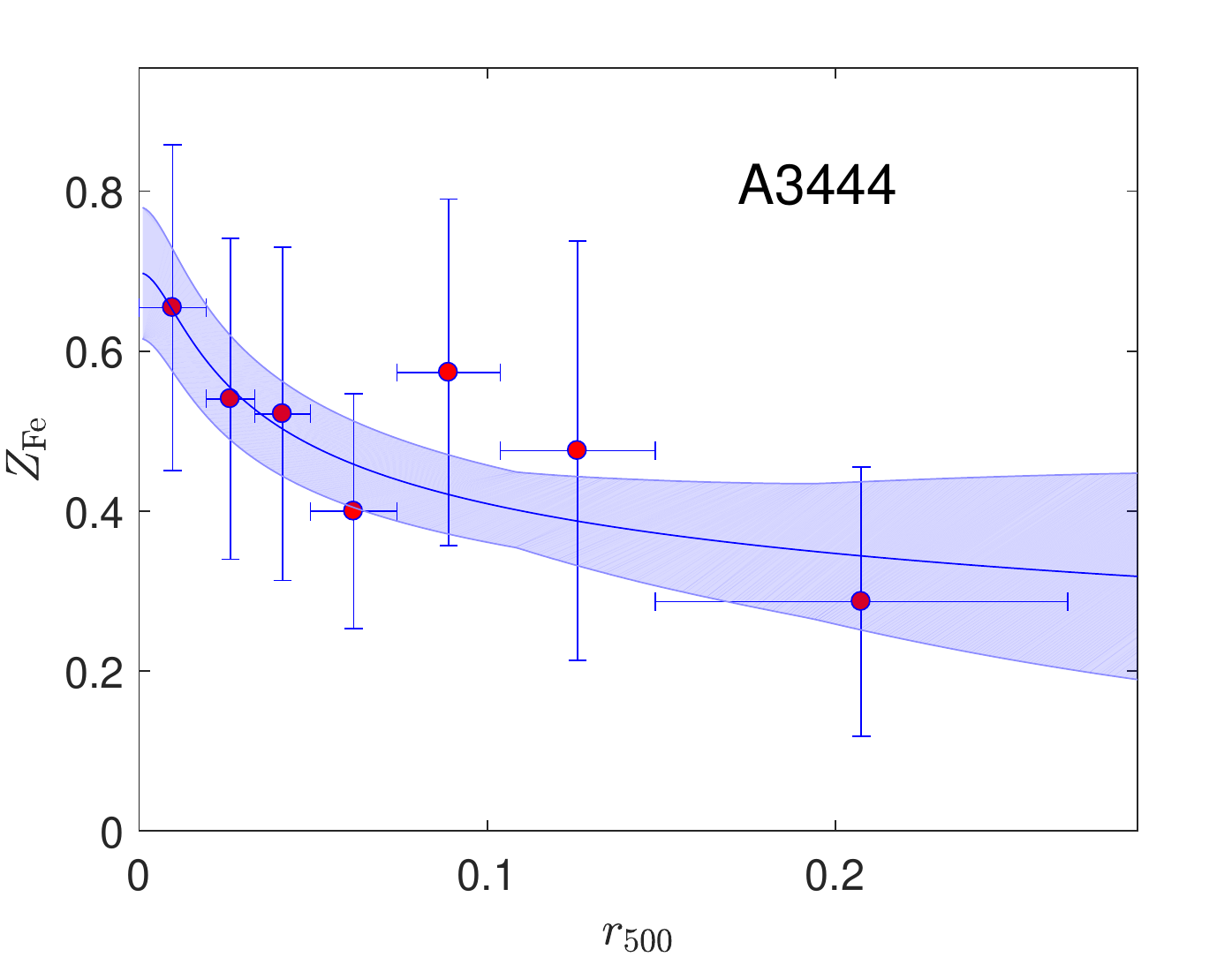}
\includegraphics[width=0.245\textwidth, trim=105 240 105 240, clip]{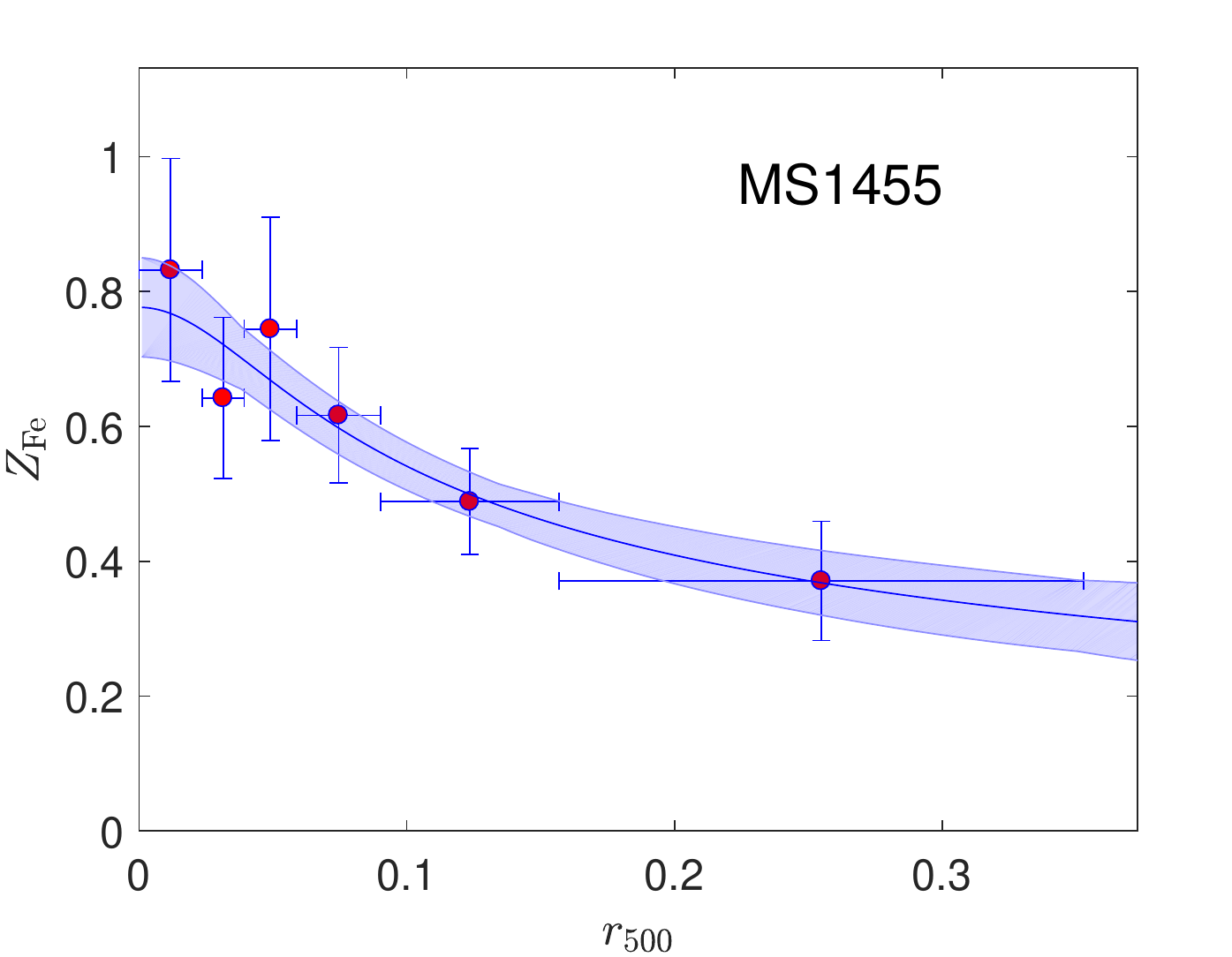}
\includegraphics[width=0.245\textwidth, trim=105 240 105 240, clip]{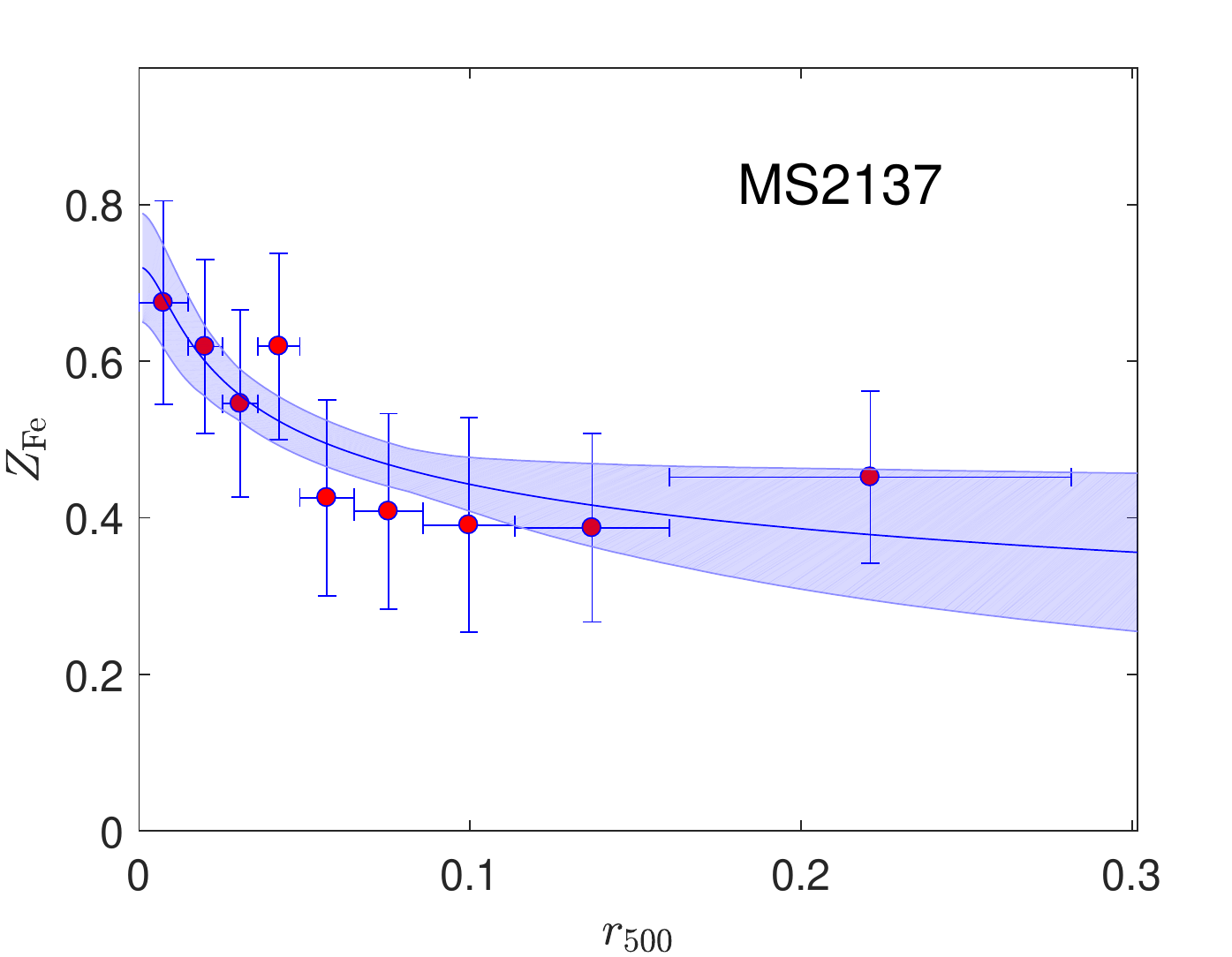}
\includegraphics[width=0.245\textwidth, trim=105 240 105 240, clip]{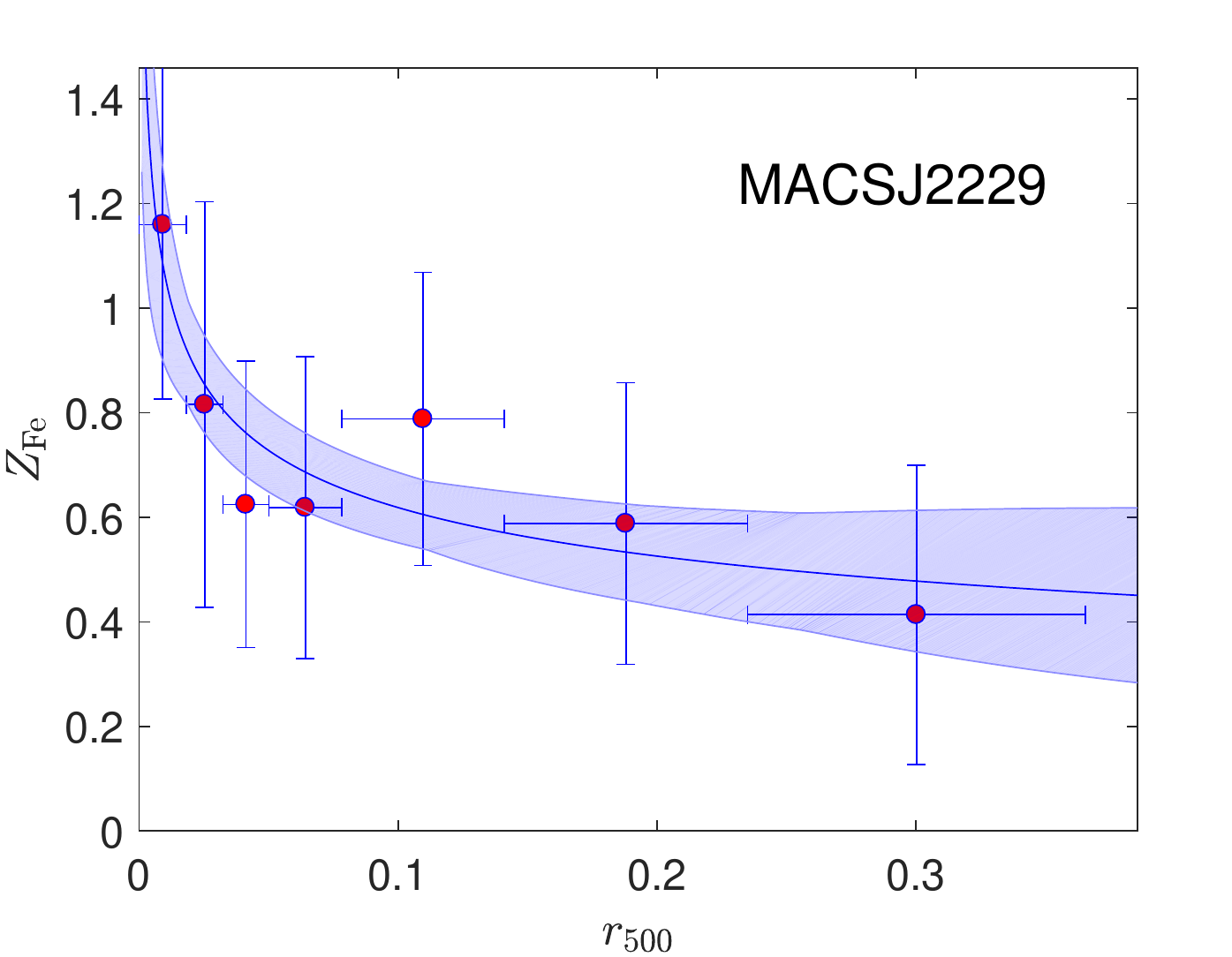}
\includegraphics[width=0.245\textwidth, trim=105 240 105 240, clip]{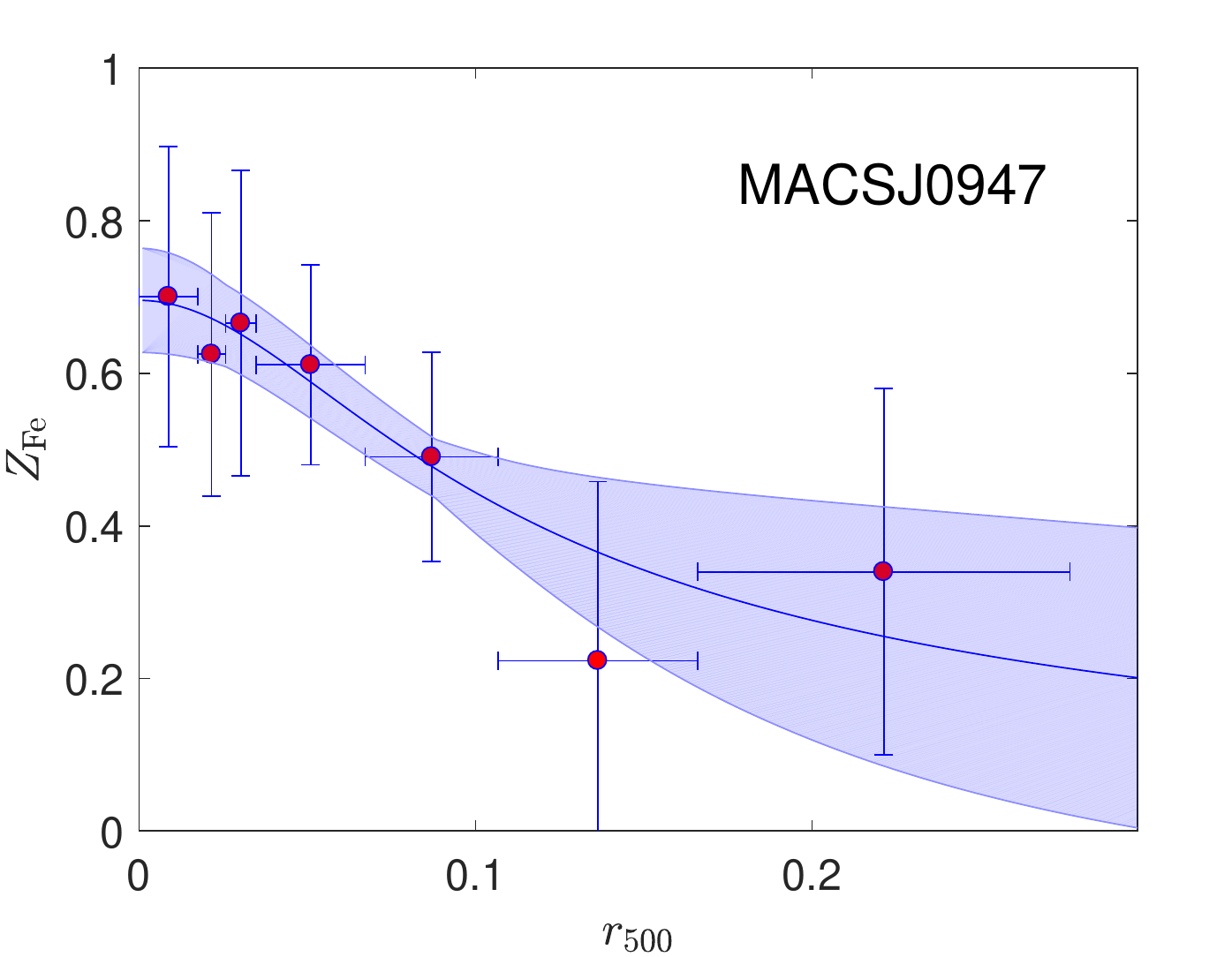}
\includegraphics[width=0.245\textwidth, trim=105 240 105 240, clip]{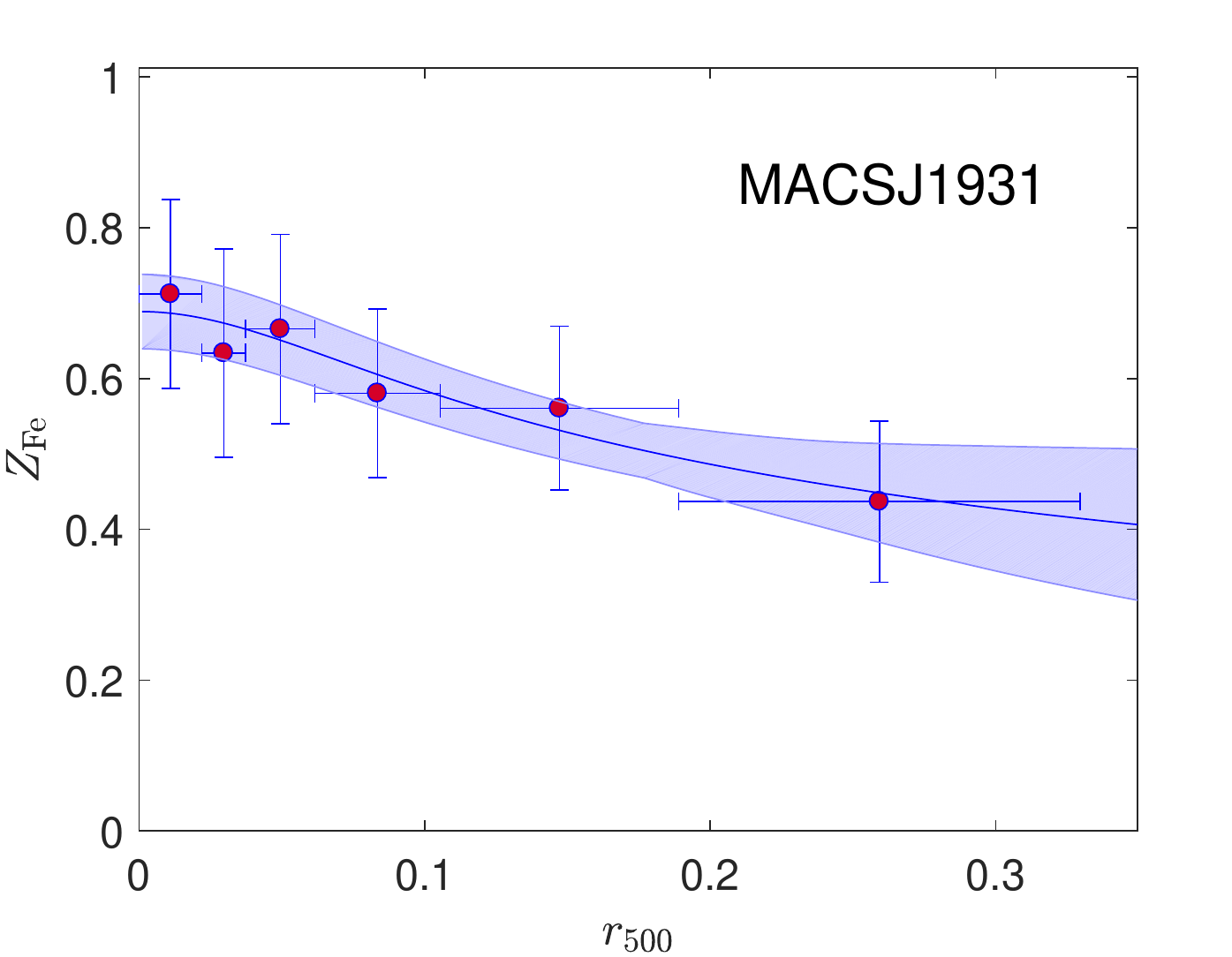}
\includegraphics[width=0.245\textwidth, trim=105 240 105 240, clip]{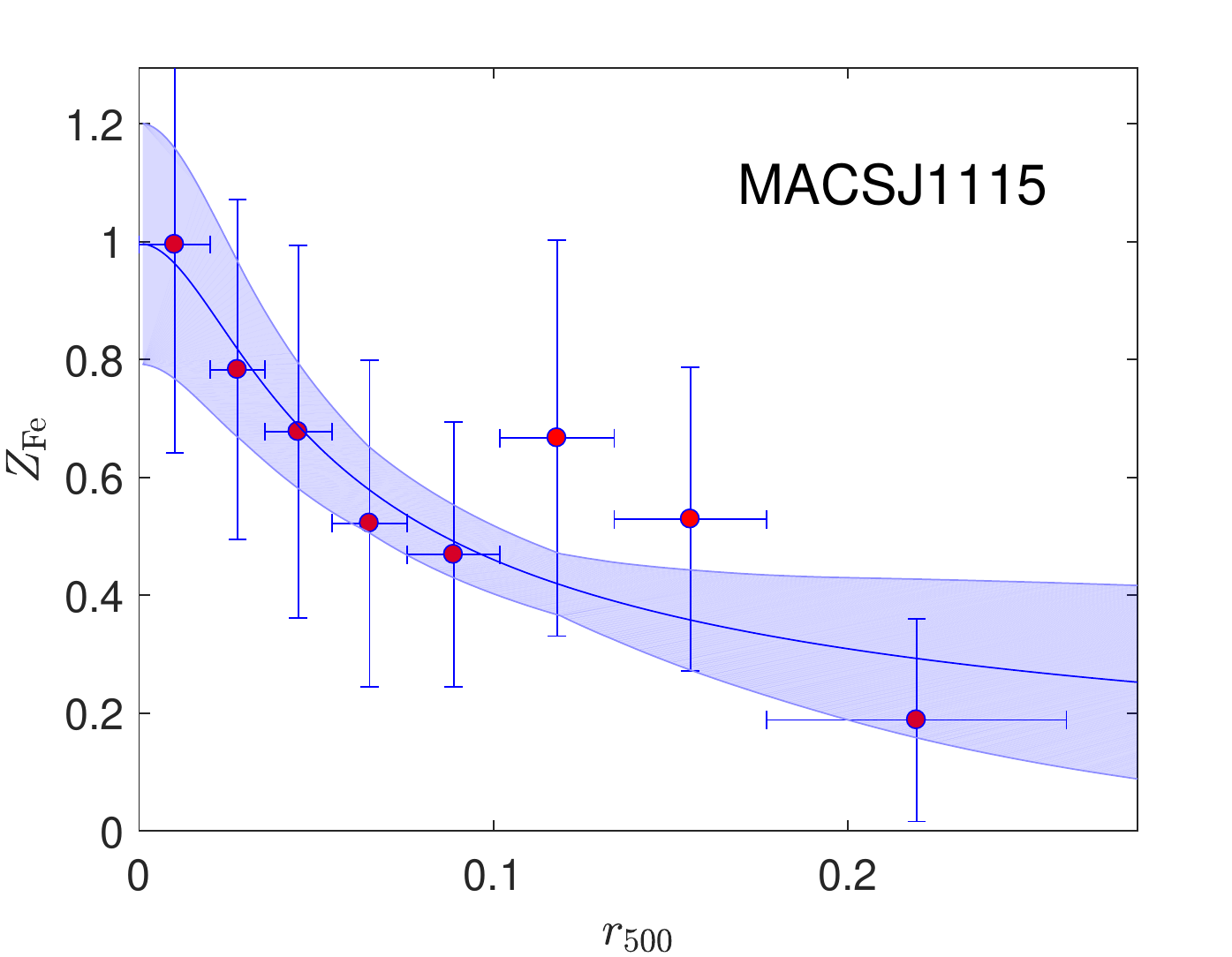}
\includegraphics[width=0.245\textwidth, trim=105 240 105 240, clip]{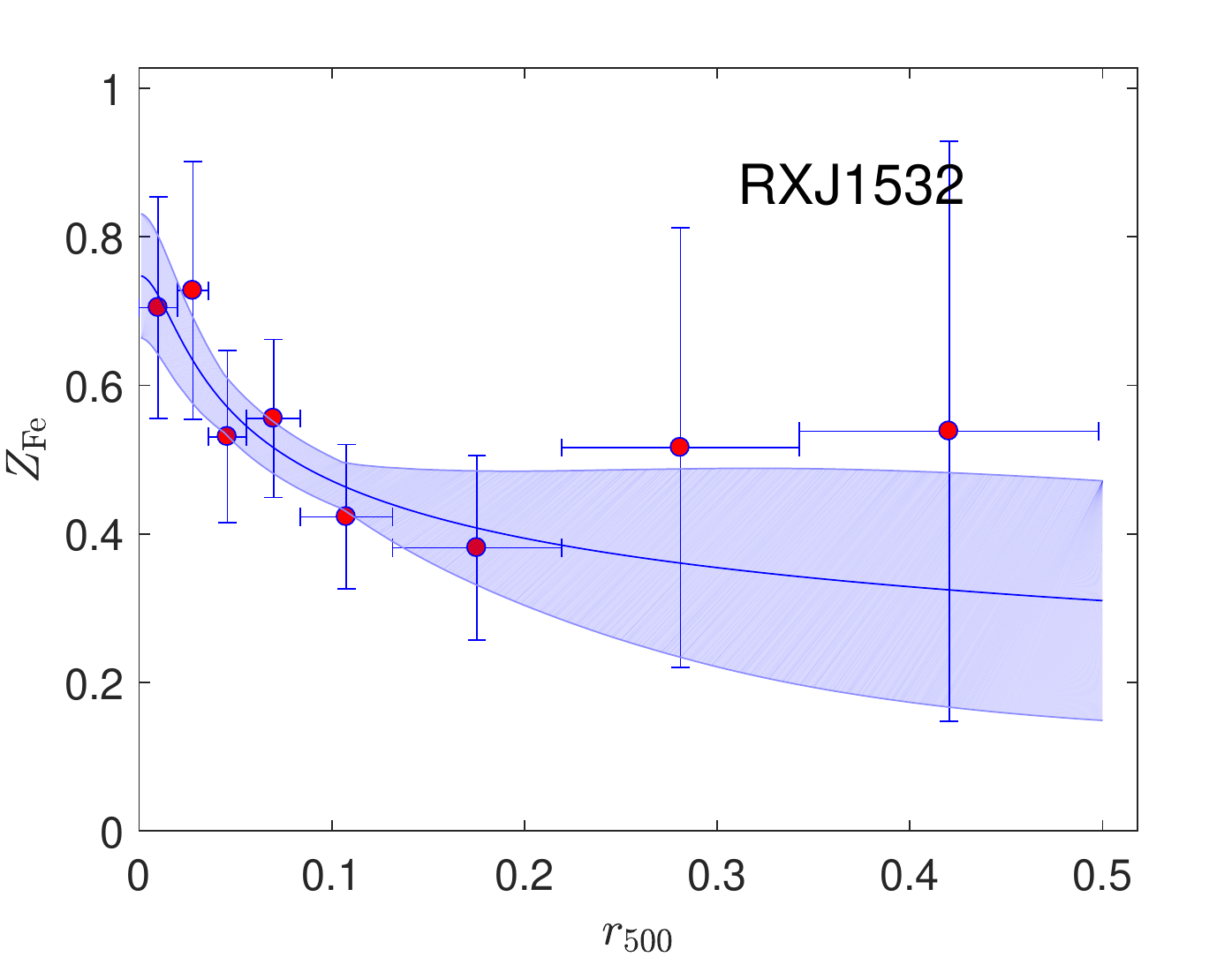}
\includegraphics[width=0.245\textwidth, trim=105 240 105 240, clip]{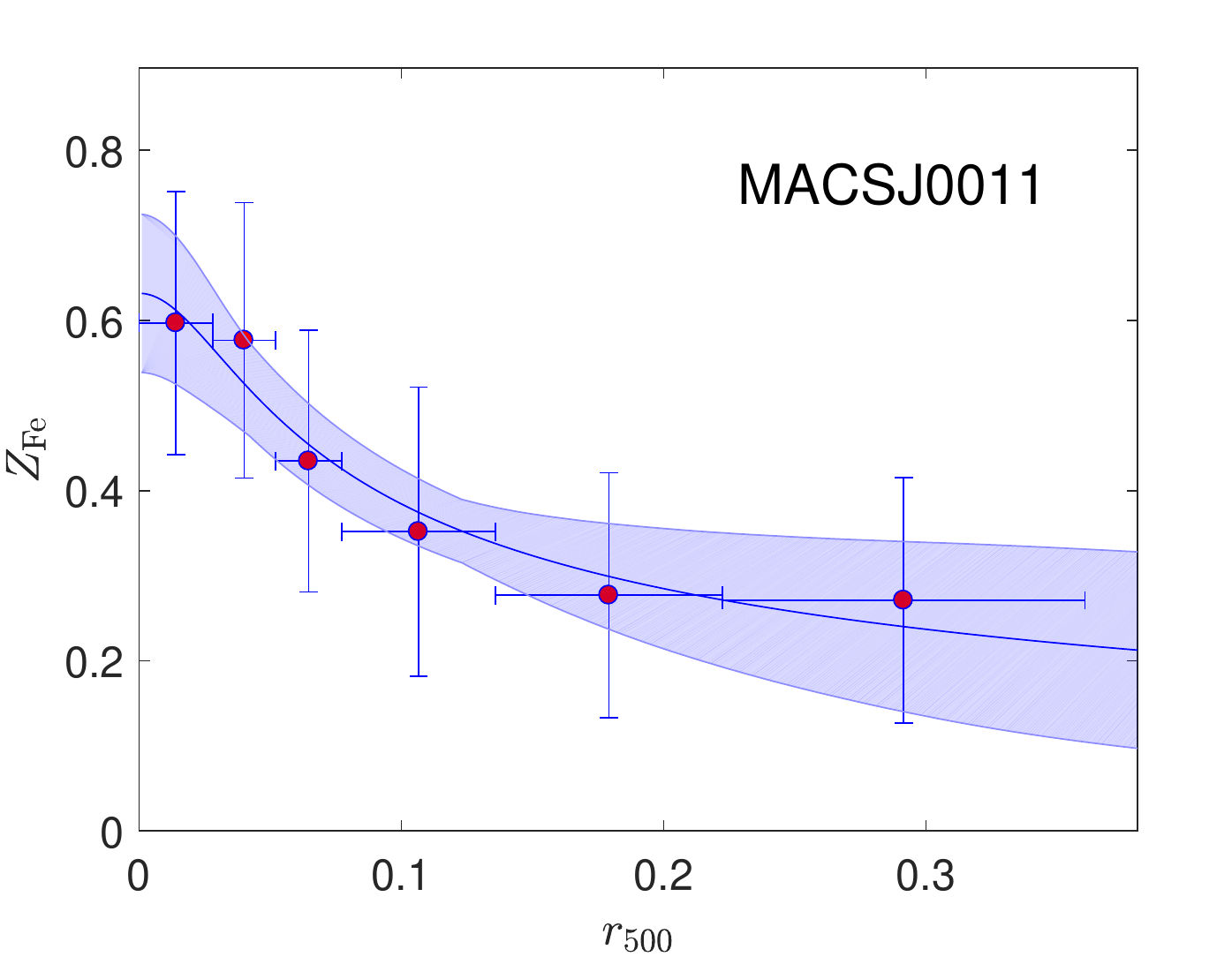}
\includegraphics[width=0.245\textwidth, trim=105 240 105 240, clip]{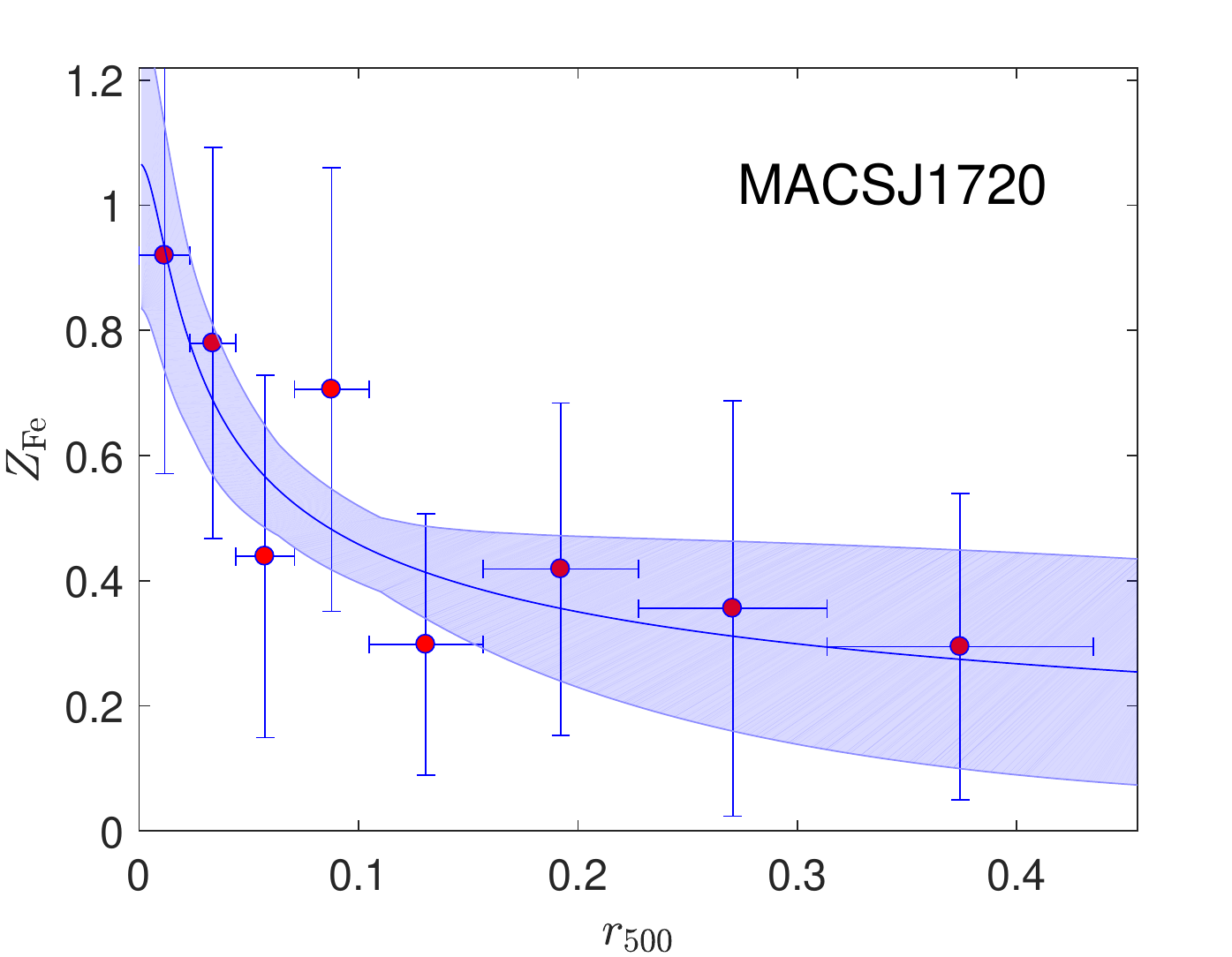}
\includegraphics[width=0.245\textwidth, trim=105 240 105 240, clip]{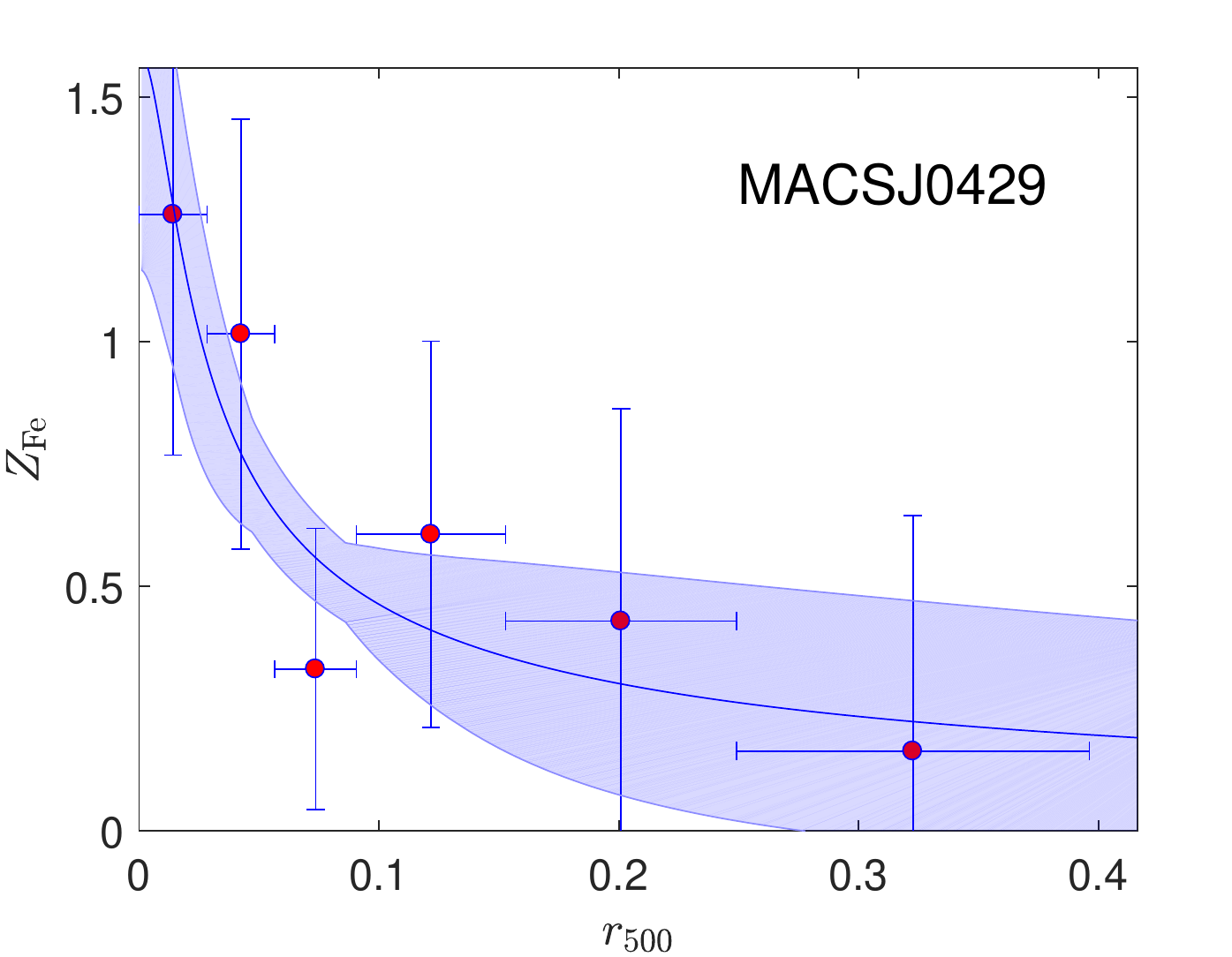}
\end{figure*}

\begin{figure*}
\label{profiles_con}
\centering
\includegraphics[width=0.245\textwidth, trim=105 240 105 240, clip]{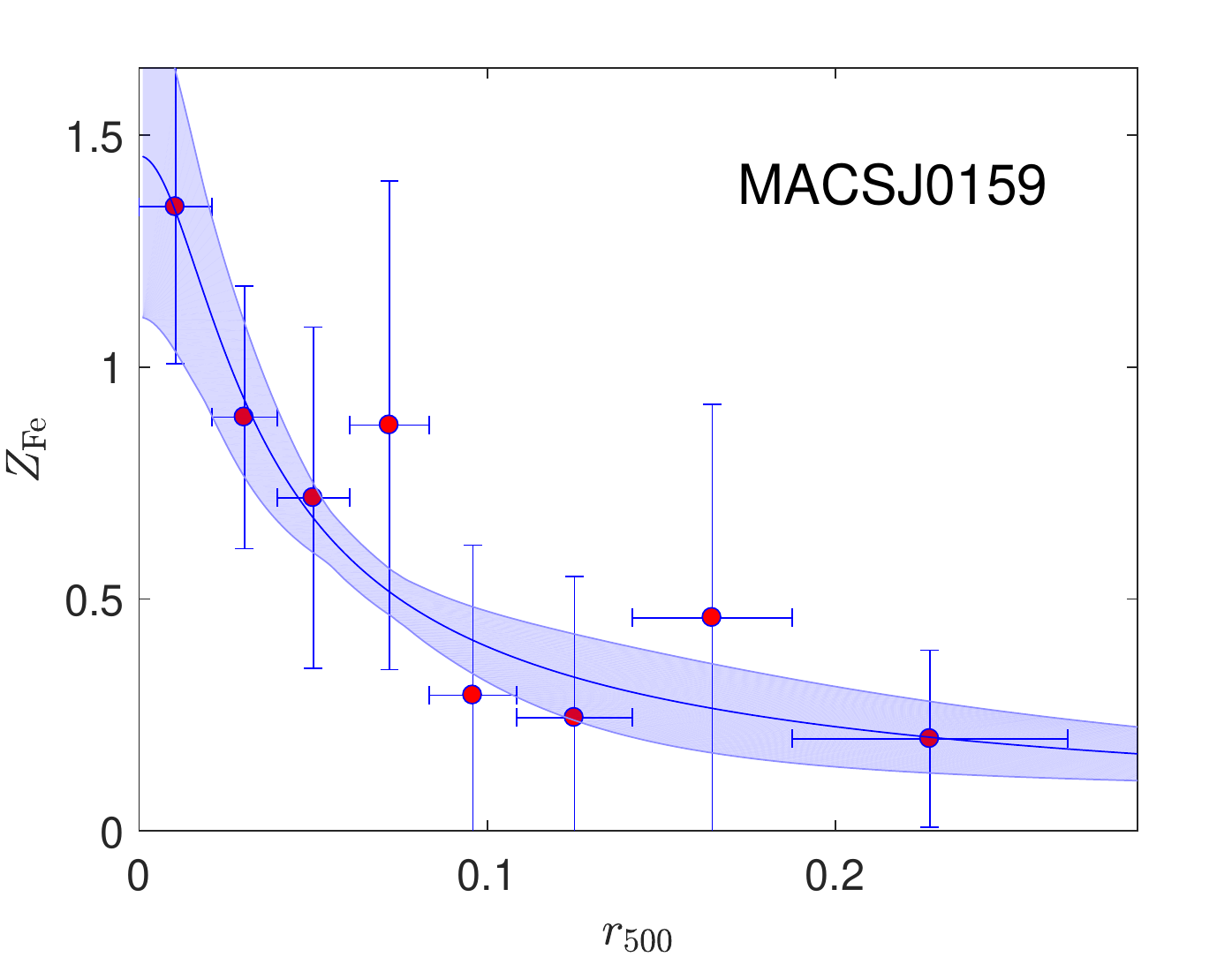}
\includegraphics[width=0.245\textwidth, trim=105 240 105 240, clip]{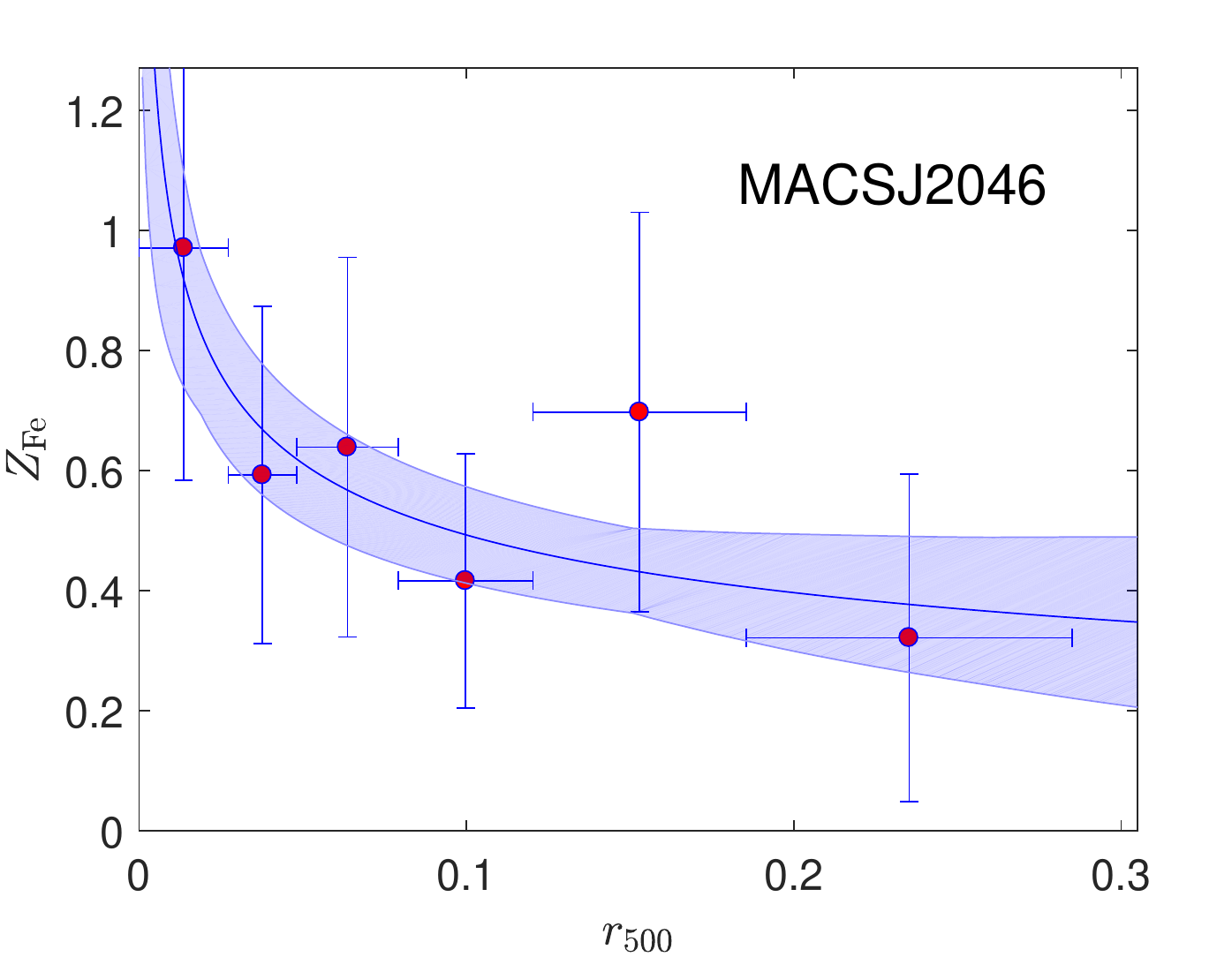}
\includegraphics[width=0.245\textwidth, trim=105 240 105 240, clip]{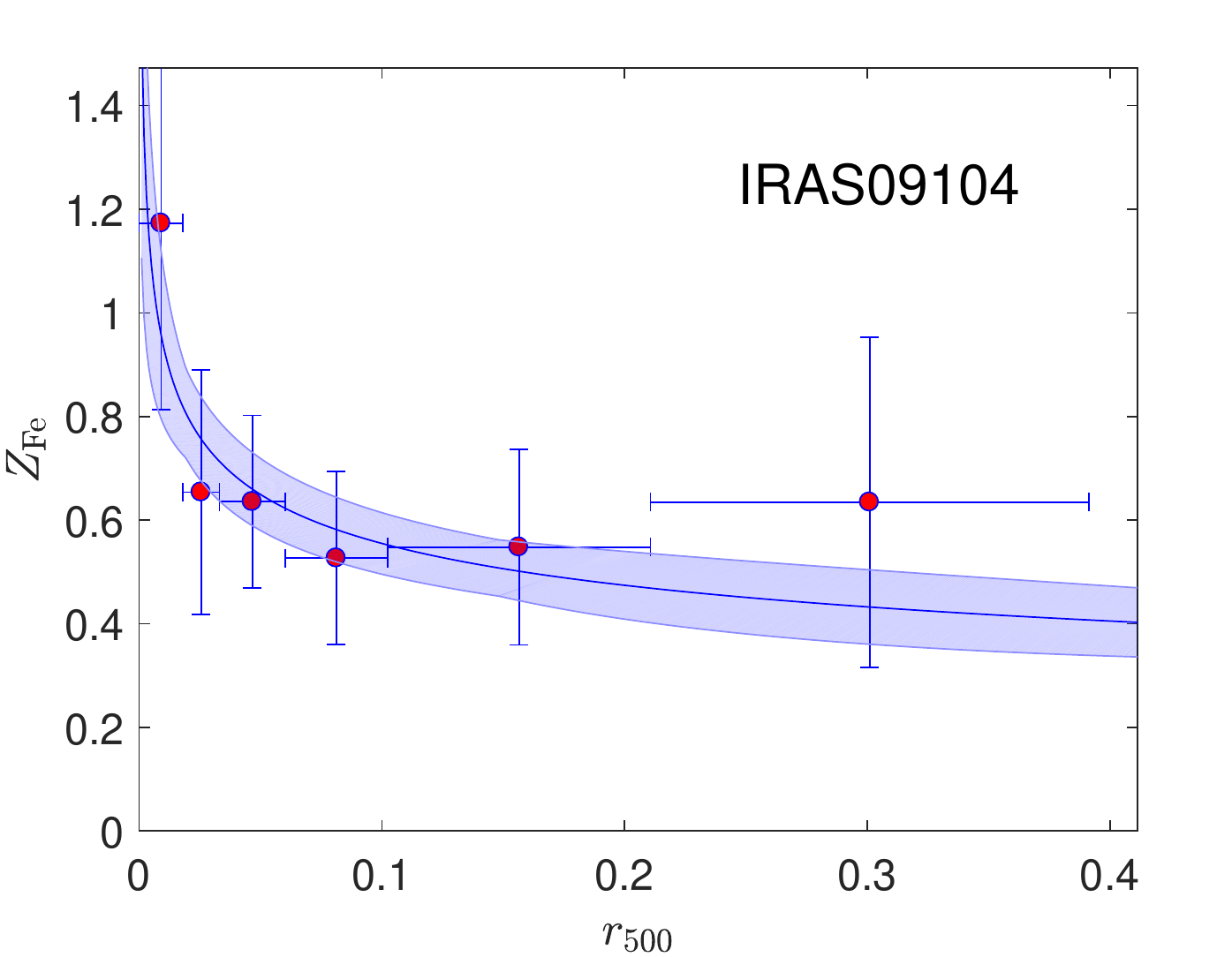}
\includegraphics[width=0.245\textwidth, trim=105 240 105 240, clip]{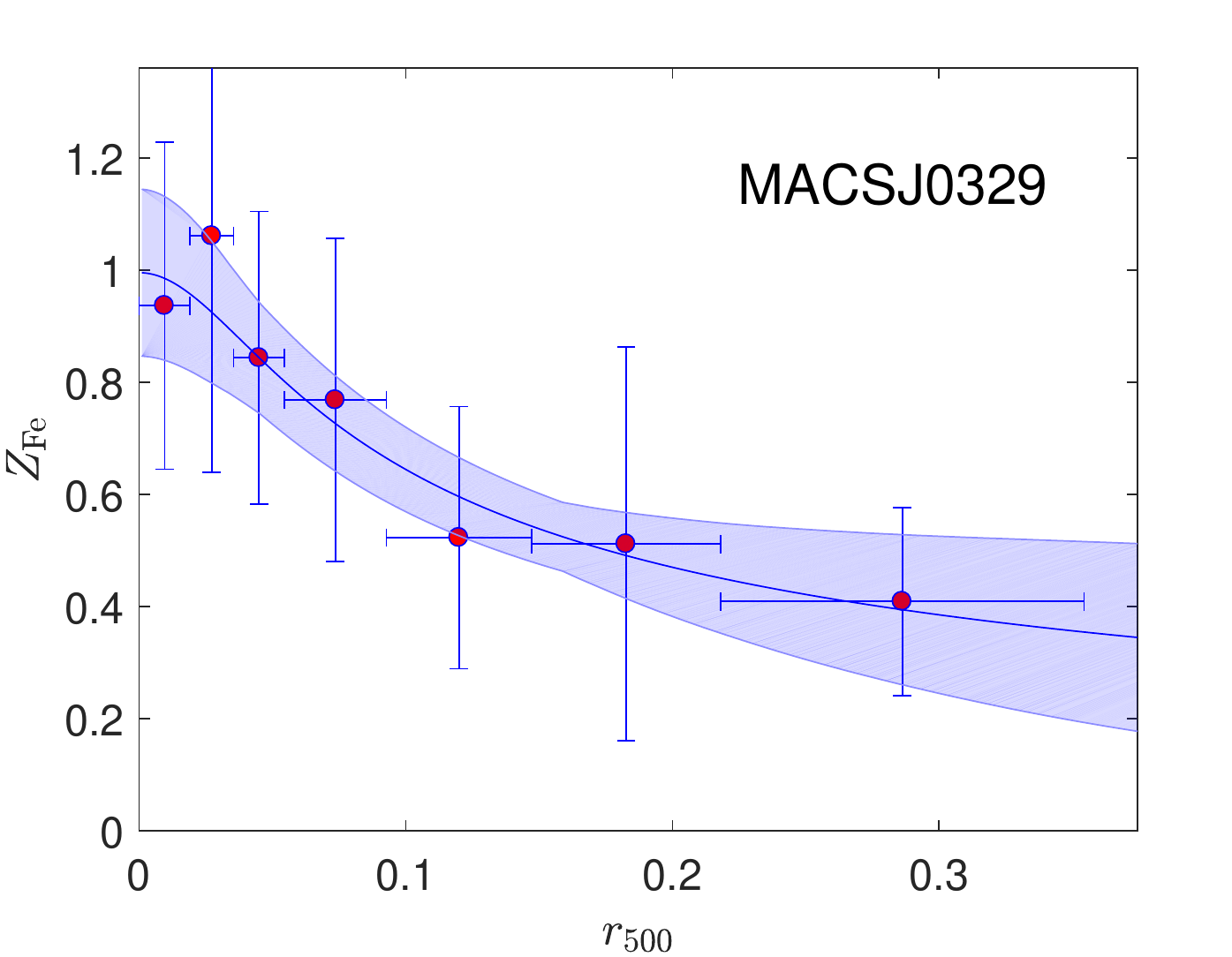}
\includegraphics[width=0.245\textwidth, trim=105 240 105 240, clip]{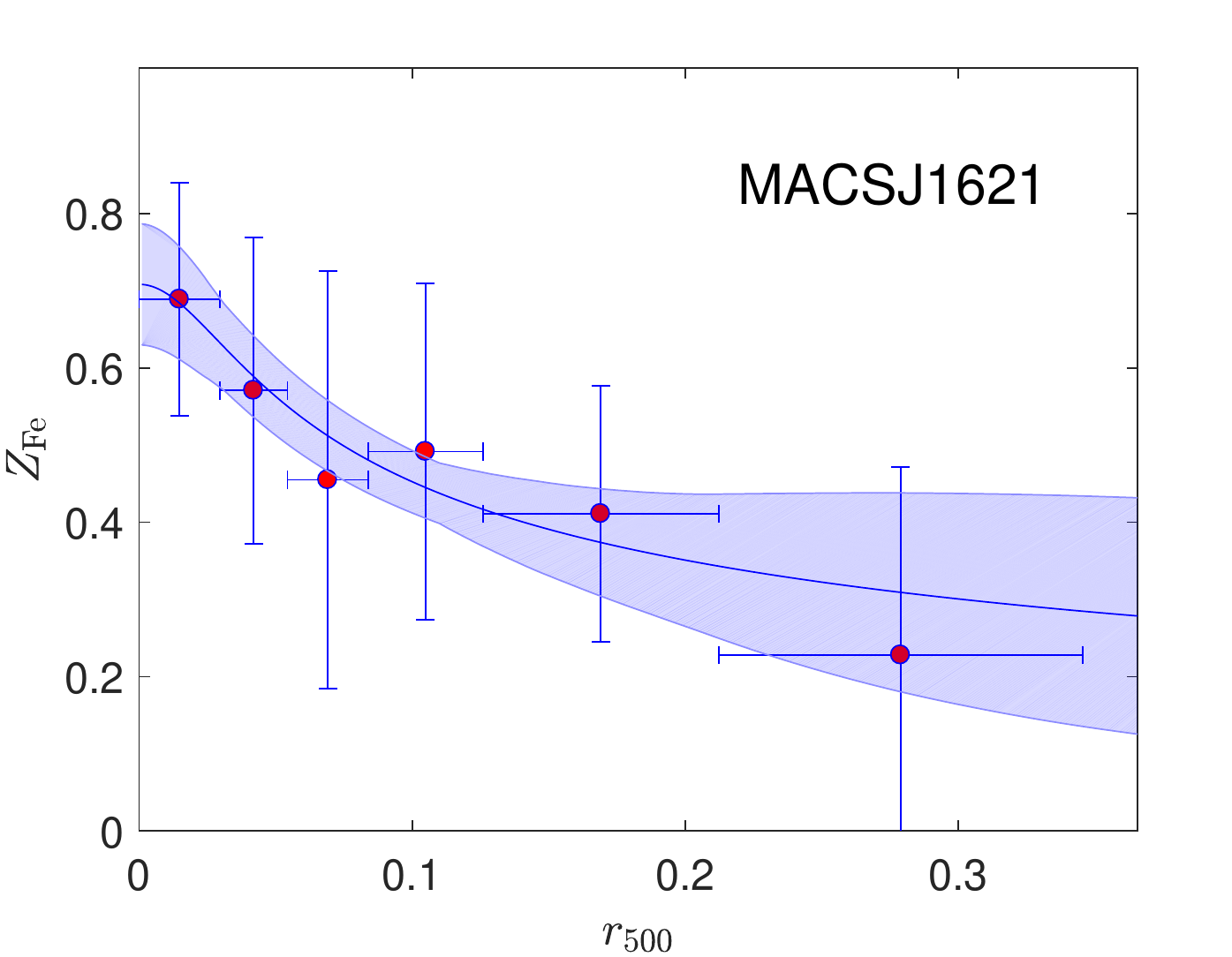}
\includegraphics[width=0.245\textwidth, trim=105 240 105 240, clip]{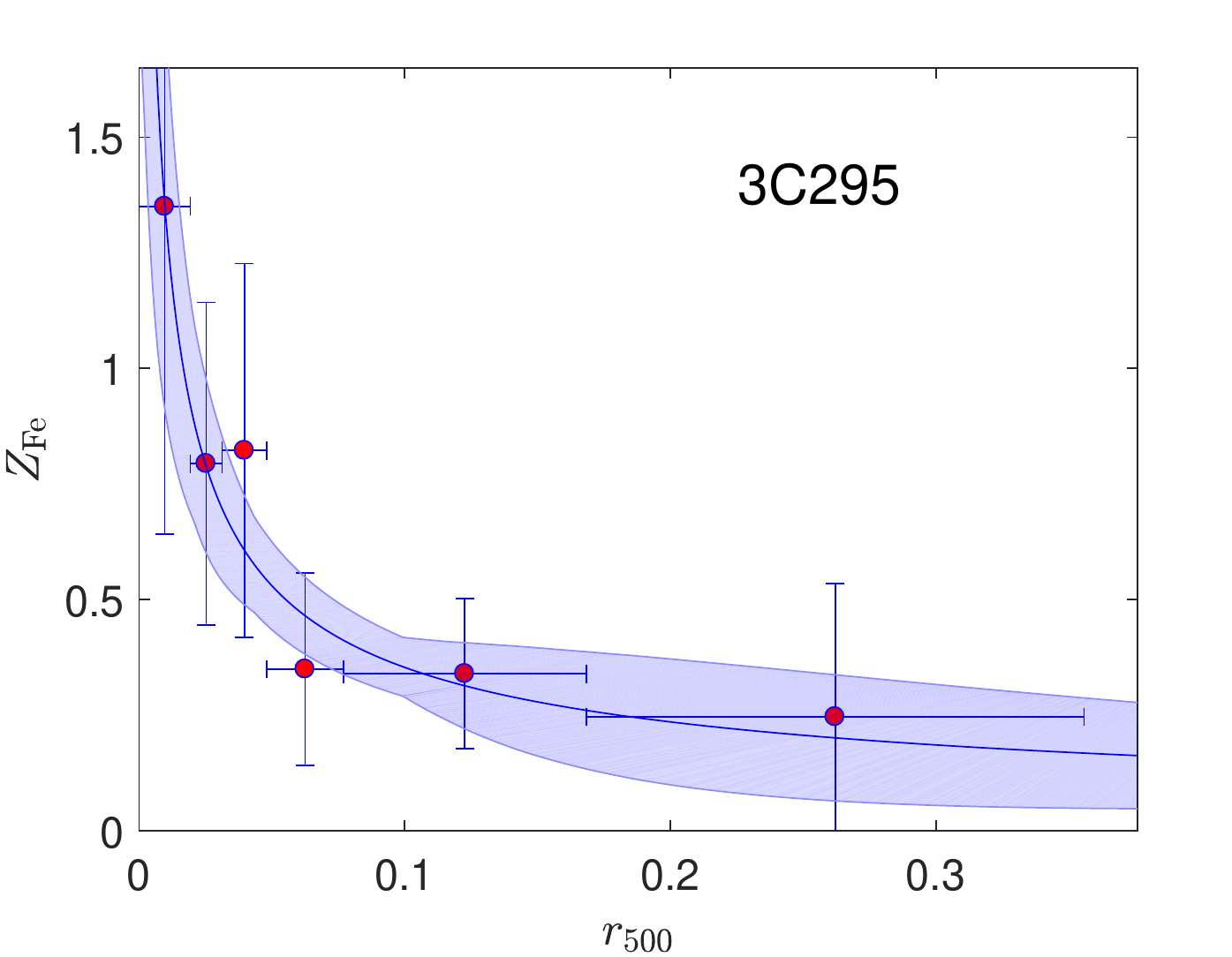}
\includegraphics[width=0.245\textwidth, trim=105 240 105 240, clip]{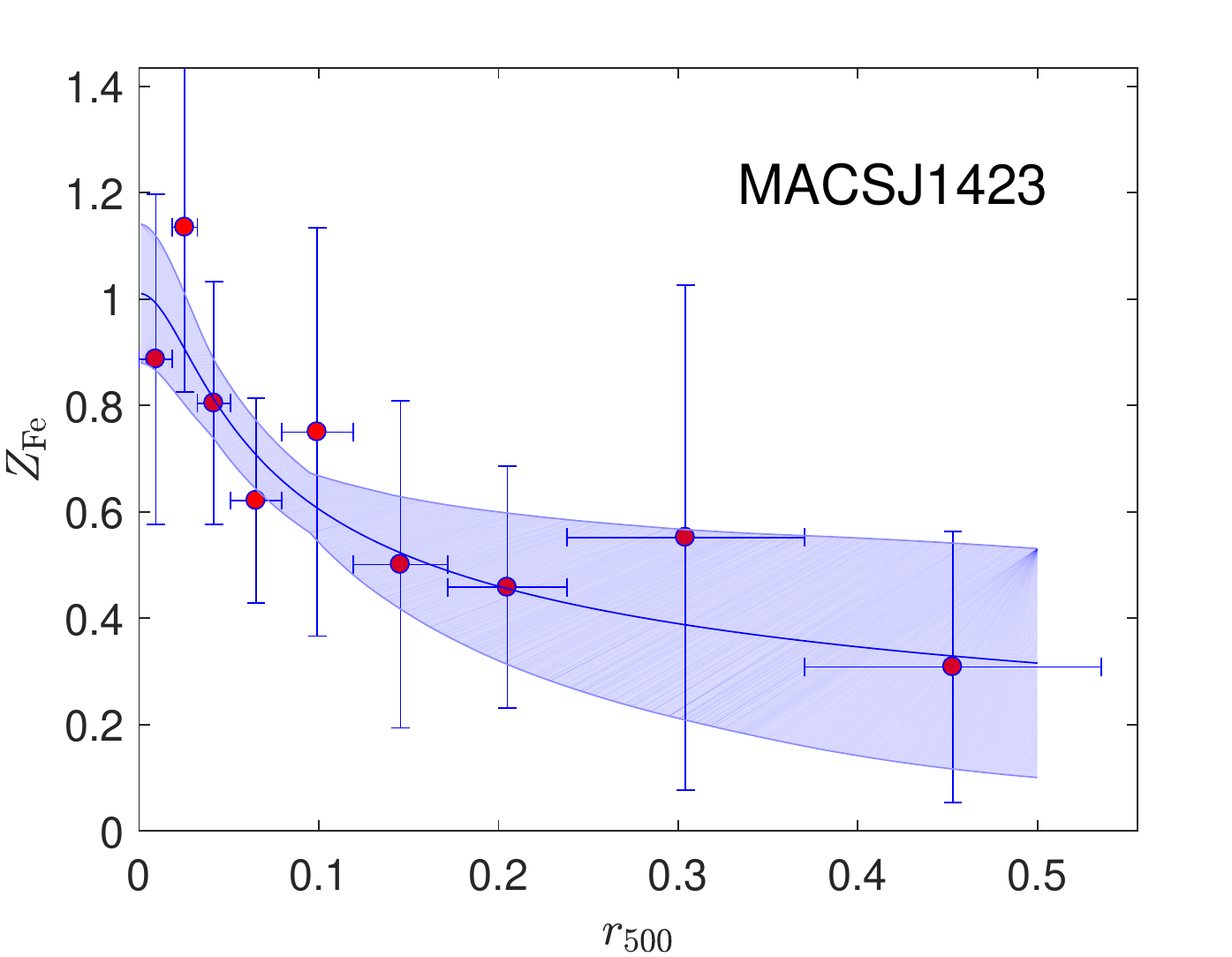}
\includegraphics[width=0.245\textwidth, trim=105 240 105 240, clip]{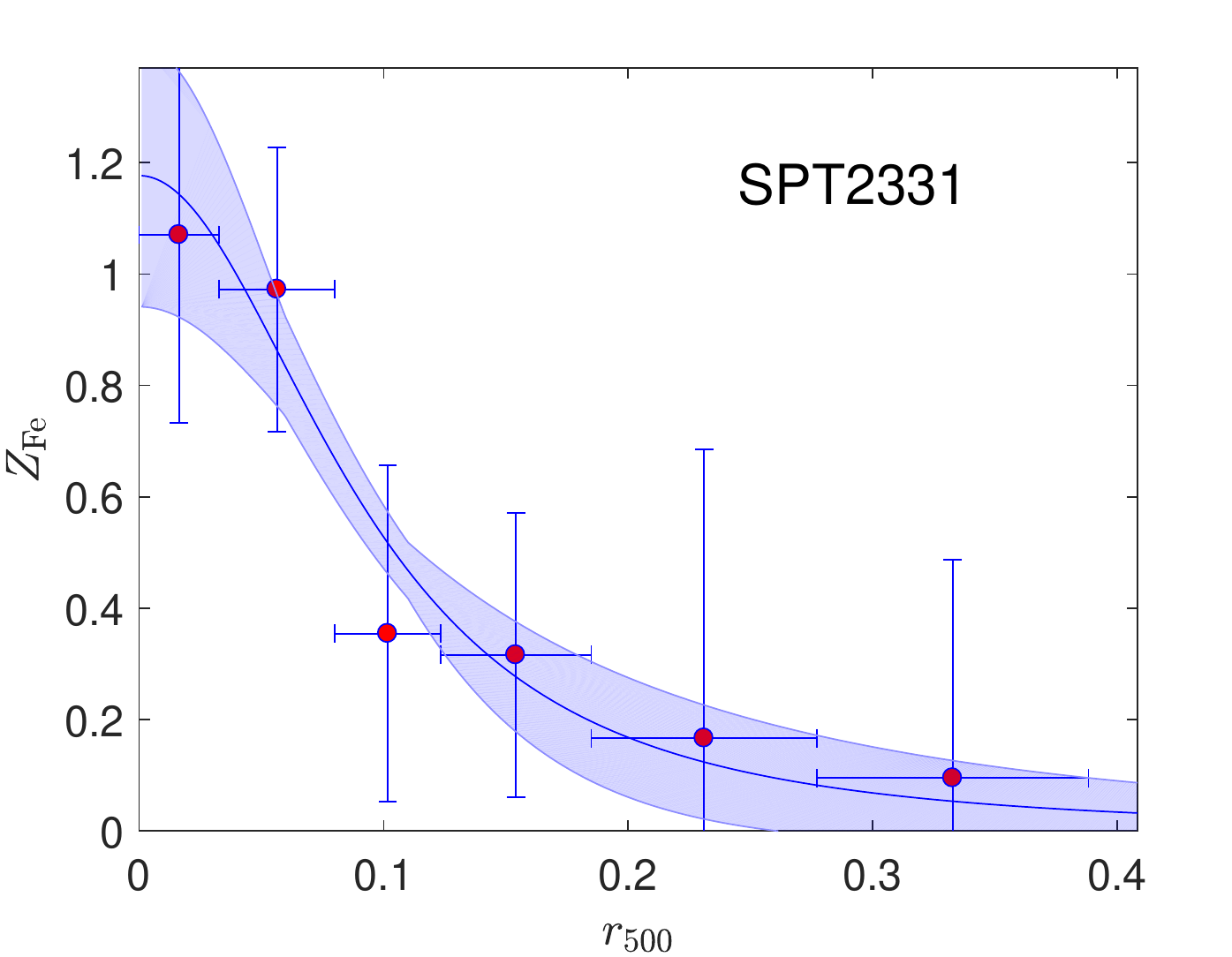}
\includegraphics[width=0.245\textwidth, trim=105 240 105 240, clip]{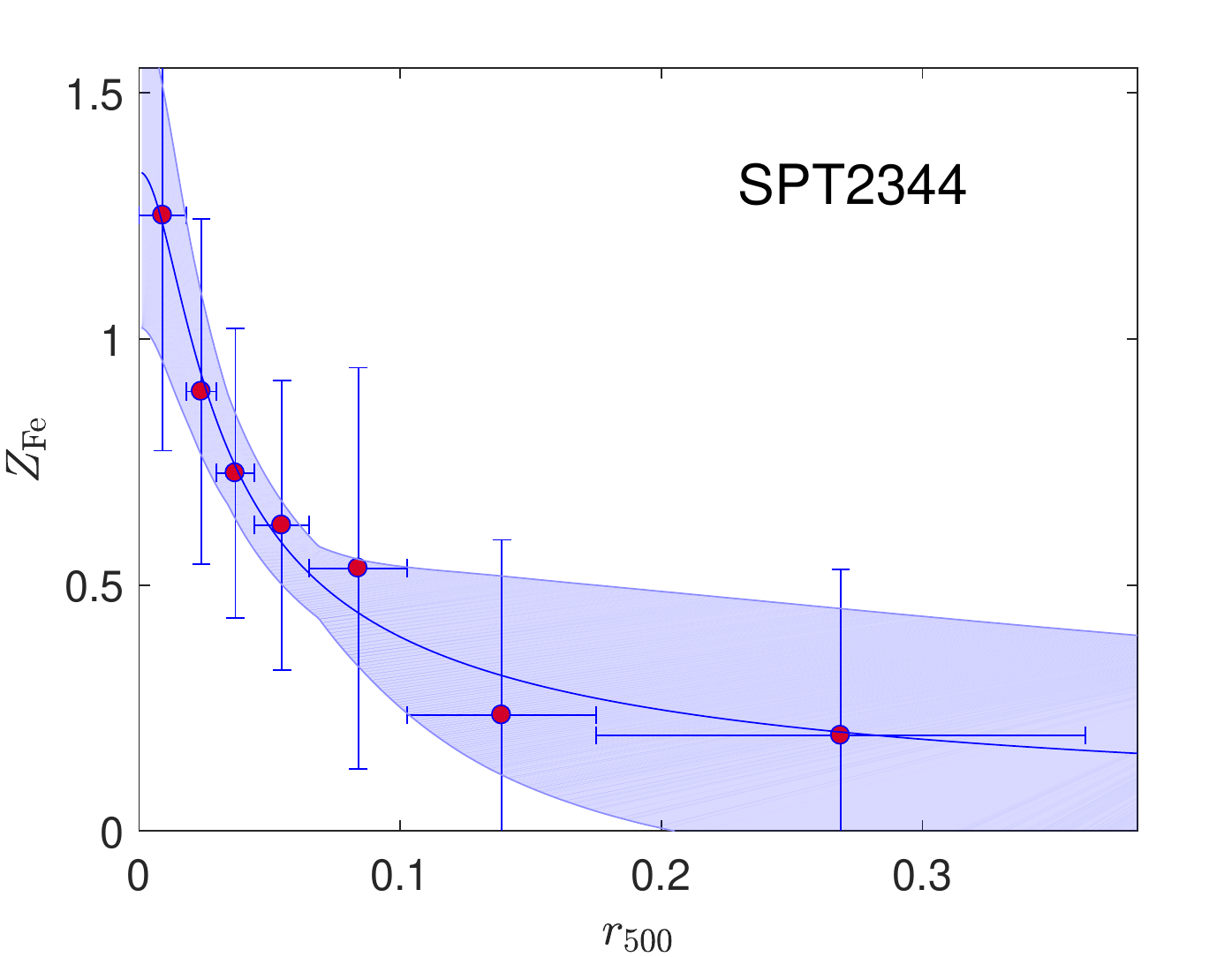}
\includegraphics[width=0.245\textwidth, trim=105 240 105 240, clip]{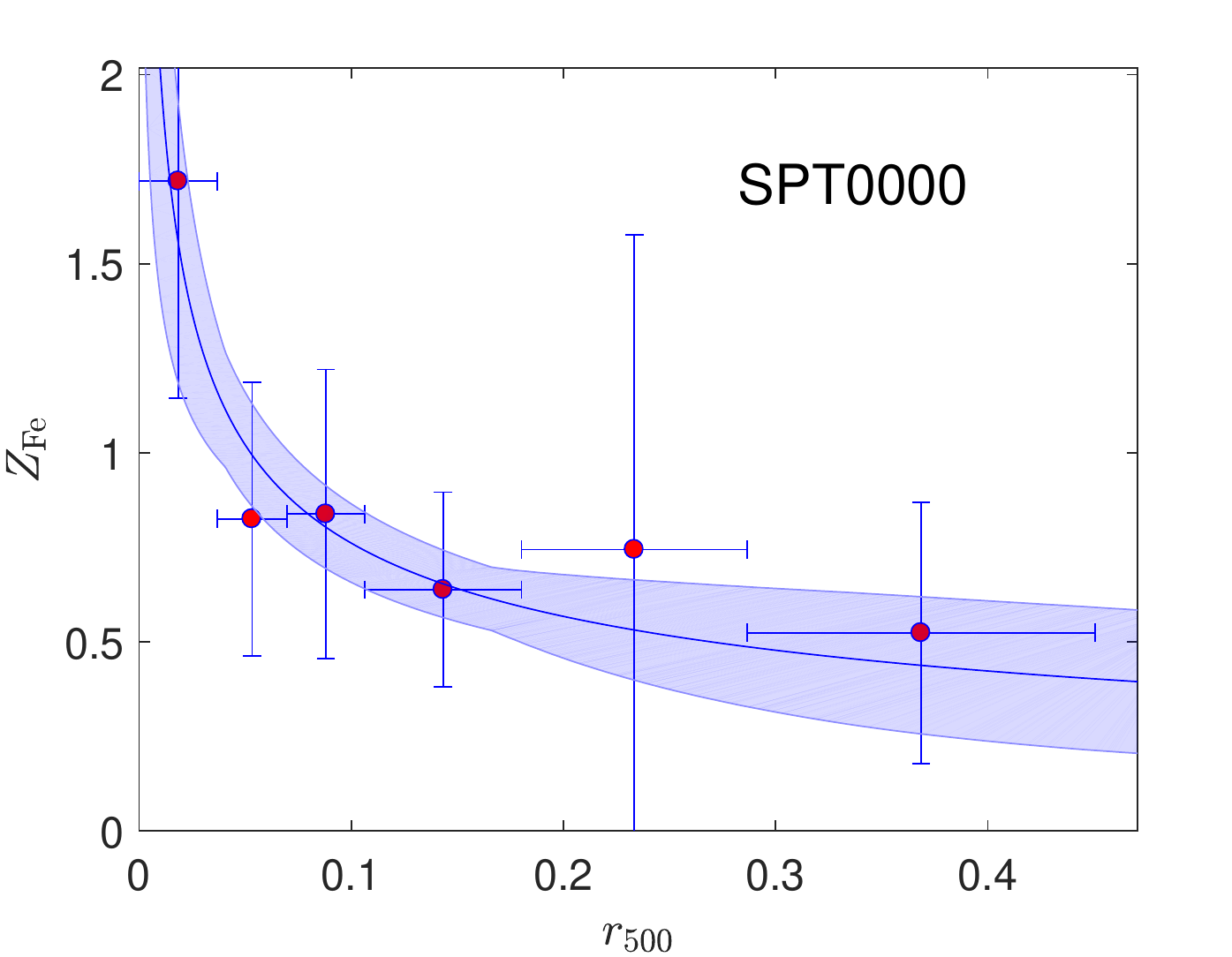}
\includegraphics[width=0.245\textwidth, trim=105 240 105 240, clip]{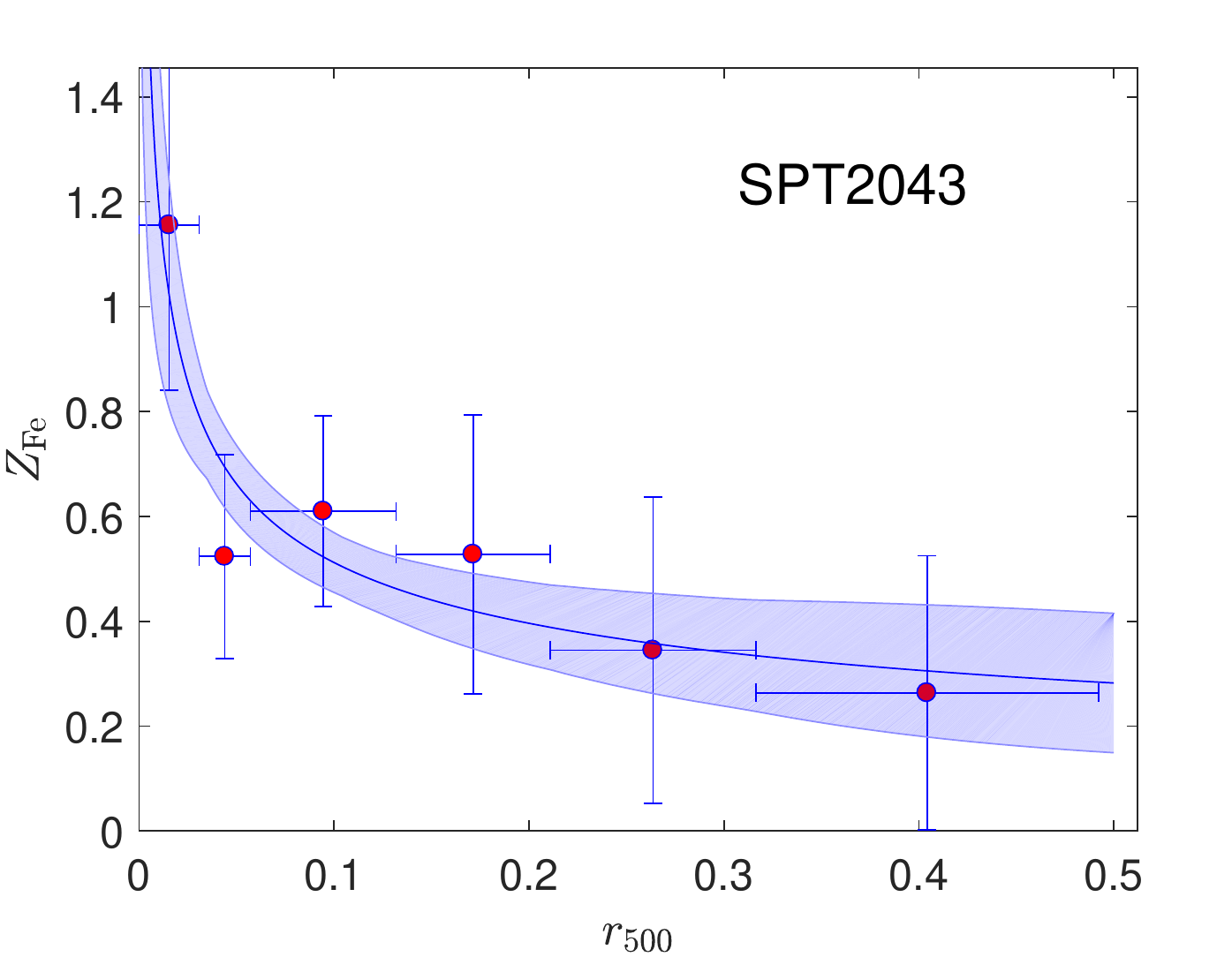}
\includegraphics[width=0.245\textwidth, trim=105 240 105 240, clip]{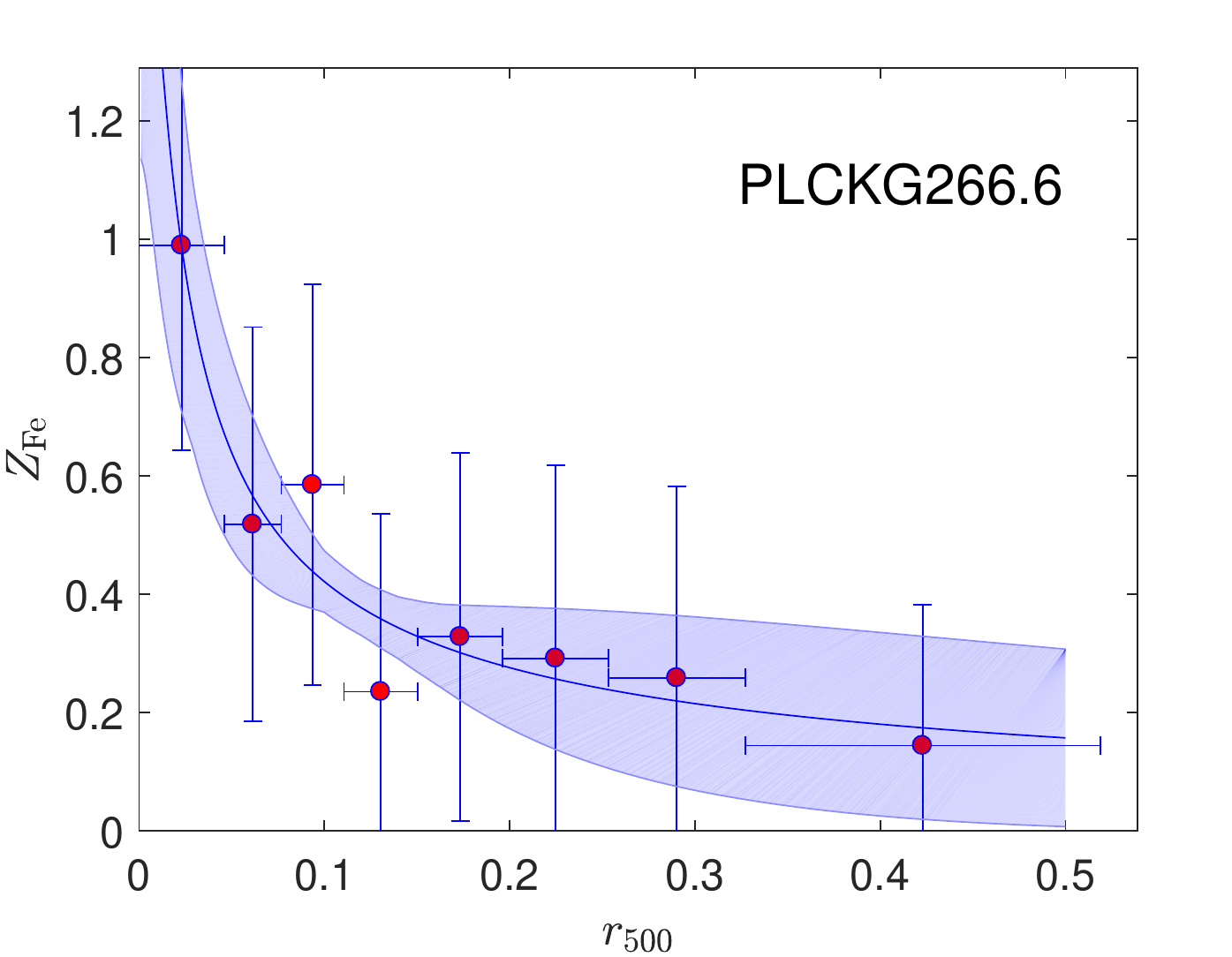}
\includegraphics[width=0.245\textwidth, trim=105 240 105 240, clip]{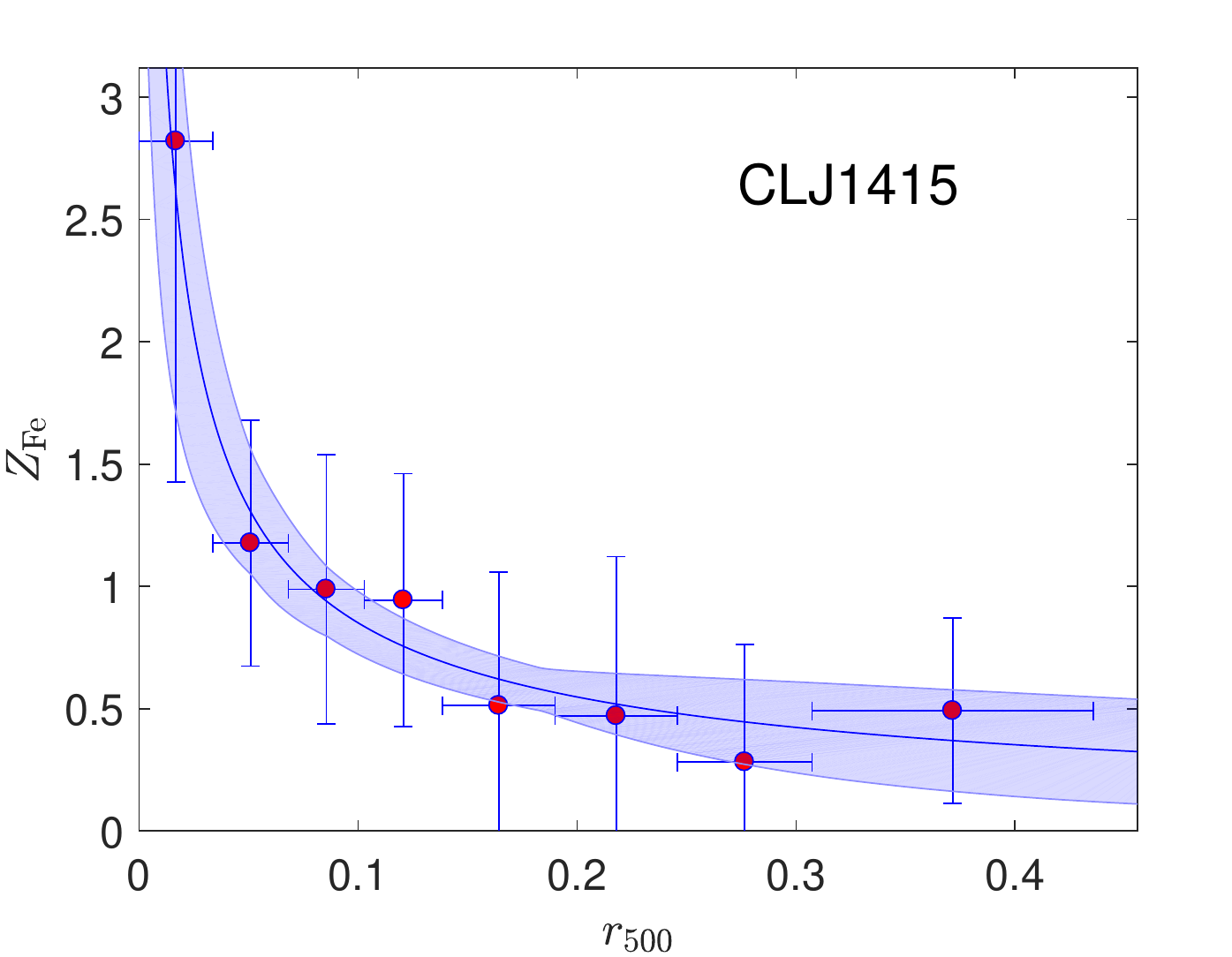}
\caption{Deprojected radial profiles of iron abundance of the clusters in our sample. The solid blue line shows the best fit model. The shaded area corresponds to 1$\sigma$ confidence interval.}
\end{figure*}

\end{document}